\documentclass[letterpaper,twocolumn,10pt]{article}
\PassOptionsToPackage{table}{xcolor}
\usepackage{hyperref}
\makeatletter
\providecommand{\headerps@out}[1]{}
\makeatother
\usepackage{usenix}

\usepackage{tikz}
\usetikzlibrary{arrows.meta,positioning,calc}
\usepackage{amsmath,amssymb}
\usepackage{booktabs}
\usepackage{graphicx}
\usepackage{url}

\usepackage{subcaption}
\usepackage{cleveref}
\usepackage{adjustbox}
\usepackage{xcolor}
\usepackage{enumitem}
\definecolor{fleetgrey}{gray}{0.92}
\newlength{\parheadskip}
\newcommand{\parhead}[1]{\vspace{\parheadskip}\noindent\textbf{#1}}

\begin{document}

\date{}

\title{No Place to Hide: An Analysis on Protected Order Flow Sandwich Attacks}

\author{
{\rm Lioba Heimbach}\\
Category Labs\\
liobaheimbach.cs@gmail.com
\and
{\rm Ozan Solmaz}\\
ETH Zurich\\
osolmaz@ethz.ch
\and
{\rm Burak \"Oz}\\
Flashbots\\
burak@flashbots.net
\and
{\rm Christof Ferreira Torres}\\
INESC-ID \& Instituto Superior Técnico (IST),
University of Lisbon \\
christof.torres@tecnico.ulisboa.pt
}

\maketitle

\begin{abstract}

Front-running has long plagued Ethereum's public mempool, earning it the nickname of a ``dark forest'', where predators lurk for profitable transactions. In response, Ethereum and other blockchain ecosystems increasingly rely on private RPCs and native protections to shield transactions from adversaries, which we refer to as protected order flow. Yet the effectiveness of these mechanisms in preventing front-running, and what trust assumptions they entail, remain poorly understood.

In this work, we conduct the first longitudinal, three-year measurement study of sandwich attacks against protected order flow across six blockchains: Ethereum, Solana, Tron, Base, Arbitrum, and Monad. We introduce detection heuristics that capture wide attacks, both within and across blocks, and filter on bot behavior to distinguish sandwiches from legitimate trading activity. We identify 28.0 million sandwich attacks on Solana, 38,567 on Tron, 30,607 on Ethereum, and 1,889 on Base against transactions intended to be protected from front-running. Reorged blocks expose a further 2,875 Ethereum victims. Unlike conventional public-mempool sandwiches, these attacks rarely occur tightly around their victims and, outside Solana, are carried out by a small number of entities.

Our analysis uncovers exposures at every layer: validator- and application-level exposure on Solana, order-flow auctions and reorged blocks on Ethereum, first-come-first-served ordering that fails to prevent latency-based front-running on Tron, and both an RPC bug that exposes pending transactions and predictable victim behavior on Base. These findings show that existing front-running protections can provide substantially weaker guarantees than users expect, highlighting the need for stronger end-to-end defenses against sandwich attacks.

\end{abstract}

\section{Introduction}

Blockchains initially propagated transactions through public peer-to-peer (P2P) networks, leaving their ordering to block proposers. Bitcoin adopted this model at launch in 2009, followed by Ethereum in 2015. While transaction ordering mattered little for their original payment use cases, the rise of DeFi transformed public mempools into a \emph{dark forest} of front-running bots, with sandwich attacks becoming particularly prominent on Ethereum around 2020~\cite{daian2019flashboys20frontrunning,dark_forest}. In response, blockchain ecosystem has gradually introduced mechanisms to protect users, ranging from private order flow that bypasses Ethereum's public mempool to alternative transaction-submission and ordering designs.

On Ethereum, private RPCs bypass the public mempool and relay transactions directly to block builders, limiting their exposure to front-running. Solana and Monad avoid a public mempool altogether, instead forwarding transactions to upcoming proposers through local mempools. While not designed specifically for sandwich protection, this design substantially reduces pre-inclusion transaction visibility. Tron retains a public mempool but orders transactions by the time the proposer first observes them, claiming to prevent around 98\% of front-running~\cite{sun2024tronmev}. Finally, L2s such as Base and Arbitrum use private mempools in which transactions are visible only to the centralized sequencer until inclusion.

Users therefore increasingly rely on mechanisms that are intended to limit sandwich attacks, yet their effectiveness remains largely unexplored.

In this work, we aim to bridge this gap. We study sandwich attacks targeting transactions protected by private submission or other mechanisms described above, which we term \emph{protected order flow sandwich attacks}. We use \emph{protected order flow} as an umbrella term for transactions subject to a submission or ordering mechanism intended to reduce front-running relative to unrestricted public-mempool ordering. Our three-year longitudinal measurement study across six chains---Ethereum, Solana, Tron, Base, Arbitrum, and Monad--- finds persistent sandwich attacks on all but Arbitrum and Monad.

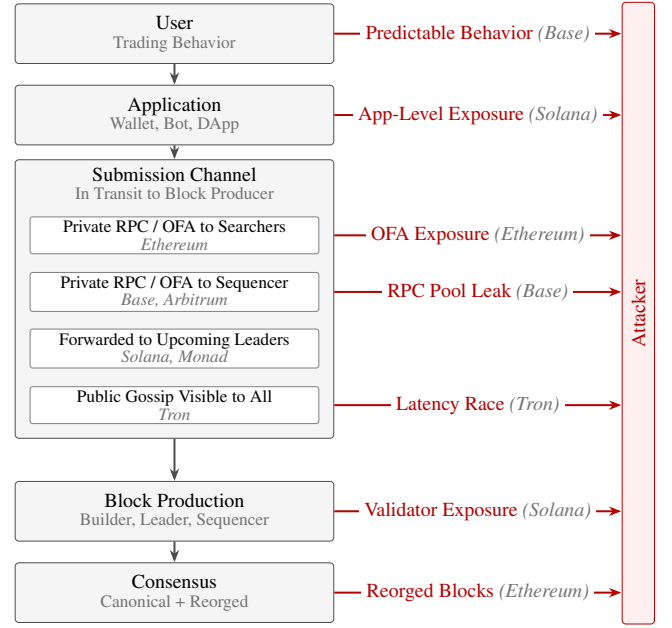
\begin{figure}[t]
  \centering
  \resizebox{\columnwidth}{!}{%
  \begin{tikzpicture}[
    font=\footnotesize,
    stage/.style={draw=black!70, rounded corners=1.5pt, fill=black!4,
      minimum height=24pt, minimum width=128pt, inner xsep=4pt, inner ysep=2pt, align=center},
    lane/.style={draw=black!50, fill=white, rounded corners=1pt,
      font=\scriptsize, inner xsep=3pt, inner ysep=1.5pt, align=center, minimum width=116pt},
    leak/.style={-{Stealth[length=5pt]}, red!70!black, thick},
    flow/.style={-{Stealth[length=5pt]}, black!70, thick},
    tag/.style={font=\footnotesize, red!70!black, align=center, fill=white,
      inner xsep=2pt, inner ysep=1pt},
  ]
  \node[stage] (user) at (0,0)
    {User\\[-2pt]{\scriptsize\textcolor{black!55}{Trading Behavior}}};
  \node[stage] (app) at (0,-1.15)
    {Application\\[-2pt]{\scriptsize\textcolor{black!55}{Wallet, Bot, DApp}}};
  \node[stage, minimum height=112pt] (subm) at (0,-3.75) {};
  \node[font=\footnotesize, align=center] at ($(subm.north)+(0,-0.34)$)
    {Submission Channel\\[-2pt]{\scriptsize\textcolor{black!55}{In Transit to Block Producer}}};
  \node[lane] (l1) at (0,-2.85)
    {Private RPC / OFA to Searchers\\[-1pt]\textcolor{black!55}{\itshape Ethereum}};
  \node[lane] (l2) at (0,-3.65)
    {Private RPC / OFA to Sequencer\\[-1pt]\textcolor{black!55}{\itshape Base, Arbitrum}};
  \node[lane] (l3) at (0,-4.45)
    {Forwarded to Upcoming Leaders\\[-1pt]\textcolor{black!55}{\itshape Solana, Monad}};
  \node[lane] (l4) at (0,-5.25)
    {Public Gossip Visible to All\\[-1pt]\textcolor{black!55}{\itshape Tron}};
  \node[stage] (prod) at (0,-6.75)
    {Block Production\\[-2pt]{\scriptsize\textcolor{black!55}{Builder, Leader, Sequencer}}};
  \node[stage] (chain) at (0,-7.9)
    {Consensus\\[-2pt]{\scriptsize\textcolor{black!55}{Canonical + Reorged}}};
  \draw[flow] (user) -- (app);
  \draw[flow] (app) -- (subm);
  \draw[flow] (subm) -- (prod);
  \draw[flow] (prod) -- (chain);
  \node[stage, draw=red!70!black, fill=red!6, text=red!70!black,
        rotate=90, minimum width=250pt, minimum height=13pt]
        (att) at (6.55,-3.95) {Attacker};
  \foreach \s in {user,app,prod,chain}
    \draw[leak] (\s.east) -- (\s.east -| 6.31,0);
  \foreach \l in {l1,l2,l4}
    \draw[leak] ($(\l.east -| subm.east)$) -- (\l.east -| 6.31,0);
  \node[tag] at ($(user.east)!0.5!(user.east -| 6.31,0)$)
    {Predictable Behavior \textcolor{black!55}{\itshape (Base)}};
  \node[tag] at ($(app.east)!0.5!(app.east -| 6.31,0)$)
    {App-Level Exposure \textcolor{black!55}{\itshape (Solana)}};
  \node[tag] at ($(l1.east -| subm.east)!0.5!(l1.east -| 6.31,0)$)
    {OFA Exposure \textcolor{black!55}{\itshape (Ethereum)}};
  \node[tag] at ($(l2.east -| subm.east)!0.5!(l2.east -| 6.31,0)$)
    {RPC Pool Leak \textcolor{black!55}{\itshape (Base)}};
  \node[tag] at ($(l4.east -| subm.east)!0.5!(l4.east -| 6.31,0)$)
    {Latency Race \textcolor{black!55}{\itshape (Tron)}};
  \node[tag] at ($(prod.east)!0.5!(prod.east -| 6.31,0)$)
    {Validator Exposure \textcolor{black!55}{\itshape (Solana)}};
  \node[tag] at ($(chain.east)!0.5!(chain.east -| 6.31,0)$)
    {Reorged Blocks \textcolor{black!55}{\itshape (Ethereum)}};
  \end{tikzpicture}%
  }
  \caption{
  Protected order flow transaction paths and identified exposure points. The submission-channel covers the four transaction-submission designs across the six chains we study. Red arrow marks an exposure we document.
  }
  \label{fig:leakage-pipeline}
\end{figure}

For each chain, we investigate how these protections fail. A transaction traverses multiple stages before inclusion: (1) the user initiates a trade, (2) an application such as a wallet, trading bot, or DApp constructs and signs the transaction, (3) a submission channel, typically an RPC provider, relays it to the block producer, and (4) the producer orders it into a block. \Cref{fig:leakage-pipeline} illustrates this pipeline, the submission designs of the six chains, and the stage at which we identify exposures.

On Ethereum, private transactions may pass through Order-Flow Auctions (OFAs), which reveal transaction details to searchers bidding to back-run trades for arbitrage. We identify exposures through which OFA information can become sufficient for sandwiching. Reorgs provide a second exposure channel: we identify 2,875 private transactions exposed through reorgs; enabling 2,576 additional sandwiches.

On Solana, transactions are shared with multiple validators, including upcoming proposers. In the first half of our dataset, a subset of validators accounts for a disproportionate share of sandwich attacks, providing evidence of validator-level exposure. From 2025 onward, this signal weakens, while victim concentration shifts toward specific applications, pointing to the application layer as the dominant exposure source.

On Tron, transactions enter a public mempool, while first-come-first-served ordering is intended to prevent a later transaction from overtaking an earlier one. The attacks we observe are consistent with bots exploiting latency races: upon observing a victim transaction, an attacker races its front-run to the block producer despite the ordering protection.

On Base, we identify two distinct exposures. One bot's activity coincides with a documented RPC-level incident that leaked pending transactions, while other attacks exploit highly predictable victim behavior and therefore do not require pending transaction visibility. More broadly, unlike conventional public-mempool sandwiches, protected order flow attacks typically span wider positions or separate blocks, consistent with attackers operating under more limited information.

\parhead{Our Contributions.} We summarize our main contributions:
\begin{itemize}[leftmargin=*,nosep,topsep=0pt]
    \item To the best of our knowledge, we provide the first longitudinal measurement of sandwich attacks across six chains with heterogeneous front-running protection mechanisms.
    \item We introduce novel heuristics to detect non-tight sandwich attacks, including attacks spanning multiple blocks, while filtering bot behavior to distinguish sandwiches from legitimate trading activity.
    \item We conduct a three-year measurement study across six chains, identifying 28.0M protected order flow sandwiches on Solana, 38,567 on Tron, 30,607 on Ethereum, and 1,889 on Base, yielding attackers over \$346M net profit. Reorgs expose a further 2,576 sandwiches on Ethereum.
    \item We trace these attacks to exposures at OFAs and reorgs on Ethereum, first-come-first-served ordering in a public mempool on Tron, validators and applications on Solana, and RPC exposure and predictable behavior on Base.
\end{itemize}

\section{Background}

\subsection{Sandwich Attacks}

We define a sandwich attack as one or more front-running transactions that trade in the same direction as the targets and execute before them, followed by one or more back-running transactions that reverse the position. The front-runs move the pool price against the targets, while the back-runs realize the profit.
We distinguish four types of sandwich attacks. \textbf{Tight} sandwiches place the front- and back-running transactions immediately around a single target transaction within the same block. \textbf{Within-block} sandwiches also occur entirely within a single block and preserve the front-run--target--back-run ordering, but have at least one intervening transaction. \textbf{Cross-block} sandwiches preserve this ordering but span multiple blocks. Finally, \textbf{evasive} sandwiches split front- or back-runs across multiple trades or distribute the attack across separate accounts linked by a token transfer. At scale, we solely observe this pattern on Solana (cf.\ Appendix~\ref{sec:appendix-sol-structure}).

\subsection{Front-Running Protections}
\label{sec:bg-protections}

\tableautorefname{} \ref{tab:front_running_protections} summarizes the studied front-running protections.

\parhead{Private Submission (Ethereum).}
  Users can bypass the public mempool via a private RPC or an OFA, so transactions reach builders without being publicly gossiped. OFAs additionally
expose transaction information to searchers for back-running and return part of
the resulting value to users, combining front-running protection with rebates.

\parhead{Centralized Sequencers (Base, Arbitrum).}
  A single sequencer orders transactions and its input queue is not
  externally observable~\cite{arbitrum_sequencer, op_stack_sequencer},
 leaving no mempool to observe.

\parhead{Local Mempools (Solana, Monad).}
  Clients forward transactions directly to validators/leaders of upcoming slots~\cite{monadlocalmempool,monadraptorcast}, rather than broadcasting them network-wide. Although this design primarily targets performance~\cite{gulfstream}, it nevertheless offers greater front-running protection by reducing the MEV surface area and narrowing searchers' extraction window~\cite{solana_mev_protection}.

\parhead{First-Come-First-Served Ordering (Tron).}
  Transactions enter a public mempool and are included in the block by arrival order~\cite{tronconsensus}, preventing attackers from purchasing priority through a fee-based auction, as on Ethereum.

\begin{table}[t!]
    \centering
    \footnotesize
    \begin{tabular}{@{}l l l l@{}}
        \toprule
        \textbf{Mechanism} & \textbf{Mempool} & \textbf{Chain} & \textbf{Trusted Party} \\
        \midrule
        Private RPC & Private & Ethereum & Builders \\
        OFA & Private & Ethereum & OFA Operator, Searchers \\
        Private RPC & Private & Arbitrum, Base & Sequencer \\
        Private RPC & Local & Monad, Solana & Validators/Leaders \\
        FCFS & Public & Tron & Ordering \\
        \bottomrule
    \end{tabular}
    \caption{Overview of front-running protection mechanisms.}
    \label{tab:front_running_protections}
\end{table}

\subsection{Order-Flow Auctions on Ethereum}
\label{sec:bg-ofas}

OFAs protect users from front-running by privately routing transactions to searchers, who bid to back-run them. Profitable back-runs return part of the proceeds to the user; otherwise, the transaction proceeds as a regular private transaction. They reject harmful bids, such as front-running and sandwiching, regardless of their value~\cite{mevblocker_docs,flashbots_mevshare_intro}, and forward the signed user transaction with the highest harmless bid to builders.
OFAs differ in the amount and type of transaction information disclosed to searchers and in searcher access to transaction feeds (cf.\ \tableautorefname{} \ref{tab:ofa_overview}). We consider four Ethereum OFAs identified by a recent benchmarking study~\cite{janicot2025privatemevprotectionrpcs}:

\parhead{MEV-Share~\cite{flashbots_mevshare_intro}.}
Users control transaction information disclosed to searchers through configurable hints, ranging from partial to full transaction data. A distinguishing feature is its use of \emph{double-hashing}~\cite{mev_share_double_hash}, which applies a second Keccak-256 hash to the transaction hash to preserve user anonymity.

\parhead{MEVBlocker~\cite{mevblocker_docs}.}
Exposes transaction information alongside synthetic or decoy transactions. In swap transactions, it additionally removes sensitive parameters, such as slippage tolerance, to mitigate sandwich attacks. Furthermore, it does not share the transaction with searchers if the likelihood of a back-run is low. Order flow access is controlled through RPC configuration~\cite{mevblocker_docs}, with transactions distributed across permissionless and curated searcher streams via the \texttt{shareAll} and \texttt{shareSafe} flags. Decoy transactions obscure genuine user transactions, reducing the risk of identification and subsequent sandwiching through alternative channels.

\parhead{Blink and Merkle~\cite{blink_docs}.}
Restricts order flow access to approved searchers who must complete an on-boarding process. Approved searchers receive full transaction data, making the on-boarding process the primary protection mechanism.

\begin{table}[t!]
    \centering
    \begin{adjustbox}{width=\columnwidth}
    \begin{tabular}{@{}l l l l@{}}
        \toprule
        \textbf{OFA} & \textbf{Entity} & \textbf{Feed Access} & \textbf{Protection} \\
        \midrule
        MEV-Share~\cite{flashbots_mevshare_intro} & Flashbots & Public & Hints \\
        MEVBlocker~\cite{mevblocker_docs} & CoW Protocol/SMG & Public/Private & Obfuscation \\
        Blink~\cite{blink_docs} & Blink Labs Ltd. & Private & Whitelisting \\
        Merkle* & Blink Labs Ltd. & Private & Whitelisting \\
        \bottomrule
        \multicolumn{4}{c}{*No longer operates as an independent entity. Merged with Blink in 2026.}
    \end{tabular}
    \end{adjustbox}
    \caption{Overview of studied OFA solutions on Ethereum.}
    \label{tab:ofa_overview}
\end{table}

\section{Data Collection}

We collect data for six of the ten largest chains by Total Value Locked (TVL) as of 17 August 2026~\cite{dl_chains}: Ethereum, Solana, Tron, Base, Arbitrum, and Monad.
We select those chains as they either natively mitigate front-running, through private mempools for instance, or expose which transactions received protection.
Our collection covers the largest AMMs on each chain~\cite{dl_dexs}, including Uniswap V2--V4 and forks such as SushiSwap, PancakeSwap, and SunSwap on EVM chains, Aerodrome V1 on Base, and Raydium, Orca Whirlpools, Meteora, and Pump.fun AMM on Solana. We release the full data collection and detection pipeline as open-source code~\cite{artifactrepo}.
Our measurement spans from 1~July 2023 until 30~June 2026, with Monad data beginning at its mainnet launch on 24~November 2025. Appendix~\ref{sec:appendix-data} details data collection, Ethereum mempool labeling, and OFA victim attribution.

\section{Detection}
\label{sec:detection-appendix-body}

At a high level, our heuristics to identify protected order flow sandwich attacks are the same on all six chains.
We detect wide and cross-block attacks alongside tight ones, since an attacker on protected order flow places its legs without seeing the victim and rarely lands directly around it. Ethereum is the main exception, where we restrict the scope to transactions that never entered the public mempool. Solana differs in smaller ways, which account for the attack patterns we observe there and for one leader controlling four consecutive slots.
We detail per-chain parameters in Appendix~\ref{sec:detection-appendix}.

\parhead{H1: Structure.}
A \textbf{candidate} is a swap address that trades in \emph{both directions} on a pool. We scan its swaps in chain order and aggregate consecutive same-direction swaps into a \textbf{front} leg and the following opposite-direction run into a \textbf{back} leg, which catches split legs~\cite{sandwichedme2025solana}. The back legs must resell what the front legs bought, i.e., $A_{\mathrm{b,in}} \in [(1-\epsilon),(1+\epsilon)]\,A_{\mathrm{f,out}}$ with $\epsilon = 0.01$, within five blocks on the EVM chains or two leader rotations on Solana. At least one swap on the same pool in the front direction, by another wallet and not itself a candidate leg, must sit strictly between the legs, and we count each such swap as a victim.

\parhead{H2: One Role.}
A transaction cannot take several roles in the same sandwich, and we disqualify a victim that also serves as a leg of any candidate. Across sandwiches, a swap cannot act as both a front and a back-run.

\parhead{H3: Persistence.}
We attribute each detected sandwich to a bot and filter for the persistent ones. On EVM chains, the bot is a single sender, a recurring sender pair or a contract (cf.\ Appendix~\ref{sec:detection-appendix}). On Solana the bot is the trading wallet that submits the front leg, including in an \textbf{evasive} sandwich where a token transfer links it to a second wallet submitting the back leg (cf.\ Appendix~\ref{sec:appendix-sol-structure}). We retain a bot with at least 100 sandwiches, at least 60\% of them profitable, whose sandwich legs make up at least 25\% of its swaps. An attack is profitable when the round trip gains in the traded token, i.e., $A_{\mathrm{b,out}} > A_{\mathrm{f,in}}$, before gas and fees.

To summarize, H1 identifies the sandwich structure, H2 keeps us from overcounting sandwiches, and H3 filters for persistence. The main difference between our heuristics and those of prior work~\cite{MancinoSandwiched2025,GogolSandwich2026} lies in H3. It keeps us from mislabeling high-frequency trading in both directions as sandwiching. When a bot makes a million swaps and fewer than 1\% of them carry the sandwich structure, those matches are more likely a by-product of its trading than genuine attacks. Appendix~\ref{sec:appendix-ablation} reports an ablation of the three heuristics.

A sandwich can enclose more than one victim when several swaps in the front direction sit between the legs. In such cases, we count everyone as a victim. Yet, we cannot tell which the attacker targeted, as it's likely that only one is the actual victim. For the Ethereum private scope, this means a sandwich may enclose several private victims. Multi-victim sandwiches are rare on the EVM chains, with only a single victim sitting in 88.3\% of sandwiches on Ethereum, 97.5\% on Tron, and 98.8\% on Base, whereas on Solana, 36.0\% of sandwiches enclose more than one victim. We address this limitation for the victim analyses that follow in the corresponding sections.

\section{Protected Order Flow Sandwiches}
\label{sec:overview}

We characterize protected order flow sandwich attacks across the six chains in our study and contrast their structure with public-mempool sandwiches previously studied on Ethereum.

Recall the protections at play. Ethereum lets users submit transactions privately, and good data coverage exists on which ones did. Unless stated otherwise, we consider only victims of these private Ethereum transactions. Solana and Monad run local mempools, i.e., a transaction goes to a small set of upcoming block proposers instead of to everyone. Tron sequences its public mempool first-come, first-served, with proposers ordering transactions by arrival time. Base and Arbitrum route transactions to a single sequencer.

Table~\ref{tab:summary} holds the summary statistics for protected order flow sandwich attacks across the six chains. The number of attackers and the number of attacks split the chains into four groups. Solana stands alone with 28.0 million sandwiches by 8,631 bots. Ethereum and Tron follow with 7 and 12 bots and tens of thousands of attacks, i.e., 30,607 sandwiches on Ethereum and 38,567 on Tron. Base forms a third group with 4 bots and 1,889 attacks. Arbitrum and Monad make up the final group without persistent sandwich attackers.

\begin{table}[t]
  \centering
  \resizebox{\columnwidth}{!}{\setlength{\tabcolsep}{5pt}
\begin{tabular}{@{}lrrrrrr@{}}
\toprule
 & \multicolumn{3}{c}{\textbf{Attacks}} & \multicolumn{3}{c}{\textbf{Profit (USD)}} \\
\cmidrule(lr){2-4}\cmidrule(l){5-7}
\textbf{Chain} & \textbf{Count} & \textbf{Entities} & \textbf{Profitable} & \textbf{Gross} & \textbf{Net} & \textbf{Priced} \\
\midrule
Ethereum & 30,607 & 7 & 91.9\% & 622,358 & 254,566 & 99.96\% \\
\quad Via reorged blocks$^{\dagger}$ & 2,576 & 93 & 98.9\% & 279,848 & 171,777 & 45.7\% \\
Tron & 38,567 & 12 & 95.4\% & 901,069 & 642,542 & 90.8\% \\
Base & 1,889 & 4 & 97.2\% & 3,031 & 2,416 & 99.7\% \\
Solana & 28,042,725 & 8,631 & 88.0\% & 383,433,932 & 345,185,014 & $>$99.99\% \\
Arbitrum & 0 & 0 & -- & -- & -- & -- \\
Monad & 0 & 0 & -- & -- & -- & -- \\
\bottomrule
\end{tabular}
}
  \caption{Protected sandwich attacks by persistent bots per chain, 1~July 2023 to 30~June 2026. \emph{Profitable} is the share whose round trip gained in the traded token, \emph{priced} the share with a USD value, over which gross and net are summed. The reorged-block row lists sandwiches on private transactions that a stale block likely made public (cf.\ Section~\ref{sec:eth-reorgs}). $^{\dagger}$Profit and profitable share cover only the 45.7\% of attacks whose fee is attributable to a single attack, the rest are multi-pool bundles (cf.\ Appendix~\ref{sec:appendix-public}).}
  \label{tab:summary}
\end{table}

Protected order flow sandwiching is rampant on Solana, with 8,631 bots extracting \$383.4 million in gross profit and retaining \$345.2 million after fees, which take 10.0\% of gross profit.

Turning to Ethereum, we find 30,607 sandwiches by 7 bots. Fees cut the profit from \$622,358 to \$254,566, i.e., 59.1\% of the gross profit goes to fees. Note that these are only the sandwiches of bots specializing in private order flow. Sandwiches originating in the public mempool are covered in Appendix~\ref{sec:appendix-public}.

A further 2,576 sandwiches executed by 93 bots land on 2,875 private transactions that a reorged block likely made public, i.e., the block carrying them was broadcast and then reorged out (cf.\ Section~\ref{sec:eth-reorgs}). Note that we price only 1,178 of the attacks, and the remaining bot transactions serve multiple attacks across pools (cf.\ Appendix~\ref{sec:appendix-public}). For the priced attacks, the gross profit is \$279,848, while the net is \$171,777.

Tron operates on a similar scale with 38,567 attacks by 12 bots, at a higher gross profit of \$901,069 and a higher net profit of \$642,542, i.e., only 28.7\% of the gross profit goes to fees. Base sits far below both, with 1,889 attacks by 4 bots and a gross profit of \$3,031, of which \$2,416 remains after fees, i.e., 20.3\% of the gross profit goes to fees.

\begin{figure*}[t]
  \centering
  \begin{subfigure}[t]{0.48\textwidth}
    \centering
    \includegraphics[width=\linewidth]{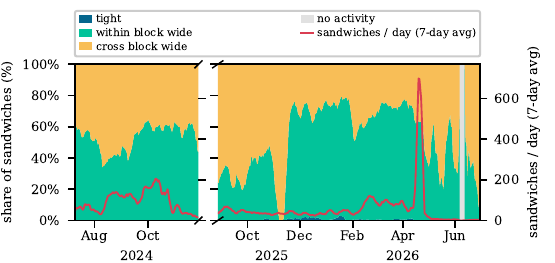}
    \caption{Ethereum}
    \label{fig:eth-per-day}
  \end{subfigure}\hfill
  \begin{subfigure}[t]{0.48\textwidth}
    \centering
    \includegraphics[width=\linewidth]{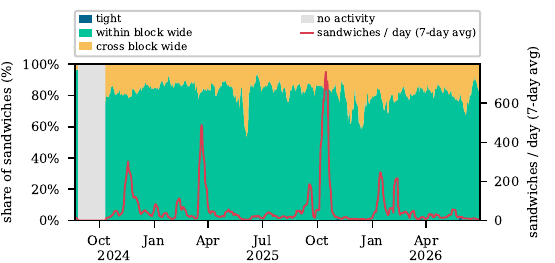}
    \caption{Tron}
    \label{fig:tron-per-day}
  \end{subfigure}

  \begin{subfigure}[t]{0.48\textwidth}
    \centering
    \includegraphics[width=\linewidth]{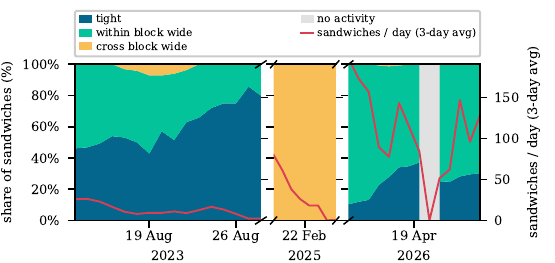}
    \caption{Base}
    \label{fig:base-per-day}
  \end{subfigure}\hfill
  \begin{subfigure}[t]{0.48\textwidth}
    \centering
    \includegraphics[width=\linewidth]{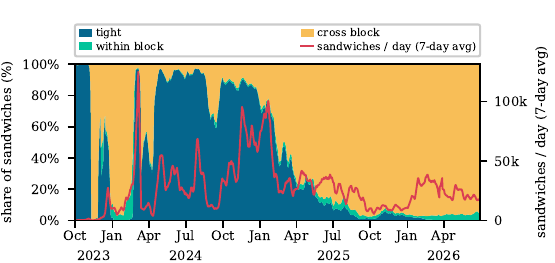}
    \caption{Solana}
    \label{fig:sol-daily}
  \end{subfigure}
  \caption{Sandwich attacks per day, per chain. The shaded area gives the attack-type composition (left axis) and the line the daily attack count (right axis), both smoothed over a rolling window. Gray marks
  days with no attacks.}
  \label{fig:per-day}
\end{figure*}

Two chains show no protected sandwiches. Arbitrum's protection design resembles Base's at a high level, with a centralized sequencer, and yet we observe no protected sandwiches on Arbitrum. Monad runs a local mempool design similar to Solana's and likewise shows none. We note that Monad has been live for less than a year, whereas Solana launched in March 2020, and the first protected sandwich we observe falls on 16 August 2023, i.e., years into its operation.\footnote{Our Solana window opens on 1~July 2023, six weeks before that first attack. Given how slowly attacks ramp up thereafter (cf.\ Figure~\ref{fig:sol-daily}), we consider attacks before our window unlikely.} Thus, even though the protections on Arbitrum and Monad mirror those on Base and Solana, respectively, we do not see them exploited in the same way. This observation points to a theme that recurs throughout our analysis: attack counts capture what attackers do today, not what is in theory possible.

The share of profitable attacks in Table~\ref{tab:summary} distinguishes protected sandwiches from Ethereum public mempool sandwiches. Across the four chains with protected sandwiches, the profitable share ranges from 88.0\% on Solana to 97.2\% on Base, with Ethereum at 91.9\%. Public mempool sandwiches on Ethereum reach 98.9\% (cf.\ Appendix~\ref{sec:appendix-public}). Figure~\ref{fig:profit-violin} in Appendix~\ref{sec:appendix-structure} further reports per-attack profit distribution per chain. Attacking protected flow likely leaves the attacker with less control over the execution order.

Thus, we turn to the structure of protected sandwiches. Figure~\ref{fig:per-day} plots the number of protected sandwiches per day together with their structure. Our analysis spans 1~July 2023 to 30~June 2026 for all chains, and we restrict each panel to the periods with sandwich attacks by real bots.

On Ethereum we observe two periods of protected sandwich activity. The first sets in at the middle of 2024 and averages 96 attacks per active day, of which 53.4\% are within-block wide and 46.6\% cross-block wide. The second averages 59 attacks per day at a comparable split of 59.1\% and 40.5\%. It contains a single spike in April 2026, where the seven-day average peaks at 698 attacks per day before falling back. Tight sandwiches are extremely rare (0.25\% of attacks).

Tron follows a similar pattern, with activity sustained from the end of 2024 to the middle of 2026. Attacks average 62 per active day, and 82.9\% of them are within-block wide against 17.0\% cross-block wide, the remainder is tight. On Tron most attacks are inside a single block, whereas half are spread across block boundaries on Ethereum.

On Base, activity is confined to three short bursts. The first one runs for 16 days in August 2023, with 13 attacks per day, and is the only period on Base in which tight sandwiches dominate, at 56.4\%. The second lasts eight days in February 2025, with 30 attacks per day, and is entirely cross-block wide. The third covers two weeks in April 2026 and accounts for 76.7\% of all Base attacks, with 76.0\% within-block-wide and 23.8\% tight. Each period carries the signature of a short-lived vulnerability and a distinct bot (cf.\ Section~\ref{sec:origins-base}).

Solana has the most sandwiches per day, averaging 25{,}586 attacks. Activity starts at the end of 2023 and peaks at 125{,}169 just before Jito shuts down their temporary public mempool. Sustained activity continues at an average of 41{,}063 attacks per day until January 2025 and is dominated by tight sandwiches. It then declines to 26{,}127 per day through 2025 and to 20{,}663 after the last wave of validator delistings in October 2025 (cf.\ Section~\ref{sec:origins-solana}), shifting to cross-slot attacks.

Overall, tight protected sandwiches are rare, with Solana before 2025 and a short period on Base being notable exceptions. In the public mempool on Ethereum, the majority of sandwiches are tight, i.e., the front-running transaction, the victim transaction, and the back-running transaction land one after the other. An attacker there sees the victim in the mempool and generally submits both legs around it in a single bundle through an MEV auction~\cite{WeintraubPrivate2022,li2023demystifying}. An attacker on protected flow, on the other hand, often places its legs blind, as we will see next. Consequently, unrelated transactions may land between them or the legs can fall in separate blocks.

\begin{figure}[t]
  \centering
  \includegraphics[width=\linewidth]{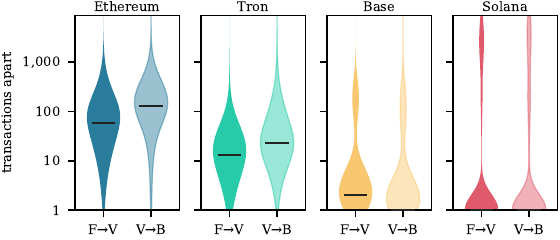}
  \caption{Transactions between the legs of a sandwich attack. F$\rightarrow$V, from the front-running transaction to the first victim; V$\rightarrow$B, from that victim to
  the back-running transaction.}
  \label{fig:txs-apart}
\end{figure}

\Cref{fig:txs-apart} reports the distance from the front-run transaction to the first victim, and from that victim to the back-run transaction. Bots on Ethereum and Tron spread their legs widely, at medians of 58 and 131 transactions on Ethereum and 13 and 23 on Tron. For Base the medians sit at 2 and 1 transactions, and on Solana at 1 and 1. Solana's median comes from the period up to 2025, when tight sandwiches dominated, and the legs move apart once the attacks turn cross-block afterwards. Both chains still carry long tails, i.e., a 99th percentile of 314 on Base and 8,651 on Solana. Thus, protected order flow sandwich attacks are in stark contrast with Ethereum public mempool sandwiches, which have a median of 1 transaction and 99.4\% of them within 6 (cf.\ Appendix~\ref{sec:appendix-public}).

We further report the number of transaction between the front and back leg in Appendix~\ref{sec:appendix-structure} which in addition to the distance also reveals whether it has an effect on the attack. Distance alone does not make an attack suboptimal. A sandwich with 100 transactions between its legs still extracts the full amount as long as none of them swaps in the same pool. Any such swap in between, however, has the potential to make the attack unprofitable. We call an attack interfered with when any trade in the attacked pool other than its first victim executes between the two legs. We find that this interference is not rare, and it is most common on Solana, where 44.6\% of attacks have an additional swap in the attacked pool between their legs, against 28.4\% on Ethereum, 10.3\% on Tron and 4.4\% on Base (cf.\ Figure~\ref{fig:pool-swaps-between}). Interference separates the failed attacks from the successful ones, i.e., a swap between the legs appears in 96.2\% of unprofitable attacks on Ethereum against 22.4\% of profitable ones, in 91.0\% against 6.5\% on Tron, and in 88.7\% against 2.0\% on Base.

To summarize, protected order flow sandwiches look nothing like their public mempool counterparts. Their legs sit far from the victim, and often in separate blocks, and their lower share of profitable attacks likely follows from the same gap, since the attacker acts on incomplete information.

\section{Exposure Paths and Attacker Competition}
\label{sec:origins}

We next examine how sandwich bots front-run seemingly protected order flow and compete with one another.

\subsection{Ethereum}
\label{sec:origins-eth}

\begin{table}[t]
  \centering
  \resizebox{\columnwidth}{!}{\begin{tabular}{@{}lrcrrrc@{}}
\toprule
\textbf{Bot} & \textbf{SWs} & \textbf{T/W/C (\%)} & \textbf{Gross (USD)} & \textbf{Fees} & \textbf{Succ.} & \textbf{Active} \\
\midrule
\rowcolor{fleetgrey}\href{https://etherscan.io/address/0x841b6ff0979f86bd76579a3cd03fa43a6a1bb697}{\texttt{0x841...697}} & 13,661 & 1 / 67 / 33 & 217,579 & 37\% & 86\% & 25-11--26-06 \\
\rowcolor{fleetgrey}\href{https://etherscan.io/address/0x8a5d8700dd19488d964d202e87b0cc85704054db}{\texttt{0x8a5...4db}} & 10,202 & 0 / 55 / 45 & 240,985 & 86\% & 77\% & 24-08--24-11 \\
\rowcolor{fleetgrey}\href{https://etherscan.io/address/0xe87a2343f12a9d812a9e87f01118b6c7c5f6b716}{\texttt{0xe87...716}} & 3,031 & 0 / 48 / 52 & 92,717 & 53\% & 74\% & 24-07--24-08 \\
\rowcolor{fleetgrey}\href{https://etherscan.io/address/0x69b03f6548f16c44b77cd974197e2b8b5ba00b78}{\texttt{0x69b...b78}} & 1,563 & 0 / 29 / 70 & 35,112 & 44\% & 81\% & 25-10--25-11 \\
\rowcolor{fleetgrey}\href{https://etherscan.io/address/0x013a9088670ffdeec705a3dc32d5229ab5ab15ea}{\texttt{0x013...5ea}} & 983 & 0 / 31 / 69 & 14,248 & 56\% & 80\% & 25-08--25-09 \\
\rowcolor{fleetgrey}\href{https://etherscan.io/address/0xe9becf8c41cee2b777dee36775613209b5273e02}{\texttt{0xe9b...e02}} & 808 & 0 / 23 / 77 & 18,462 & 35\% & 82\% & 25-09--25-10 \\
\rowcolor{fleetgrey}\href{https://etherscan.io/address/0xc72a44de529163b3ecd26e24f8eda50d7943859b}{\texttt{0xc72...59b}} & 359 & 0 / 55 / 45 & 3,255 & 30\% & 91\% & 26-04--26-06 \\
\bottomrule
\end{tabular}
}
  \caption{Ethereum bots, sandwiches (SWs), tight (T) / within-block-wide (W) / cross-block-wide split (C), gross profit in USD, the share of it consumed by fees, success rate (\%), and active period. Rows with a gray background mark the bots we attribute to a single entity.}
  \label{tab:bots-eth}
\end{table}

On Ethereum, we observe seven sandwich attack bots that specialize in front-running transactions that never enter the public mempool. Table~\ref{tab:bots-eth} reports per-bot statistics. The two largest bots dominate, with 13,661 and 10,202 sandwiches, totaling 78.0\% of the 30,607 attacks. Tight sandwiches are almost absent. The within-block share ranges from 23\% to 67\%, with the remainder falling across block boundaries. Success rates fall within a similar range, from 74\% to 91\%. Profit follows the attack count only loosely. For example, \href{https://etherscan.io/address/0x8a5d8700dd19488d964d202e87b0cc85704054db}{\texttt{0x8a5...4db}} earns a higher gross profit than \href{https://etherscan.io/address/0x841b6ff0979f86bd76579a3cd03fa43a6a1bb697}{\texttt{0x841...697}} on 25\% fewer sandwiches. The fee share separates the bots by period rather than by size. The bots active in 2024 give up the most, at 86\% and 53\% of their gross profit, whereas the two active into 2026 give up 37\% and 30\%. Ethereum fees fell over this period, which likely accounts for the difference.

\begin{figure}[t]
    \centering
    \includegraphics[width=1.0\linewidth]{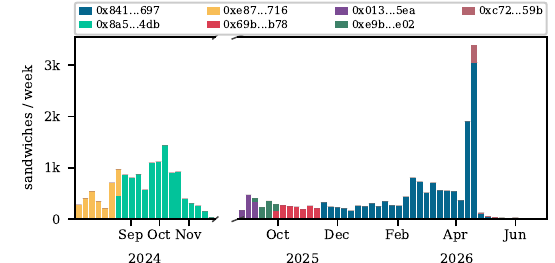}
    \caption{Weekly sandwiches on Ethereum by attacker bot.}
    \label{fig:eth-weekly-by-bot}
\end{figure}

Finally, the bots are generally active over non-overlapping windows. \Cref{fig:eth-weekly-by-bot} plots the weekly sandwich count with each bot shown separately. The bots hand over from one to the next inside a single week, and no two of them run alongside each other for longer. The exceptions are \href{https://etherscan.io/address/0x841b6ff0979f86bd76579a3cd03fa43a6a1bb697}{\texttt{0x841...697}} and \href{https://etherscan.io/address/0xc72a44de529163b3ecd26e24f8eda50d7943859b}{\texttt{0xc72...59b}}, which are active simultaneously for one week without following the handover pattern of the others. The handover pattern alone suggests that the first six bots are controlled by one entity. Bytecode similarity and deployment history indicate that, in fact, all seven bots belong to a single entity (cf.\ Appendix~\ref{sec:appendix-eth-clustering}). MEV is already known to be highly concentrated, yet for protected order flow sandwiches, a single entity is often the dominant case. This concentration is repeatedly observed throughout the paper.

\subsubsection{Ruling Out Block Builders}

On Ethereum, proposer-builder separation (PBS) separates block construction from block proposal~\cite{HeimbachKTW23}. Builders assemble transactions into candidate blocks and compete by submitting bids to proposers, who select the highest-bidding block for inclusion. As builders control transaction ordering, they occupy a priveleged position for MEV extraction, including sandwich attacks.
To understand how transactions that never entered the public mempool were nevertheless front-run, we first examine whether protected sandwich attacks concentrate in the blocks of particular builders, given the emergence of integrated searcher-builder constellations~\cite{HeimbachNonAtomic2024}. Because builders receive private transactions in plaintext, they are an obvious potential exposure point in the trust model. However, no builder exhibits the clear over-representation that would be expected from a builder-specific exposure source (cf.\ Appendix~\ref{sec:appendix-eth-builders}). We therefore move upstream in the transaction-submission pipeline of \Cref{fig:leakage-pipeline} and examine the channels through which private transactions reach builders.

\subsubsection{OFA Attribution}\label{sec:ofa-attribution}
We turn to OFAs and examine the proportion of victim transactions that can be attributed to them.
Among the 34{,}360 victims of 28{,}128 successful private sandwich attacks, 66.8\% appear in at least one OFA dataset, with 35.5\% appearing in multiple ones. We additionally identify 5{,}101 victims across 2{,}479 unsuccessful private sandwich attempts, of which 35.5\% appear in at least one OFA dataset. These attempts satisfy the same structural criteria as successful sandwiches, but the attacker's front--back sequence fails our profitability criterion. We include successful and unsuccessful attempts in the following exposure analysis because, regardless of realized profit, an attempted sandwich indicates that the attacker was able to act around the victim transaction.

\begin{figure}[t]
  \centering
  \includegraphics[width=\linewidth]{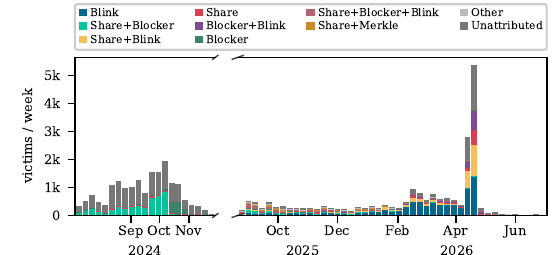}
  \caption{
  Weekly evolution of sandwich victims by observed OFA label combination for the persistent bots in Table~\ref{tab:bots-eth}. \emph{Share} denotes MEV-Share and \emph{Blocker} denotes MEVBlocker. Victims may appear in multiple datasets; for example, Share+Blocker denotes victims observed in both. \emph{Unattributed} denotes victims not matched by any of the four datasets, while \emph{Other} aggregates all remaining combinations, each accounting for less than 1\% of victims overall.
  }
  \label{fig:eth-ofa-evolution}
\end{figure}

Across all sandwich attempts, 62.7\% of victims appear in at least one OFA dataset. Through February 2025, we attribute 6{,}870 of 17{,}614 victims to at least one OFA, of which 70.4\% appear jointly in MEV-Share and MEVBlocker. From August 2025 onward, we observe 21{,}847 victims, of which 17{,}889 are attributed to at least one OFA. Of these, 86.2\% appear in the Blink data, while 46.2\% are seen exclusively in Blink.
\Cref{fig:eth-ofa-evolution} shows the weekly evolution of OFA labels, with combinations accounting for less than 1\% grouped as \emph{Other}.

These temporal patterns should be interpreted in light of our data coverage. In particular, the Blink data begins only in August 2025, while 61.0\% of victims up to February 2025 remain unattributed. More generally, the 37.3\% of victims without an OFA attribution over the full period may still have passed through an OFA that is not captured by our external datasets. As discussed in Appendix~\ref{par:ofa-coverage}, these datasets provide positive evidence of observed OFA exposure but cannot establish that a transaction did not appear elsewhere. We therefore use the observed changes in OFA composition to motivate the mechanism-level analysis below, rather than as evidence of changes in underlying OFA usage. The full breakdown of observed OFA label combinations is reported in \Cref{tab:ofa-involvement-breakdown}.

A further limitation is the presence of multi-victim sandwiches. Our methodology treats each qualifying transaction between the attacker legs as a victim, although in a multi-victim sandwich only one transaction may have been deliberately targeted and the others may have been included incidentally. Overall, 68.5\% of victims on Ethereum are the sole target of their sandwich, while the remainder appear in multi-victim sandwiches. This may explain part of the unattributed population.
Among victims without an observed OFA label, 56.5\% occur in multi-victim sandwiches and therefore may not themselves have been the transaction whose exposure enabled the attack.
Conversely, an OFA-attributed transaction in a multi-victim sandwich may also have been included incidentally rather than exploited through its OFA submission path. We report the single-victim share for each OFA label combination in \Cref{tab:ofa-involvement-breakdown} and account for this distinction in the mechanism-level analysis below.

\subsubsection{OFA Exposure Mechanisms}\label{sec:ofa-exposure-mechanisms}
In the following, we study how sandwich bots may exploit information exposed through OFAs.

\parhead{Cross-OFA Correlation.}
Overall, 53.0\% of all attributed sandwich victims appear in multiple OFA datasets. This is particularly relevant for MEV-Share and MEVBlocker, whose disclosure mechanisms can be correlated to reveal additional information. MEV-Share selectively discloses transaction information together with a double-hashed transaction identifier~\cite{mev_share_double_hash}. MEVBlocker mixes genuine transactions with decoys, and for swap transactions, withholds sensitive information. Transactions unlikely to yield a back-run are not exposed~\cite{mevblocker_docs}. A searcher observing both feeds can double-hash transaction hashes exposed by MEVBlocker and compare them against the MEV-Share stream. A match identifies a genuine MEVBlocker transaction and can reveal information unavailable from either feed in isolation.

Among attributed victims, 28.3\% appear in both MEV-Share and MEVBlocker, of which 75.3\% are in no other OFA (cf.~\Cref{tab:ofa-involvement-breakdown}). Importantly, 88.6\% of victims in this MEV-Share--MEVBlocker-only category occur in single-victim sandwiches, making incidental inclusion unlikely. Together, the prevalence of this combination and its concentration among sole-target victims make cross-OFA correlation a plausible exposure mechanism for a substantial class of attacks.

\parhead{Full-Transaction Disclosure.}
Blink and Merkle follow a different security model. Rather than obscuring transaction contents from searchers, they restrict order flow access to an approved set of participants, to whom the full transaction is disclosed. Thus, once a searcher has access to the feed, no additional cross-feed inference is needed to identify or size a potential sandwich. Instead, protection depends on admitted searchers' behavior and the access policy enforcement.

This model is relevant for a substantial fraction of attributed victims. Blink appears in 62.2\% of attributed victims, and 33.4\% of attributed victims are observed only in Blink (cf.~\Cref{tab:ofa-involvement-breakdown}). Among Blink-only victims, 89.8\% occur in single-victim sandwiches, making incidental inclusion unlikely.

Blink also appears jointly with other OFAs in several large categories, including Blink--MEV-Share (14.3\% of attributed victims), Blink--MEVBlocker (6.9\%), and Blink--MEV-Share--MEVBlocker (5.8\%). In these cases, Blink's full-transaction disclosure is already sufficient to expose the information needed for a potential sandwich. The additional OFA observations are therefore not necessary to explain informational sufficiency. They may nevertheless reflect other routing or exposure paths. These observations do not establish that the attacker obtained the transaction from Blink, since other unobserved exposure paths may exist. Nevertheless, they show that a large class of victim transactions was available in full to Blink's approved searchers before inclusion.

Merkle follows the same disclosure model, although it appears in only 5.1\% of attributed victims and was subsequently merged into Blink. Only 104 victims are observed exclusively in Merkle, corresponding to 0.4\% of attributed victims. Among these Merkle-only victims, 43.3\% occur in multi-victim sandwiches, making the evidence for Merkle as the exposure source less direct in these cases.

\parhead{Unresolved Single-OFA Exposure.}
We also observe victims associated with MEV-Share or MEVBlocker in isolation in our attribution data: 1{,}727 MEV-Share-only victims (7\% of attributed) and 1{,}553 MEVBlocker-only victims (6.3\%;
cf.\ \Cref{tab:ofa-involvement-breakdown}). \Cref{fig:mevshare_mevblocker} shows that these cases are strongly concentrated in time. MEVBlocker-only victims exhibit a short-lived spike around October--November 2024 and are nearly absent thereafter. MEV-Share-only victims are more dispersed, but show a pronounced spike in April 2026.

For these isolated labels, we do not identify an exposure mechanism analogous to those described above. An additional exposure source may therefore be missing from our data. For the MEVBlocker-only cases, we cannot rule out exposure through another OFA for which we lack a corresponding label. The MEV-Share-only spike in April 2026, in turn, coincides with a period of high Blink-associated sandwich activity. Because our Blink labels are derived from bundles observed by BuilderNet, Blink bundles not observed by BuilderNet would be absent from our attribution data.

Multi-victim sandwiches provide another possible explanation, particularly for the MEV-Share-only cases. We find that 70.5\% of MEV-Share-only victims occur in multi-victim sandwiches, and may therefore have been included incidentally rather than being the transaction whose exposure enabled the attack. In contrast, only 32.6\% of MEVBlocker-only victims occur in multi-victim sandwiches, leaving the majority of these cases unexplained by incidental inclusion alone.

\subsubsection{Attacker Behavior}

\Cref{tab:bot-by-ofa} shows that OFA association also differs across bot contracts, despite our evidence that they are likely controlled by the same entity. The largest bot, \href{https://etherscan.io/address/0x841b6ff0979f86bd76579a3cd03fa43a6a1bb697}{\texttt{0x841...697}}, is predominantly associated with Blink, which appears among 74\% of its victims. In contrast, \href{https://etherscan.io/address/0x8a5d8700dd19488d964d202e87b0cc85704054db}{\texttt{0x8a5...4db}} and \href{https://etherscan.io/address/0xe87a2343f12a9d812a9e87f01118b6c7c5f6b716}{\texttt{0xe87...716}} have no observed Blink-associated victims and are instead concentrated in MEV-Share and MEVBlocker order flow. Smaller attackers active in the later period exhibit more heterogeneous OFA associations, often spanning several providers. Together with the temporal patterns in \Cref{fig:eth-weekly-by-bot} and \Cref{fig:eth-ofa-evolution}, this suggests that the entity adapts its exposure or routing strategy as the observed OFA environment changes, rather than using a single strategy throughout the measurement period.

\begin{table}[htbp]
  \centering
  \footnotesize
  \resizebox{\columnwidth}{!}{%
  % Ethereum entities by observed OFA association. Generated by
% build_eth_bot_ofa_table.py -- do not edit by hand.
\begin{tabular}{@{}lrrrrrr@{}}
\toprule
\textbf{Bot} & \textbf{Victims} & \textbf{MEV-Share} & \textbf{MEVBlocker} & \textbf{Blink} & \textbf{Merkle} & \textbf{Unattributed} \\
\midrule
\href{https://etherscan.io/address/0x841b6ff0979f86bd76579a3cd03fa43a6a1bb697}{\texttt{0x841...697}} & 17,355 & 30\% & 14\% & 74\% & 3\% & 19\% \\
\href{https://etherscan.io/address/0x8a5d8700dd19488d964d202e87b0cc85704054db}{\texttt{0x8a5...4db}} & 13,400 & 30\% & 38\% & 0\% & 0\% & 59\% \\
\href{https://etherscan.io/address/0xe87a2343f12a9d812a9e87f01118b6c7c5f6b716}{\texttt{0xe87...716}} & 4,214 & 30\% & 29\% & 0\% & 0\% & 67\% \\
\href{https://etherscan.io/address/0x69b03f6548f16c44b77cd974197e2b8b5ba00b78}{\texttt{0x69b...b78}} & 1,781 & 53\% & 24\% & 64\% & 20\% & 14\% \\
\href{https://etherscan.io/address/0x013a9088670ffdeec705a3dc32d5229ab5ab15ea}{\texttt{0x013...5ea}} & 1,101 & 72\% & 59\% & 59\% & 10\% & 10\% \\
\href{https://etherscan.io/address/0xe9becf8c41cee2b777dee36775613209b5273e02}{\texttt{0xe9b...e02}} & 995 & 67\% & 34\% & 47\% & 27\% & 11\% \\
\href{https://etherscan.io/address/0xc72a44de529163b3ecd26e24f8eda50d7943859b}{\texttt{0xc72...59b}} & 615 & 33\% & 16\% & 58\% & 0\% & 35\% \\
\bottomrule
\end{tabular}
}
    \caption{Observed OFA association by Ethereum attacker bot, for the
    persistent bots of Table~\ref{tab:bots-eth}. Each OFA
    column reports the share of the bot's victims observed in the corresponding
    OFA dataset; columns need not sum to 100\% because a victim may appear in
    multiple datasets. \emph{Unattributed} denotes victims with no observed OFA label.}
  \label{tab:bot-by-ofa}
\end{table}

\subsubsection{Back-Running Opportunities}

OFAs expose transaction information to searchers, enabling arbitrage back-runs that generate value for users. Yet, nearly two in three sandwich victims are attributed to at least one OFA. This raises a natural question: why were these transactions attractive sandwich targets despite being exposed to competing back-runners? An arbitrage back-run executed between the victim transaction and the sandwicher's back-run can reduce or even eliminate the sandwich profit (cf. Appendix~\ref{sec:appendix-backrun} for a benchmark). However, such an opportunity exists only when sufficient alternative liquidity is available. Thus, we examine whether these victims offered profitable arbitrage back-running opportunities.

A direct arbitrage against the sandwiched $X$--$Y$ pool requires at least another active $X$--$Y$ pool, while additional liquidity in $X$ and $Y$ may enable multihop routes. For each of the 39{,}461 victim transactions across our 30{,}607 sandwiches, we therefore compute two measures: the number of active $X$--$Y$ pools (\emph{direct count}) and, as a proxy for potential multihop routes, the minimum number of active pools containing $X$ or $Y$ (\emph{endpoint proxy}). Both measures include the source pool and indicate only potential routing opportunities, but not whether an arbitrage route is executable or profitable.

\iffalse
\begin{table}[t]
  \centering
  \footnotesize
  \begin{tabular}{@{}rrlrl@{}}
    \toprule
    \textbf{Pool Count} & \multicolumn{2}{c}{\textbf{Direct $X$--$Y$}} & \multicolumn{2}{c}{\textbf{Endpoint Proxy}} \\
    \midrule
    1 & 26,887 & (68.1\%) & 22,824 & (57.8\%) \\
    2 & 8,333 & (21.1\%) & 5,893 & (14.9\%) \\
    $>2$ & 4,241 & (10.7\%) & 10,744 & (27.2\%) \\
    \bottomrule
  \end{tabular}
   \caption{Victim transactions by source-inclusive, back-run relevant two-token-pool count at the end of the block preceding each victim ($N=39{,}461$).}
  \label{tab:backrun-pool-counts}
\end{table}
\fi

At the end of the preceding block, the direct pool count is one for 26,887 victims (68.1\%), two for 8,333 (21.1\%), and greater than two for 4,241 (10.7\%). The endpoint proxy is one pool for 22,824 victims (57.8\%), two for 5,893 (14.9\%), and greater than two for 10,744 (27.2\%). Thus, 31.9\% have at least one active direct alternative, while 42.2\% satisfy the broader criterion for a potential multihop route.

Appendix~\ref{sec:appendix-backrun} evaluates a victim-only direct-route counterfactual for the 31,259 victims whose front-run occurs in the same block. We classify a victim as \emph{not back-runnable under our coverage} when either (i) its endpoint proxy is one, or (ii) its endpoint proxy equals its direct count and every direct alternative has a conclusive, non-positive gross-profit optimum. In the first case, at least one endpoint has no active non-source pool, ruling out a source-pool-disjoint route. In the second, all alternatives incident to the lower-degree endpoint are direct $X$--$Y$ pools, exhausting the possible simple source-pool-disjoint routes, and each is conclusively unprofitable.

Of the 31{,}259 same-block cases, 17{,}993 satisfy the first condition. Among the remainder, 3{,}475 have direct alternatives but no gross-profitable direct back-run, and 1{,}417 of these also have an endpoint proxy equal to their direct count. Hence, we rule out a profitable back-run for 19{,}410 victims (62.1\%) under our coverage. Of the remaining 11{,}849 victims, 3{,}905 have a gross-profitable direct route, 2{,}504 have an inconclusive direct-route result, and 5{,}440 have no detected profitable direct route but remain unresolved because of possible multihop paths or an unmodeled source state, such as those of V4 hooks. Profit is ETH-evaluable for 3,863 of the gross-positive cases, of which 484 (1.5\% of the full same-block cohort) remain profitable under our gas assumptions. Thus, profitable direct back-runs appear rare in our counterfactual, which may help explain why these transactions remained attractive sandwich targets despite their exposure to OFA searchers.

\subsubsection{Reorged Blocks}
\label{sec:eth-reorgs}

In addition to OFA exposure, we examine reorged blocks as a second way in which private transactions can become exposed. Ethereum blocks are occasionally reorged, when a block is proposed but does not become part of the canonical chain~\cite{xatu-data,beaconchain-light}. This generally happens because the block was not proposed in time and failed to gather enough votes before the deadline. Any private transaction in such a block loses its privacy since the block carrying it has been broadcast.

We look for transactions that were private when the reorged block was proposed, not part of a sandwich in that block, and later a victim of a sandwich on the canonical chain. Across the 9{,}297 reorged blocks in our data, covering January 2024 to June 2026, we find 140{,}854 swap transactions, of which 56{,}730 were private when the stale block was proposed. Publication of the stale block reveals all of these transactions. In addition, 62.0\% (35{,}193) are later observed individually in the public mempool before their canonical inclusion. Upon re-inclusion, 2{,}875 are sandwiched: 72.4\% (2{,}082) are republished in the public mempool beforehand, while 27.6\% (793) are not. The counts keep H3's persistence filter but also includes the bots specializing on the public mempool.

\begin{figure}[t]
  \centering
  \includegraphics[width=\linewidth]{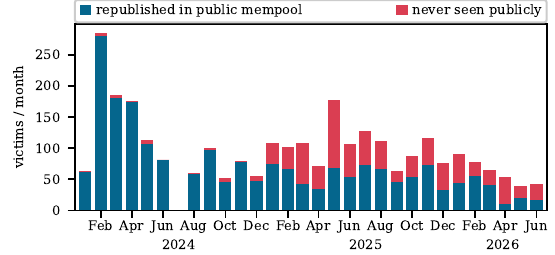}
  \caption{Reorged private transactions sandwiched after canonical re-inclusion, per month, split by whether the transaction was republished in the public mempool between the reorg and its re-inclusion or never observed publicly.}
  \label{fig:reorg-monthly}
\end{figure}

Figure~\ref{fig:reorg-monthly} plots the monthly number of these sandwiches. Until the start of 2025, nearly all of the victims had been republished in the public mempool. From the beginning of 2025, we observe sandwiches on victims that never appeared in the public mempool individually and became visible only as part of the reorged block. This suggests that some bots specialize in monitoring stale blocks and sandwiching the swaps they reveal. The attacker populations differ accordingly. In total, 93 bots attack these victims. Victims republished in the public mempool before they are sandwiched in the canonical block fall to 87 of them, whereas the extraction on victims never seen individually is highly concentrated, with one bot attacking 55\% of the victims and the top five 82\%.

Victims not observed individually in the public mempool face a narrower attacker set than republished victims, which are exposed to the broader public-mempool searcher population. The profit, thus, shows a clear difference (cf.\ Table~\ref{tab:reorg-profit}). The median attack nets \$37 on a victim never seen publicly and one cent on a republished one, where competition pushes fees from 11\% of the gross profit to 76\%.

In summary, reorged blocks provide a second exposure avenue: transactions originally protected from front-running can become targets as if they had initially been submitted publicly. In principle, an attacker who could induce such a reorg might profit from the resulting exposure, for example by incentivizing validators to build on a competing branch. Whether such a strategy is economically or operationally feasible depends on the consensus conditions and the cost of inducing the reorg, and our empirical evidence is not sufficient to determine whether any of the observed reorgs were triggered for this purpose. Note that our estimates are conservative, as we observe only Xatu-recorded reorgs and recover 65.9\% of stale block bodies (cf. Appendix~\ref{sec:data-reorg}).

\begin{table}[t]
  \centering
  \footnotesize
  \resizebox{\columnwidth}{!}{\begin{tabular}{@{}lrrrr@{}}
\toprule
\textbf{Victim Exposure} & \textbf{SWs} & \textbf{Gross (USD)} & \textbf{Fees} & \textbf{Median Net (USD)} \\
\midrule
Never seen publicly & 592 & \$160,672 & 11\% & \$37.31 \\
Republished & 586 & \$119,176 & 76\% & \$0.01 \\
\midrule
\textbf{All} & \textbf{1,178} & \textbf{\$279,848} & \textbf{39\%} & \textbf{\$6.85} \\
\bottomrule
\end{tabular}
}
  \caption{Profit of reorg-exposed sandwiches by exposure mechanism,
  restricted to the 1{,}178 of 2{,}576 attacks whose fee is attributable to a single attack. The rest are multi-pool bundles, often carrying an arbitrage cycle, whose per-attack net profit would understate the return (cf.\ Appendix~\ref{sec:appendix-public}).}
  \label{tab:reorg-profit}
\end{table}

\subsection{Tron}
\label{sec:origins-tron}

Next, we turn to Tron. \Cref{tab:tron-summary} summarizes the per-bot statistics. As on Ethereum, activity is highly concentrated: two bots account for 19{,}247 and 10{,}438 sandwiches, respectively. Together, they conduct 77.0\% of the 38{,}567 attacks. Between 72\% and 89\% of each entity's attacks are within-block wide, with the remainder falling across block boundaries. Profit tracks the attack count only loosely. The largest entity grosses \$540,247, more than three times the second largest, and under twice the attacks. We note that it also has the lowest success rate, at 74\%, suggesting a higher risk tolerance. The remaining success rates range from 80\% to 96\%. The largest bot has also been active the longest, from August 2024 to June 2026. It faced no competition until February 2025; a second possible explanation for its profit.

\begin{table}[t]
  \centering
  \resizebox{\columnwidth}{!}{% Tron persistent entities, one column-width table. Generated by
% build_tron_fleet_table.py -- do not edit by hand.
% Shaded rows: the nine entities first funded by the shared hub wallet.
\begin{tabular}{@{}lrcrrrc@{}}
\toprule
\textbf{Bot} & \textbf{SWs} & \textbf{T/W/C (\%)} & \textbf{Gross (USD)} & \textbf{Fees} & \textbf{Succ.} & \textbf{Active} \\
\midrule
\href{https://tronscan.org/\#/address/TUE2Sq295N2jgRenGKC81AHE77jBNexTJc}{\texttt{0xc83...28b}} & 19,247 & 0 / 84 / 16 & 540,247 & 35\% & 74\% & 24-08--26-06 \\
\href{https://tronscan.org/\#/address/TEtPcNXwPj1PEdsDRCZfUvdFHASrJsFeW5}{\texttt{0x35e...9a1}} & 10,438 & 0 / 86 / 14 & 169,339 & 28\% & 88\% & 25-07--26-06 \\
\rowcolor{fleetgrey}\href{https://tronscan.org/\#/address/TLQgkULsFzihsvbCh5tLssYBRBhf87coLa}{\texttt{0x728...6e9}} & 2,218 & 0 / 77 / 23 & 35,727 & 9\% & 90\% & 25-02--26-05 \\
\rowcolor{fleetgrey}\href{https://tronscan.org/\#/address/TCoLAx59RB89dbmHrVhaLCJt9nACS2xd97}{\texttt{0x1f0...600}} & 1,731 & 0 / 75 / 25 & 85,159 & 11\% & 88\% & 25-03--26-06 \\
\rowcolor{fleetgrey}\href{https://tronscan.org/\#/address/TNdCoLAooNdYa27KGUnCqYtVsFXvJd3XFR}{\texttt{0x8ad...267}} & 1,355 & 0 / 78 / 22 & 24,270 & 13\% & 87\% & 25-03--26-04 \\
\rowcolor{fleetgrey}\href{https://tronscan.org/\#/address/TCoLazogrq7LH7dUHh8MMMpgJeUrPYGufM}{\texttt{0x1f0...f9e}} & 1,033 & 0 / 72 / 27 & 7,321 & 45\% & 81\% & 25-03--26-06 \\
\rowcolor{fleetgrey}\href{https://tronscan.org/\#/address/TCoLAwNYd4icfQCrk65evwt3cdJe5kKZqh}{\texttt{0x1f0...bd4}} & 749 & 0 / 79 / 21 & 17,118 & 23\% & 83\% & 25-03--26-06 \\
\rowcolor{fleetgrey}\href{https://tronscan.org/\#/address/TCoLa4ZPnWRQduaxYCG9iGVMX45KpFVird}{\texttt{0x1f0...e9f}} & 650 & 0 / 82 / 18 & 9,549 & 9\% & 93\% & 25-03--25-10 \\
\rowcolor{fleetgrey}\href{https://tronscan.org/\#/address/TEoG6ZYpWh7YSfGWx1rqcKu6TcZQMucoLA}{\texttt{0x34f...978}} & 374 & 0 / 83 / 16 & 6,469 & 5\% & 89\% & 25-03--26-06 \\
\href{https://tronscan.org/\#/address/TZBL7fQ3i4SbP3Usf5Fr3V5rWAstYfCoLA}{\texttt{0xfe9...ef4}} & 322 & 1 / 73 / 26 & 1,111 & 62\% & 80\% & 25-10--26-06 \\
\rowcolor{fleetgrey}\href{https://tronscan.org/\#/address/TQU1hX6G71i9dzLGciU2xoryM9avE9coLA}{\texttt{0x9f0...70d}} & 302 & 0 / 82 / 18 & 3,883 & 7\% & 91\% & 25-03--26-05 \\
\rowcolor{fleetgrey}\href{https://tronscan.org/\#/address/TSGXM5pZKgfrddcoLAjMaYmDmz5D8Jyyy9}{\texttt{0xb2c...713}} & 148 & 0 / 89 / 11 & 878 & 9\% & 96\% & 25-02--26-06 \\
\bottomrule
\end{tabular}
}
  \caption{Tron bots, sandwiches (SWs), tight (T) / within-block-wide (W) / cross-block-wide split (C), gross profit in USD, the share of it consumed by fees, success rate (\%), and active period. Gray rows mark the nine bots we attribute to one entity (cf.\ Appendix~\ref{sec:appendix-tron-funding}).}
  \label{tab:tron-summary}
\end{table}

As on Ethereum, we cluster the Tron bots into entities. Table~\ref{tab:tron-summary} marks in gray the nine bots we attribute to one entity. Three observations support the grouping (cf.\ Appendix~\ref{sec:appendix-tron-funding}). First, all nine received their first on-chain funding from a single shared wallet. Second, they partition the pools they attack, sharing no pool across any of their 36 pairs. Third, four of them further share an address prefix that only brute-forcing produces, so one operator created them together.
Thus, protected order flow sandwiching on Tron rests with four entities behind twelve bots.

\subsubsection{Ruling Out Block Producers}

On Tron, transactions become visible as they are gossiped through the public P2P network. Front-running protection comes from the ordering policy of the block producer, which is intended to provide first-come-first-served inclusion. This ordering rule is implemented at the client level and could, in principle, be replaced by a custom policy. As on Ethereum, we first look at the block producers (Super Representatives on Tron), as a potential source of attacks on protected order flow. However, attacks are distributed almost uniformly across producers (cf.\ Appendix~\ref{sec:appendix-tron-producers}), making producer misbehavior an unlikely explanation. Thus, we turn to potential weaknesses in Tron's front-running-protection mechanism itself.

\subsubsection{Front-Running Through Latency Races}

A plausible explanation is that attackers obtain their front-running position through a latency race, although our data cannot establish this directly. One possible strategy is to query pending transaction pool, which nodes expose through the \texttt{gettransactionlistfrompending} endpoint.\footnote{\url{https://developers.tron.network/reference/gettransactionlistfrompending}} An attacker polling this endpoint across multiple nodes can learn of a victim swap shortly after it reaches one of the monitored nodes, immediately submits the front leg, and race the victim transaction to the block producer. Peering with many nodes and minimizing latency to producers could improve the chances of the attacker of winning this race often enough to make the strategy profitable. The back leg can then be submitted with a short delay.
Tron states that its ordering mechanism blocks around 98\% of front-running transactions and attributes the remainder to bots that mostly fail~\cite{sun2024tronmev}. Our data does not support this: the twelve bots we identify land 38{,}567 sandwiches with high per-entity success rates between 74\% and 96\%.

\subsection{Base}
\label{sec:origins-base}
On Base, our detector identifies four sandwich bot contracts that form three distinct groups, each active during a separate period (cf.\ Table~\ref{tab:bots-base}). \href{https://basescan.org/address/0x000000224f8364bb91259b24435881efc3752b3e}{\texttt{0x000...b3e}} operates almost entirely within a single block where 56\% of its sandwiches are tight, 42\% are within-block wide, and only 1\% are cross-block wide. Despite conducting the fewest sandwich attempts (202), it
earns the largest gross profit (\$1{,}885) and has the highest success rate (99\%). We attribute these sandwiches to transaction-pool leakage (cf.~Section~\ref{sec:base-ofa-leak}). In contrast, \href{https://basescan.org/address/0x14d676cbaac800c5b0297cd1226770b5d2d2bb75}{\texttt{0x14d...b75}} operates exclusively across blocks and has the lowest success rate, at 51\%. These attacks point to the exploitation of predictable victim behavior (cf.\ Section~\ref{sec:base-predictable}). Finally, \href{https://basescan.org/address/0xc9cc0e1ecf0ae9af9439933b0c38ee843977b41b}{\texttt{0xc9c...41b}} and \href{https://basescan.org/address/0xfcf103349cb306c0234c4171559150f385745a26}{\texttt{0xfcf...a26}} account for the most sandwiches, with 750 and 698, respectively, and have the highest fee shares, at 64\% and 23\%. All but three of their sandwiches occur within a single block. We likewise attribute this group's attacks to predictable victim
behavior (cf.~Section~\ref{sec:base-predictable}).

\begin{table}[t]
  \centering
  \resizebox{\columnwidth}{!}{\begin{tabular}{@{}lrcrrrc@{}}
\toprule
\textbf{Bot} & \textbf{SWs} & \textbf{T/W/C (\%)} & \textbf{Gross (USD)} & \textbf{Fees} & \textbf{Succ.} & \textbf{Active} \\
\midrule
\rowcolor{fleetgrey}\href{https://basescan.org/address/0xc9cc0e1ecf0ae9af9439933b0c38ee843977b41b}{\texttt{0xc9c...41b}} & 750 & 17 / 83 / 0 & 673 & 64\% & 80\% & 26-04--26-04 \\
\rowcolor{fleetgrey}\href{https://basescan.org/address/0xfcf103349cb306c0234c4171559150f385745a26}{\texttt{0xfcf...a26}} & 698 & 31 / 69 / 0 & 268 & 23\% & 84\% & 26-04--26-05 \\
\href{https://basescan.org/address/0x14d676cbaac800c5b0297cd1226770b5d2d2bb75}{\texttt{0x14d...b75}} & 239 & 0 / 0 / 100 & 206 & 4\% & 51\% & 25-02--25-03 \\
\href{https://basescan.org/address/0x000000224f8364bb91259b24435881efc3752b3e}{\texttt{0x000...b3e}} & 202 & 56 / 42 / 1 & 1,885 & 6\% & 99\% & 23-08--23-08 \\
\bottomrule
\end{tabular}
}
  \caption{Base bots, sandwiches (SWs), tight (T) / within-block-wide (W) / cross-block-wide split (C), gross profit in USD, the share of it consumed by fees, success rate (\%), and active period. We believe the two shaded rows to be controlled by the same operator (cf.\ Appendix~\ref{sec:appendix-base-clustering}).}
  \label{tab:bots-base}
\end{table}

\subsubsection{Transaction-Pool Leakage}
\label{sec:base-ofa-leak}

\begin{figure}[t]
  \centering
  \includegraphics[width=\linewidth]{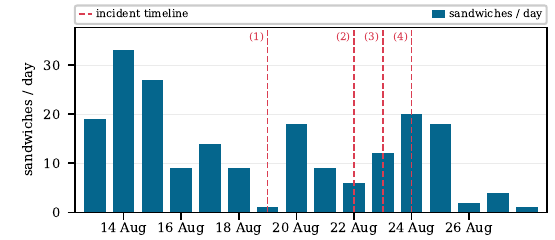}
  \caption{Daily sandwich attacks by \href{https://basescan.org/address/0x000000224f8364bb91259b24435881efc3752b3e}{\texttt{0x000...b3e}} on Base against the public timeline of the August 2023 transaction-pool leak: (1) incident reported, (2) op-geth fix \#118, (3) op-geth fix \#122 / Base fix \#100, (4) incident resolved. The bot's detected sandwich activity spans these 16 days.}
  \label{fig:base-000-incident}
\end{figure}

The sandwich attempts by \href{https://basescan.org/address/0x000000224f8364bb91259b24435881efc3752b3e}{\texttt{0x000...b3e}} exhibit two patterns consistent with transaction leakage through an RPC or other order flow provider. First, the bot's front- and back-run legs are positioned tightly around the victim, and their fee settings closely track those of the victim even when victim fees vary substantially. Second, the bot's active period coincides with a documented Base incident in which transactions from local pending pools were exposed~\cite{base_incident_report_1,base_tx_pool_leak,base_incident_report_2,op_tx_pool_leak} (cf.\ \Cref{fig:base-000-incident}). Together, these observations point to the documented transaction-pool leak as the likely exposure source for this bot.

\subsubsection{Predictable Behavior}
\label{sec:base-predictable}

The victims of \href{https://basescan.org/address/0x14d676cbaac800c5b0297cd1226770b5d2d2bb75}{\texttt{0x14d...b75}}, \href{https://basescan.org/address/0xc9cc0e1ecf0ae9af9439933b0c38ee843977b41b}{\texttt{0xc9c...41b}} and \href{https://basescan.org/address/0xfcf103349cb306c0234c4171559150f385745a26}{\texttt{0xfcf...a26}} exhibit recurring trading patterns that can make their
future transactions predictable. The clearest case is \href{https://basescan.org/address/0x14d676cbaac800c5b0297cd1226770b5d2d2bb75}{\texttt{0x14d...b75}}, which conducts sandwiches exclusively across blocks and targets a concentrated group of victims that follow a fixed trading loop.
Three wallets, one of them \href{https://basescan.org/address/0xd3fcD6bFD7Ab2613Cc0275644b74Dcbe7BEb0F17}{\texttt{0xd3f...f17}}, sell a similar amount of WETH roughly every two minutes for a different token, over 230 times in total, and the bot places its sandwich legs in the neighboring blocks (cf.\ Table~\ref{tab:base-14d-cycle}). The combination of this highly regular victim behavior and exclusively cross-block sandwiches points to prediction rather than pending-transaction visibility: recurring trade becomes predictable, thus the attacker need not observe the transaction before submitting its front leg.
\begin{table}[t]
  \centering
  \footnotesize
  \resizebox{\columnwidth}{!}{\begin{tabular}{@{}lllll@{}}
\toprule
\textbf{Block} & \textbf{Role} & \textbf{Account} & \textbf{Trade} & \textbf{Transaction} \\
\midrule
26{,}721{,}580 & Front-Run & \texttt{0x14d...b75} & 0.000688 WETH $\rightarrow$ 2.6M ANNON & \href{https://basescan.org/tx/0x0bd7aef39aac4c74576000f8be5b2962c3bffcdc945d60792cb5256573dd7c24}{\texttt{0x0bd7aef3}} \\
26{,}721{,}581 & Victim & \texttt{0xd3f...f17} & 0.075800 WETH $\rightarrow$ 275.7M ANNON & \href{https://basescan.org/tx/0x723adb6efe0cf0160c2b5ce0c085396cfbea135cba614d40cc2d0ceca8c3f1ab}{\texttt{0x723adb6e}} \\
26{,}721{,}583 & Back-Run & \texttt{0x14d...b75} & 2.6M ANNON $\rightarrow$ 0.000726 WETH & \href{https://basescan.org/tx/0xe89200b4b047f299d8559a70c61446b6df29a60fd08eae4511f5c31350f70577}{\texttt{0xe89200b4}} \\
\midrule
26{,}721{,}644 & Front-Run & \texttt{0x14d...b75} & 0.000726 WETH $\rightarrow$ 2.8M GROK3 AI & \href{https://basescan.org/tx/0x740cb25056259127f9e6a8698dbe323177a90848cb40c79fc528dab5ebd004e0}{\texttt{0x740cb250}} \\
26{,}721{,}646 & Victim & \texttt{0xd3f...f17} & 0.070300 WETH $\rightarrow$ 264.4M GROK3 AI & \href{https://basescan.org/tx/0x8d66df63df6d23517438d4e2f25e1e0081401f371695b589f9dfc52f99c5b554}{\texttt{0x8d66df63}} \\
26{,}721{,}647 & Back-Run & \texttt{0x14d...b75} & 2.8M GROK3 AI $\rightarrow$ 0.000763 WETH & \href{https://basescan.org/tx/0x1a223921bfb384ba54261c316ac8b5cd564f441c49977f1ae9c4337ba8677c1a}{\texttt{0x1a223921}} \\
\midrule
26{,}721{,}700 & Front-Run & \texttt{0x14d...b75} & 0.000763 WETH $\rightarrow$ 2.6M DEEPSEEK & \href{https://basescan.org/tx/0x8d1e99142baff506fc3e624c9fdfbf2f047b4cb66c2c369e91d046a103d91ef4}{\texttt{0x8d1e9914}} \\
26{,}721{,}702 & Victim & \texttt{0xd3f...f17} & 0.098500 WETH $\rightarrow$ 317.3M DEEPSEEK & \href{https://basescan.org/tx/0xc6171d668c21df6a58ce4c46215ebc98fc6490ba98809ce4803dafdcad4a0a56}{\texttt{0xc6171d66}} \\
26{,}721{,}703 & Back-Run & \texttt{0x14d...b75} & 2.6M DEEPSEEK $\rightarrow$ 0.000819 WETH & \href{https://basescan.org/tx/0x1f93db480d877b2ab49bd6bc02bb50a5941262939280bc37682ebb0208873def}{\texttt{0x1f93db48}} \\
\bottomrule
\end{tabular}
}
  \caption{Three consecutive cycles of the scripted loop that \href{https://basescan.org/address/0x14d676cbaac800c5b0297cd1226770b5d2d2bb75}{\texttt{0x14d...b75}} sandwiches. The near-identical cycle repeats over 230 times. Roughly every two minutes, one of three wallets, among them \href{https://basescan.org/address/0xd3fcD6bFD7Ab2613Cc0275644b74Dcbe7BEb0F17}{\texttt{0xd3f...f17}}, sells a similar WETH amount for a different token. The repetition makes the next swap, so the bot needs no pending-transaction visibility.}
  \label{tab:base-14d-cycle}
\end{table}

The victims of \href{https://basescan.org/address/0xc9cc0e1ecf0ae9af9439933b0c38ee843977b41b}{\texttt{0xc9c...41b}} and \href{https://basescan.org/address/0xfcf103349cb306c0234c4171559150f385745a26}{\texttt{0xfcf...a26}} likewise exhibit regular, funding-linked trading patterns, although less rigidly than in the fixed loop above. At the same time, both bots closely match their victims' fee settings, including the few transactions with irregular fees. Predictable victim behavior may account for part of these attacks, while some additional transaction visibility may also have been available to set fees appropriately (cf.~Appendix~\ref{sec:appendix-base-details}).

Overall, Base illustrates two distinct ways in which front-running protection can fail. Transactions may become exposed within the submission pipeline, as in the documented leakage incident, while sufficiently regular victim behavior can make future trades predictable without requiring pending-transaction visibility.

\subsection{Solana}
\label{sec:origins-solana}

Finally, we turn to Solana, which accounts for by far the largest number of protected sandwich attacks in our dataset. We note that the degree of protection varies across chains. Base relies on a centralized sequencer whose pending transaction queue is not visible externally. Solana instead forwards transactions to a small set of upcoming leaders rather than broadcasting them network-wide, limiting their pre-inclusion visibility. As the Solana Foundation notes, this design ``reduces the MEV surface area compared to chains with public mempools''~\cite{solana_mev_protection}, while not eliminating it entirely.

On Solana, attack volume is substantially larger than on the other chains, accompanied by fiercer competition among attackers. Moreover, the nature of sandwich attacks changes over time in response to interventions by the ecosystem (cf.\ Table~\ref{tab:solana-eras}). Thus, we organize our Solana analysis by era rather than the in-depth per-bot analysis used for the other chains.

The first era covers the Jito mempool, which operated through Jito's Block Engine until March~2024. A validator running Jito-Solana connected to a Jito relayer, which received the transactions destined for that validator. The relayer held each transaction for up to 200\,ms before forwarding it, while searchers subscribed to the stream could observe transactions before execution~\cite{jito_relayer_docs,jito_searcher_examples}. In effect, this created a public mempool for a subset of Solana transactions.

\begin{table}[t]
  \centering
  \resizebox{\columnwidth}{!}{\begin{tabular}{@{}llrrrrrr@{}}
\toprule
\textbf{Era} & \textbf{From} & \textbf{SWs} & \textbf{Daily} & \textbf{T/W/C (\%)} & \textbf{Bots} & \textbf{Supp/Elev} & \textbf{Elev} \\
\midrule
Jito mempool & 07/23 & 2.3M & 9k & 54 / 7 / 39 & 386 & 1,022/282 & 50\% \\
Mempool closed & 03/24 & 2.6M & 28k & 87 / 1 / 13 & 287 & 61/27 & 7\% \\
SFDP delist & 06/24 & 11.1M & 46k & 84 / 1 / 15 & 2,082 & 40/172 & 16\% \\
Marinade MIP.9 & 02/25 & 6.7M & 26k & 18 / 3 / 79 & 4,256 & 13/115 & 9\% \\
JitoSOL delist & 10/25 & 5.2M & 20k & 1 / 3 / 96 & 3,115 & 6/8 & 2\% \\
\bottomrule
\end{tabular}
}
  \caption{Solana sandwiching per era: daily sandwiches, tight (T) / within-block-wide (W) / cross-block-wide split (C), suppression (supp) / elevation (elev) counts leaders admitting under $0.5\times$ or over $1.5\times$ their share of the era's attacks given the blocks they produced, counted over leaders with at least 5{,}000 blocks in the era, and elev representing the share of attacks landing in the latter's slots.}
  \label{tab:solana-eras}
\end{table}

Our collection window begins in July 2023, with the first notable attack volume appearing in October (cf.\ Table~\ref{tab:solana-eras}). Activity increases sharply around February 2024 and exceeds 100{,}000 sandwiches per day (cf.\ Figure~\ref{fig:sol-daily}). It then drops abruptly when Jito suspends the mempool, citing ``negative externalities impacting users on Solana''~\cite{jito_mempool_suspend}. Sandwich activity is also concentrated among a subset of validators. We find that 50\% of attacks occur in slots led by validators that are over-represented relative to their share of blocks (cf.\ Table~\ref{tab:solana-eras}). This is consistent with only a subset of validators receiving the transaction flow exposed through the Jito mempool.

The second, third, and fourth eras end each with a validator-delisting event. In June~2024, the Solana Foundation removed validators from its delegation program~\cite{sfdp2024delist}. Large stake pools, such as Marinade, subsequently took similar measures, blocklisting validators from its stake auction in February 2025~\cite{marinade2025mip9}, and Jito's stake pool (JitoSOL) delisted additional validators in October 2025~\cite{jito2025blacklist}. These interventions targeted validators identified as facilitating sandwich attacks and removed delegated stake from them.

Across these eras, the share of sandwiches landing in slots led by over-represented validators falls sharply, from 50\% under the Jito mempool era, to 2\% in the final era. Daily attack volume, however, does not follow. It climbs to roughly 46{,}000 attacks per day and later settles around 20{,}000 (cf.\ Figure~\ref{fig:sol-daily}). What changes substantially is the structure of the attacks. Tight sandwiches comprise between 54\% and 87\% of attacks in the eras before the delisting, but only 18\% in the Marinade era and 1\% after the JitoSOL delisting.

The final era begins after the JitoSOL delisting. Same block sandwiches collapse to 4\% of attacks: 1\% of attacks are tight and 3\% within-block wide. Thus, 96\% of attacks span blocks.
Daily attack volume settles at roughly 20{,}000, while the share of attacks landing in slots led by over-represented leaders drops to 2\%. These patterns make leader-specific exposure an unlikely explanation for most attacks in this era. Most sandwiches span slots produced by different leaders, and only six leaders are under- and eight are over-represented. Instead, the victims concentrate among the users of a few applications.

We thus observe two distinct exposure signals over time on Solana. The first appears at the validator level, which we examine next (cf.\ Section~\ref{sec:solana-leaders}). Their role is most apparent during the Jito mempool era and recedes over time as the ecosystem interventions accumulate, from the closure of the mempool to the successive validator delistings. The second appears at the application layer (cf.\ Section~\ref{sec:solana-apps}) and becomes the strongest signal in the later period.

\subsubsection{Validator-Level Exposure}\label{sec:solana-leaders}
\begin{figure}[t]
  \centering
  \includegraphics[width=\linewidth]{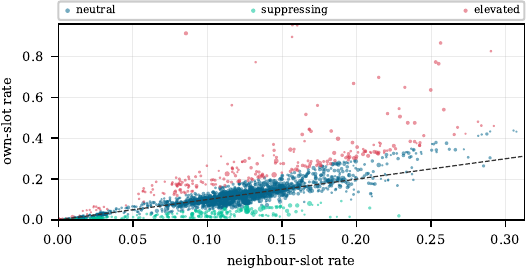}
  \caption{Sandwiches per slot in each Solana leader's own slots against the rate in neighboring slots of other validators.}
  \label{fig:sol-leader-admission}
\end{figure}

Recall that the first four eras show elevated sandwich activity in the slots of a subset of validators. \Cref{fig:sol-leader-admission} compares, per leader, the sandwich rate in its own slots with the rate in neighboring slots led by other validators, revealing a clear separation between groups. The leaders in blue lie close to the equality line, i.e., the attack rate in their slots is similar to the surrounding rate. The leaders in red carry far more attacks than their neighbors, whereas those shown in green carry substantially fewer. This heterogeneity is consistent with validator differences in how protected order flow or sandwiches are handled. The over-represented validators provide evidence of validator-level exposure, while the under-represented validators appear to suppress or exclude such attacks.

The concentration fades over time. During the Jito mempool era, the ten most over-represented leaders account for 26.4\% of sandwiches while producing only 15.7\% of blocks, corresponding to an excess of 1.68 times. This excess falls to 1.01 by the Marinade era. The distribution of per-leader sandwich rates, i.e., the share of each leader's blocks containing a sandwich, tells the same story. When sandwich activity is concentrated among specific leaders, their blocks carry substantially more sandwiches than those of other leaders. Under the Jito mempool, this distribution is bimodal, as the mempool was opt-in per validator: a validator either connected to the Jito relayer or not. By the Marinade era, the distribution becomes unimodal and the two groups largely merge, indicating that sandwich activity is no longer concentrated among a distinct subset of leaders. Appendix~\ref{sec:appendix-solana-details} tracks this excess weekly alongside the ecosystem interventions and reports the corresponding Lorenz curves and per-leader sandwich-rate distributions (cf.\ Figures~\ref{fig:sol-leader-conc}, \ref{fig:sol-leader-conc-lorenz}, and \ref{fig:sol-leader-bimodal}). Attackers also adapt to these interventions. Evasive sandwiches increase after the Jito mempool closure, with bots increasingly splitting their legs as validator-specific exposure declines (cf.\ Appendix~\ref{sec:appendix-sol-structure}).

To summarize, Solana forwards each transaction to only a small set of upcoming leaders, and leader-side confidentiality would substantially limit the opportunity for sandwiching in the absence of upstream exposure. Yet, particular validators show clear excess attack rates in their slots, most prominently in the early eras, providing evidence of validator-level exposure during the first part of our observations. This signal largely disappears around the start of 2025, while sandwich attacks persist. Their continued presence thus points to an exposure further upstream in the transaction-submission pipeline. We next examine applications as a potential source.

\subsubsection{Application-Level Exposure}\label{sec:solana-apps}

\begin{figure}[t]
  \centering
  \includegraphics[width=\linewidth]{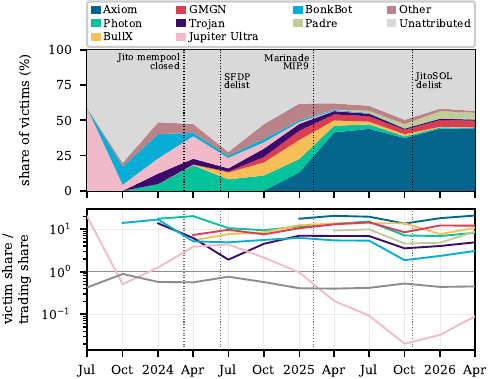}
  \caption{Quarterly victim share per application on Solana (top) and its ratio to swap share (bottom, from 80{,}896 sampled slots). Dotted rules mark interventions.}
  \label{fig:sol-app-share-excess}
\end{figure}

\Cref{fig:sol-app-share-excess} shows how victim concentration across applications evolves over time, for the eight applications with the largest number of victims. An application here is the interface a user trades through, such as a Telegram bot or a DApp like Axiom~\cite{axiom}, Photon~\cite{photon}, BullX~\cite{bullx}, Trojan~\cite{trojan}, or GMGN~\cite{gmgn}, that constructs and submits the transaction on the user's behalf. The bottom panel normalizes each application's victim share by its share of all swap transactions. A value above one therefore indicates that the application's users are sandwiched more often than their share of trading activity would imply.

All applications except Jupiter Ultra~\cite{j_ultra} are over-represented among sandwich victims. This may partly reflect differences in the order flow they route. Telegram bots and DApps handle more long-tail token volume through thinner liquidity pools, where sandwich opportunities can be more attractive, whereas Jupiter Ultra routes less of this flow. Unattributed transactions are under-represented among victims. One possible explanation is that users submitting transactions directly, rather than through an identified application, are less exposed to the application layer providers. Our data, however, does not allow us to determine whether this is due to differences in user sophistication, submission infrastructure, or other characteristics of the underlying order flow.

Axiom emerges as validator-level exposure declines, accounting for 41.4\% of victims in Q2~2025 and remaining above 37\% thereafter. Its excess ratio ranges from 13.6 to 20.9. Photon, GMGN, and BullX exhibit similar but smaller patterns, with excess ratios peaking at 20.3, 14.5, and 13.9, respectively. In multi-victim sandwiches, however, we cannot determine which of the enclosed transactions was deliberately targeted (cf.\ Section~\ref{sec:detection-appendix-body}). Thus, we recompute the excess using only single-victim sandwiches. The ratios decrease, by roughly half for Axiom, but remain well above one for every over-represented application (cf.\ Table~\ref{tab:sol-app-excess-sole} in Appendix~\ref{sec:appendix-solana-details}).

Many of these transactions nevertheless use explicit front-running-protection mechanisms. Jito's \texttt{jitodontfront} feature lets a transaction opt-out of being front-run within a bundle. When the flag is set, Jito's Block Engine accepts only bundles that place the flagged transaction first, preventing another transaction in the same bundle from executing ahead of it. The guarantee therefore applies specifically to front-running within the Jito bundle. Appendix~\ref{sec:appendix-solana-details} reports the per-application victim shares and adoption rates of \texttt{jitodontfront}, with adoption near universal among several of the most heavily sandwiched applications.

Axiom, the application associated with the most victims, makes the limits of this protection explicit. It offers three settings. \emph{Off} provides no dedicated MEV protection. \emph{Reduced} routes the transaction through Jito bundles and thus inherits exactly the guarantee above, i.e., no front-running within a bundle, but no protection beyond that path. \emph{Secure}, the recommended setting, additionally restricts submission to a whitelist of validators. However, our measurements are hard to reconcile with interpreting these
settings as providing end-to-end protection. Axiom sets \texttt{jitodontfront} on 98.5\% of its 21.5M sandwiched victim transactions since the flag launched, while its victim share stays above 37\% and its excess ratio never falls below 13. Near-total flag adoption therefore does not prevent substantial sandwich exposure in our data. The attack structure helps explain this limitation. Of the 5.02M sandwiches involving flagged Axiom victims, only 16.5\% carry a Jito tip. Thus, the large majority do not exhibit the bundle-path signal governed by \texttt{jitodontfront}, consistent with these transactions reaching the leader through another submission path.

In conclusion, exposure on Solana shifts rather than disappears. From the end of 2023 to the beginning of 2025, we observe a strong validator-level signal, with a subset of validators carrying a substantially larger share of sandwiches than of blocks produced. This concentration fades as the ecosystem progressively delists involved validators. In the later period, the strongest signal instead appears at the application layer, although our data does not establish whether the applications themselves are the exposure source or whether correlated order flow characteristics contribute to the excess. Each intervention therefore appears to remove or weaken one source of exposure without eliminating the attacks.

\section{Related Work}

\noindent
\textbf{Quantifying MEV.}
Prior work has quantified MEV predominantly on Ethereum~\cite{daian2019flashboys20frontrunning,TorresFrontrunner2021,QinBEV2022,ZhouHFT2021,WangCyclic2022,QinLiquidations2021,HeimbachNonAtomic2024,GuptaPrivateOrderFlow2023}, with Torres et al.~\cite{TorresRolling2024} extending the analysis to L2s and finding no evidence of sandwiching through August 2023. Weintraub et al.~\cite{WeintraubPrivate2022} show that MEV increasingly shifted to private mempools, while Öz et al.~\cite{oz2025cross} document cross-chain MEV through arbitrage. Mancino et al.~\cite{MancinoSandwiched2025} report that 40\% of sandwich victims migrate to private routing within 60 days and identify 2,932 private sandwiches in November--December 2024. We observe substantially fewer, partly due to Mempool Dumpster outages that misclassified victims as private. Gogol et al.~\cite{GogolSandwich2026} study sandwiching across L2s but lack systematic evidence, whereas our stricter detection captures multi-block attacks and requires persistent attacker activity. Sandwiched.me~\cite{sandwichedme2025solana} reports Solana sandwiching without examining leader or application exposure, while Gerzon et al.~\cite{GerzonJito2025} study Jito bundles over four months. Our measurement spans three years and is not limited to Jito. Pahari et al.~\cite{PahariExclusive2025} show that private transactions can later become public, exposing them to sandwiching. Finally, recent work examines speculative MEV and its implications for the ecosystem~\cite{SolmazOptimistic2025,gogol2025priority,sevim2026signals,WangBlockspace2026,PahariSpeculative2026,wu2026waitprobe}.

\noindent
\textbf{MEV Protections.}
A range of mechanisms mitigate front-running by modifying transaction ordering and limiting MEV extraction~\cite{EskandariSoK2019,YangCountermeasures2022,HeimbachPreventing2022,HeimbachEliminating2022,BaumMitigation2022,ZhouA2MM2021,ZhangFlashFreezing2022}. Fair-ordering schemes provide arrival-order guarantees~\cite{KelkarOrderFairness2020,KelkarThemis2023,KelkarPermissionless2022,KursaweWendy2020,ZhangByzantineOrdered2020,CachinQuickOrderFairness2022,ConstantinescuClock2023,KiayiasBounded2024,BurakFCFS2024}, while blind ordering hides transaction contents until ordering is finalized~\cite{KavousiBlindPerm2023,LiFairness2023}. Other approaches distribute sequencing through systems such as SUAVE~\cite{SUAVE2023}, protect transaction contents using TEEs~\cite{BentovTesseract2019,StathakopoulouTEE2021}, or employ batched threshold encryption~\cite{BEATMEV2025,BEASTMEV2025}. Protocol-level proposals such as PEPC~\cite{MonnotPEPC2022}, ePBS~\cite{EthereumEIP7732}, and FOCIL~\cite{ThieryFOCIL2024} constrain proposer ordering power, while OFAs~\cite{janicot2025privatemevprotectionrpcs} such as MEV-Blocker~\cite{mevblocker_docs} and MEV-Share~\cite{flashbots_mevshare_intro} provide private order flow and MEV redistribution. PROF~\cite{BabelPROF2026} further combines private ordering with incentives for profit-seeking validators. Despite these defenses, their effectiveness and underlying trust assumptions remain largely untested. We provide the first large-scale empirical study of sandwiching across multiple protection mechanisms, revealing widespread violations of these assumptions.

\section{Concluding Discussion}

Our study shows that sandwiches against protected order flow occur across four chains, spanning L1s and L2s with substantially different transaction-submission and mempool designs. Outside Solana, however, we observe only a handful of entities exploiting the exposures we identify, possibly reflecting limited profitability or a less mature extraction ecosystem. On Solana, we observe no considerable sandwich activity until late 2023, after which attacks grow rapidly and adapt as successive exposure paths are mitigated.

More broadly, our findings highlight the need for protection mechanisms that account for information exposure across the full transaction-submission pipeline rather than at a single layer. Thus, the attacks we observe may represent only a lower bound on the risk posed by such exposure if these mechanisms become more widely exploited.

\parhead{OFAs.} Ethereum's transaction-submission pipeline is fragmented across components whose threat models are not necessarily designed to compose. If a transaction originator submits the same signed transaction through multiple RPCs or OFAs, information revealed along one path can weaken the protections of another. Such multi-path submission should therefore be treated as a composition risk: OFAs and RPC providers should ensure that their disclosure mechanisms remain safe when combined. This is particularly important because wallets and applications may route transactions through multiple providers to improve inclusion guarantees or rebates, leaving users with limited control over the resulting submission path.

OFAs can further reduce unnecessary exposure by withholding transactions that are unlikely to generate meaningful back-run value. Simple heuristics, such as avoiding disclosure for trades in tokens with no alternative liquid trading pool, could eliminate many cases in which users are exposed without a corresponding arbitrage opportunity. Finally, OFAs increasingly operate on L2s such as Base, Arbitrum, Robinhood chain, where signed transactions may be provided directly to integrated searchers~\cite{blink_docs}. Such designs can reintroduce pre-inclusion transaction visibility into environments whose sequencer-based submission model would otherwise keep pending order flow private.

\parhead{Encrypted Mempools and TEEs.}
Encrypted mempools can mitigate many of the exposure paths identified in this study~\cite{passerat2026sok,EPRINT:DziFauLuh24,EPRINT:MomGorZha22}, with recent constructions making such designs increasingly practical~\cite{EPRINT:GaiMelRom23,FC:DHMW23,AC:CDKKN22,EPRINT:CCNPS23,USENIX:CGPP24,USENIX:CGPW25,BEATMEV2025,BEASTMEV2025,EPRINT:ABDG+25,EPRINT:BonLauTas25,EPRINT:BNRT26,agarwal2026btx,EPRINT:FPTX25,EPRINT:GWWW25,EPRINT:XioZhaChe25,EPRINT:SueAshShi24,C:AgaFerPin25}. By hiding transaction contents until inclusion, they can prevent bots and block producers from front-running based on pre-inclusion transaction information. However, exposure before encryption, for example at
the RPC or application layer, remains possible~~\cite{garimidi2025limits}. Delayed disclosure can also hinder timely price discovery and back-running, potentially reducing or delaying OFA rebates. TEE-based designs offer a different tradeoff: they allow searchers to operate on full transaction contents inside an isolated execution environment~\cite{flashbots2025teeprotect}, preserving back-running opportunities while limiting information exposed outside the TEE for front-running.

\parhead{Predictable Behavior.} As we have seen on Base, predictable transaction patterns can be front-run even without any exposure through the submission-pipeline. As long as past on-chain behavior remains observable, sufficiently regular activity can reveal information about future transactions. Although users may in principle avoid such patterns, doing so consistently is often impractical. More comprehensive protection may therefore require stronger forms of transaction privacy, potentially including designs that conceal transaction intent or relevant on-chain activity until execution.

Users ultimately need protection that holds in practice. We hope that exposing this darker corner of the dark forest, where supposedly protected transactions can still be front-run, helps drive the development of more effective defenses.

\section*{Acknowledgments}

We would like to thank DataAlways, Jonathan Passerat-Palmbach, Arthur Bagourd, Antony Denyer, and Albin Mamuti for their valuable comments and feedback.

\bibliographystyle{plainurl}
\bibliography{references}

@misc{jito_mempool_suspend,
  title        = {{Jito Labs suspends the mempool offered through the Jito Block Engine}},
  author       = {{Jito Labs}},
  howpublished = {\url{https://x.com/jito_labs/status/1766228889888514501}},
  note         = {Accessed: 2026-08-17},
  year         = {2024}
}

@misc{wu2026waitprobe,
  title         = {To Wait or To Probe: Arbitrage Competition on High-Throughput Blockchains},
  author        = {Fei Wu and Burak {\"O}z},
  year          = {2026},
  eprint        = {2606.00720},
  archivePrefix = {arXiv},
  primaryClass  = {cs.CE},
  url           = {https://arxiv.org/abs/2606.00720}
}

@misc{jito_searcher_examples,
  title        = {{jito-labs/searcher-examples: mempool subscription and its removal}},
  author       = {{Jito Labs}},
  howpublished = {\url{https://github.com/jito-labs/searcher-examples}},
  note         = {Commit \texttt{fc04568} (2022-07-26) ships a bot calling
                  \texttt{subscribe\_pending\_transactions}; commit
                  \texttt{367b7fe} ``Remove Mempool from CLI'' is authored
                  2024-03-08. Accessed: 2026-08-17},
  year         = {2024}
}

@misc{jito_relayer_docs,
  title        = {{Jito Relayer and Block Engine}},
  author       = {{Jito Labs}},
  howpublished = {\url{https://jito-labs.gitbook.io/mev/searcher-services/recommendations}},
  note         = {Accessed: 2026-08-17},
  year         = {2024}
}

@misc{flashbotsmempooldumpster,
  author       = {{Flashbots}},
  title        = {{Mempool Dumpster}},
  howpublished = {\url{https://github.com/flashbots/mempool-dumpster}},
  note         = {Accessed: 2026-06-25},
  year         = {2024}
}

@online{flashbots2025teeprotect,
  author  = {{Flashbots}},
  title   = {Enabling TEE Searching for Protect Fast Mode},
  year    = {2025},
  month   = aug,
  day     = {28},
  url     = {https://collective.flashbots.net/t/enabling-tee-searching-for-protect-fast-mode/5219},
  note    = {The Flashbots Collective, accessed 2026-08-25}
}

@misc{monadlocalmempool,
  author       = {{Monad Labs}},
  title        = {{Monad Documentation: Local Mempool and Transaction Forwarding}},
  howpublished = {\url{https://docs.monad.xyz}},
  note         = {Accessed: 2026-06-25},
  year         = {2025}
}

@misc{monadraptorcast,
  author       = {{Monad Labs}},
  title        = {{RaptorCast: Monad's Block Propagation Protocol}},
  howpublished = {\url{https://docs.monad.xyz}},
  note         = {Accessed: 2026-06-25},
  year         = {2025}
}

@misc{tronconsensus,
  author       = {{TRON DAO}},
  title        = {{TRON Whitepaper: DPoS Consensus and Transaction Ordering}},
  howpublished = {\url{https://tron.network}},
  note         = {Accessed: 2026-06-25},
  year         = {2024}
}

@misc{sandwichedme2025solana,
  author       = {{sandwiched.me}},
  title        = {{Sandwich Attacks on Solana}},
  howpublished = {\url{https://sandwiched.me}},
  note         = {Accessed: 2026-06-25},
  year         = {2025}
}

@misc{janicot2025privatemevprotectionrpcs,
      title={{Private MEV Protection RPCs: Benchmark Study}}, 
      author={Paul Janicot and Alex Vinyas},
      year={2025},
      eprint={2505.19708},
      archivePrefix={arXiv},
      primaryClass={econ.GN},
      url={https://arxiv.org/abs/2505.19708}, 
}

@misc{flashbots_mevshare_intro,
  title        = {{MEV-Share Introduction}},
  author       = {{Flashbots}},
  howpublished = {\url{https://docs.flashbots.net/flashbots-mev-share/introduction}},
  note         = {Accessed: 2026-08-05},
  year         = {2026}
}

@misc{mevblocker_docs,
  title        = {{MEVBlocker Documentation}},
  author       = {{MEVBlocker}},
  howpublished = {\url{https://docs.mevblocker.io/}},
  note         = {Accessed: 2026-08-05},
  year         = {2026}
}

@misc{blink_docs,
  title        = {{Blink Documentation}},
  author       = {{Blink Labs}},
  howpublished = {\url{https://docs.blinklabs.xyz/blink}},
  note         = {Accessed: 2026-08-05},
  year         = {2026}
}

@misc{gulfstream,
  title        = {{Gulf Stream: Solana's Mempool-less Transaction Forwarding Protocol}},
  author       = {Anatoly Yakovenko},
  howpublished = {\url{https://solana.com/news/gulf-stream--solana-s-mempool-less-transaction-forwarding-protocol}},
  note         = {Accessed: 2026-08-16},
  year         = {2019}
}

@misc{solana_mev_protection,
  title        = {{MEV Protection with Jito DontFront}},
  author       = {{Solana Foundation}},
  howpublished = {\url{https://solana.com/docs/defi/mev-protection}},
  note         = {Accessed: 2026-08-13},
  year         = {2026}
}

@misc{arbitrum_sequencer,
  title        = {{The Arbitrum Sequencer}},
  author       = {{Offchain Labs}},
  howpublished = {\url{https://docs.arbitrum.io/how-arbitrum-works/sequencer}},
  note         = {Accessed: 2026-08-05},
  year         = {2026}
}

@misc{op_stack_sequencer,
  title        = {{OP Stack: Sequencing and the Transaction Pool}},
  author       = {{Optimism Foundation}},
  howpublished = {\url{https://docs.optimism.io/}},
  note         = {Accessed: 2026-08-05},
  year         = {2026}
}

@misc{mev_share_double_hash,
  title        = {{MEV-Share: Understanding double-hash}},
  author       = {{Flashbots}},
  howpublished = {\url{https://docs.flashbots.net/flashbots-mev-share/searchers/event-stream#understanding-double-hash}},
  note         = {Accessed: 2026-08-16},
  year         = {2026}
}

@misc{base_incident_report_1,
  title        = {{Base Incident Report 1}},
  author       = {{Base}},
  howpublished ={\url{https://status.base.org/incidents/m9fnx4p3bhp5}},
  note         = {Accessed: 2026-08-16},
  year         = {2026}
}

@misc{base_tx_pool_leak,
  title        = {{Base Tx Pool Leak Github}},
  author       = {{Base Github}},
  howpublished ={\url{https://github.com/base/node/pull/96}},
  note         = {Accessed: 2026-08-16},
  year         = {2026}
}

@misc{op_tx_pool_leak,
  title        = {{OP-GETH Tx Pool Leak Github}},
  author       = {{OP-GETH Github}},
  howpublished ={\url{https://github.com/ethereum-optimism/op-geth/pull/118}},
  note         = {Accessed: 2026-08-16},
  year         = {2026}
}

@misc{base_incident_report_2,
  title        = {{Base Incident Report 2}},
  author       = {{Base}},
  howpublished ={\url{https://status.base.org/incidents/czj7m0j86m09}},
  note         = {Accessed: 2026-08-16},
  year         = {2026}
}

@article{MancinoSandwiched2025,
  author       = {Davide Mancino and
                  Davide Rezzoli},
  title        = {{Sandwiched and Silent: Behavioral Adaptation and Private Channel Exploitation
                  in Ethereum {MEV}}},
  journal      = {CoRR},
  volume       = {abs/2512.17602},
  year         = {2025},
  url          = {https://doi.org/10.48550/arXiv.2512.17602},
  doi          = {10.48550/ARXIV.2512.17602},
}

@article{GogolSandwich2026,
  author       = {Krzysztof Gogol and
                  Manvir Schneider and
                  Jan Gorzny and
                  Claudio J. Tessone},
  title        = {{How to Serve Your Sandwich? {MEV} Attacks in Private {L2} Mempools}},
  journal      = {CoRR},
  volume       = {abs/2601.19570},
  year         = {2026},
  url          = {https://doi.org/10.48550/arXiv.2601.19570},
  doi          = {10.48550/ARXIV.2601.19570},
  eprinttype   = {arXiv},
  eprint       = {2601.19570},
  bibsource    = {dblp computer science bibliography, https://dblp.org}
}

@inproceedings{YangCountermeasures2022,
  author    = {Sen Yang and Fan Zhang and Ken Huang and Xi Chen and
               Youwei Yang and Feng Zhu},
  title     = {{{SoK}: {MEV} Countermeasures}},
  booktitle = {Proceedings of the 2024 Workshop on Decentralized Finance and
               Security, DeFi 2024, Salt Lake City, UT, USA, October 14-18, 2024},
  pages     = {21--30},
  publisher = {{ACM}},
  year      = {2024},
  doi       = {10.1145/3689931.3694911}
}

@inproceedings{HeimbachPreventing2022,
  author       = {Lioba Heimbach and
                  Roger Wattenhofer},
  editor       = {Maurice Herlihy and
                  Neha Narula},
  title        = {{SoK: Preventing Transaction Reordering Manipulations in Decentralized
                  Finance}},
  booktitle    = {Proceedings of the 4th {ACM} Conference on Advances in Financial Technologies,
                  {AFT} 2022, Cambridge, MA, USA, September 19-21, 2022},
  pages        = {47--60},
  publisher    = {{ACM}},
  year         = {2022},
  url          = {https://doi.org/10.1145/3558535.3559784},
  doi          = {10.1145/3558535.3559784},
  bibsource    = {dblp computer science bibliography, https://dblp.org}
}

@inproceedings{HeimbachEliminating2022,
  author       = {Lioba Heimbach and
                  Roger Wattenhofer},
  editor       = {Yuji Suga and
                  Kouichi Sakurai and
                  Xuhua Ding and
                  Kazue Sako},
  title        = {{Eliminating Sandwich Attacks with the Help of Game Theory}},
  booktitle    = {{ASIA} {CCS} '22: {ACM} Asia Conference on Computer and Communications
                  Security, Nagasaki, Japan, 30 May 2022 - 3 June 2022},
  pages        = {153--167},
  publisher    = {{ACM}},
  year         = {2022},
  url          = {https://doi.org/10.1145/3488932.3517390},
  doi          = {10.1145/3488932.3517390},
  bibsource    = {dblp computer science bibliography, https://dblp.org}
}

@inproceedings{BaumMitigation2022,
  author       = {Carsten Baum and
                  James Hsin{-}yu Chiang and
                  Bernardo David and
                  Tore Kasper Frederiksen and
                  Lorenzo Gentile},
  editor       = {Shin'ichiro Matsuo and
                  Lewis Gudgeon and
                  Ariah Klages{-}Mundt and
                  Daniel Perez Hernandez and
                  Sam Werner and
                  Thomas Haines and
                  Aleksander Essex and
                  Andrea Bracciali and
                  Massimiliano Sala},
  title        = {{SoK: Mitigation of Front-Running in Decentralized Finance}},
  booktitle    = {Financial Cryptography and Data Security. {FC} 2022 International
                  Workshops - CoDecFin, DeFi, Voting, WTSC, Grenada, May 6, 2022, Revised
                  Selected Papers},
  series       = {Lecture Notes in Computer Science},
  volume       = {13412},
  pages        = {250--271},
  publisher    = {Springer},
  year         = {2022},
  url          = {https://doi.org/10.1007/978-3-031-32415-4\_17},
  doi          = {10.1007/978-3-031-32415-4\_17},
  bibsource    = {dblp computer science bibliography, https://dblp.org}
}

@inproceedings{TorresRolling2024,
  author       = {Christof Ferreira Torres and
                  Albin Mamuti and
                  Ben Weintraub and
                  Cristina Nita{-}Rotaru and
                  Shweta Shinde},
  editor       = {Bo Luo and
                  Xiaojing Liao and
                  Jun Xu and
                  Engin Kirda and
                  David Lie},
  title        = {{Rolling in the Shadows: Analyzing the Extraction of {MEV} Across Layer-2
                  Rollups}},
  booktitle    = {Proceedings of the 2024 on {ACM} {SIGSAC} Conference on Computer and
                  Communications Security, {CCS} 2024, Salt Lake City, UT, USA, October
                  14-18, 2024},
  pages        = {2591--2605},
  publisher    = {{ACM}},
  year         = {2024},
  url          = {https://doi.org/10.1145/3658644.3690259},
  doi          = {10.1145/3658644.3690259},
  bibsource    = {dblp computer science bibliography, https://dblp.org}
}

@inproceedings{WeintraubPrivate2022,
  author       = {Ben Weintraub and
                  Christof Ferreira Torres and
                  Cristina Nita{-}Rotaru and
                  Radu State},
  editor       = {Chadi Barakat and
                  Cristel Pelsser and
                  Theophilus A. Benson and
                  David R. Choffnes},
  title        = {{A Flash(bot) in the Pan: Measuring Maximal Extractable Value in Private
                  Pools}},
  booktitle    = {Proceedings of the 22nd {ACM} Internet Measurement Conference, {IMC}
                  2022, Nice, France, October 25-27, 2022},
  pages        = {458--471},
  publisher    = {{ACM}},
  year         = {2022},
  url          = {https://doi.org/10.1145/3517745.3561448},
  doi          = {10.1145/3517745.3561448},
  bibsource    = {dblp computer science bibliography, https://dblp.org}
}

@inproceedings{TorresFrontrunner2021,
  author       = {Christof Ferreira Torres and
                  Ramiro Camino and
                  Radu State},
  editor       = {Michael D. Bailey and
                  Rachel Greenstadt},
  title        = {{Frontrunner Jones and the Raiders of the Dark Forest: An Empirical
                  Study of Frontrunning on the Ethereum Blockchain}},
  booktitle    = {30th {USENIX} Security Symposium, {USENIX} Security 2021, August 11-13,
                  2021},
  pages        = {1343--1359},
  publisher    = {{USENIX} Association},
  year         = {2021},
  url          = {https://www.usenix.org/conference/usenixsecurity21/presentation/torres},
  bibsource    = {dblp computer science bibliography, https://dblp.org}
}

@inproceedings{GerzonJito2025,
  author       = {Nicole Gerzon and
                  Ben Weintraub and
                  Junbeom In and
                  Alan Mislove and
                  Cristina Nita{-}Rotaru},
  editor       = {Paul Barford and
                  Aruna Balasubramanian and
                  Alberto Dainotti},
  title        = {{Quantifying the Threat of Sandwiching {MEV} on Jito: {A} Measurement
                  of Solana's Leading Validator Client}},
  booktitle    = {Proceedings of the 2025 {ACM} Internet Measurement Conference, {IMC}
                  2025, Madison, WI, USA, 28-31 October 2025},
  pages        = {937--943},
  publisher    = {{ACM}},
  year         = {2025},
  url          = {https://doi.org/10.1145/3730567.3764493},
  doi          = {10.1145/3730567.3764493},
  bibsource    = {dblp computer science bibliography, https://dblp.org}
}

@inproceedings{BurakFCFS2024,
  author       = {Burak {\"{O}}z and
                  Jonas Gebele and
                  Parshant Singh and
                  Filip Rezabek and
                  Florian Matthes},
  title        = {{Playing the {MEV} Game on a First-Come-First-Served Blockchain}},
  booktitle    = {{IEEE} International Conference on Blockchain and Cryptocurrency,
                  {ICBC} 2024, Dublin, Ireland, May 27-31, 2024},
  pages        = {220--224},
  publisher    = {{IEEE}},
  year         = {2024},
  url          = {https://doi.org/10.1109/ICBC59979.2024.10634397},
  doi          = {10.1109/ICBC59979.2024.10634397},
  bibsource    = {dblp computer science bibliography, https://dblp.org}
}

@article{PahariExclusive2025,
  author       = {Vabuk Pahari and
                  Andrea Canidio},
  title        = {{How Exclusive Are Ethereum Transactions? Evidence from Non-Winning
                  Blocks}},
  journal      = {CoRR},
  volume       = {abs/2509.16052},
  year         = {2025},
  url          = {https://doi.org/10.48550/arXiv.2509.16052},
  doi          = {10.48550/ARXIV.2509.16052},
  eprinttype   = {arXiv},
  eprint       = {2509.16052},
  bibsource    = {dblp computer science bibliography, https://dblp.org}
}

@article{ZhouA2MM2021,
  author       = {Liyi Zhou and
                  Kaihua Qin and
                  Arthur Gervais},
  title        = {{{A2MM:} Mitigating Frontrunning, Transaction Reordering and Consensus
                  Instability in Decentralized Exchanges}},
  journal      = {CoRR},
  volume       = {abs/2106.07371},
  year         = {2021},
  url          = {https://arxiv.org/abs/2106.07371},
  eprinttype   = {arXiv},
  eprint       = {2106.07371},
  bibsource    = {dblp computer science bibliography, https://dblp.org}
}

@inproceedings{QinBEV2022,
  author       = {Kaihua Qin and
                  Liyi Zhou and
                  Arthur Gervais},
  title        = {{Quantifying Blockchain Extractable Value: How Dark Is the Forest?}},
  booktitle    = {43rd {IEEE} Symposium on Security and Privacy, {SP} 2022, San Francisco,
                  CA, USA, May 22-26, 2022},
  pages        = {198--214},
  publisher    = {{IEEE}},
  year         = {2022},
  url          = {https://doi.org/10.1109/SP46214.2022.9833734},
  doi          = {10.1109/SP46214.2022.9833734},
  bibsource    = {dblp computer science bibliography, https://dblp.org}
}

@inproceedings{BabelPROF2026,
  author       = {Kushal Babel and
                  Nerla Jean{-}Louis and
                  Yan Ji and
                  Ujval Misra and
                  Mahimna Kelkar and
                  Kosala Yapa Mudiyanselage and
                  Andrew Miller and
                  Ari Juels},
  title        = {{{PROF:} Protected Order Flow in a Profit-Seeking World}},
  booktitle    = {11th {IEEE} European Symposium on Security and Privacy, EuroS{\&}P
                  2026, Lisbon, Portugal, July 6-10, 2026},
  pages        = {398--418},
  publisher    = {{IEEE}},
  year         = {2026},
  url          = {https://doi.org/10.1109/EuroSP68448.2026.00034},
  doi          = {10.1109/EUROSP68448.2026.00034},
  bibsource    = {dblp computer science bibliography, https://dblp.org}
}

@inproceedings{li2023demystifying,
  author       = {Zihao Li and Jianfeng Li and Zheyuan He and Xiapu Luo and
                  Ting Wang and Xiaoze Ni and Wenwu Yang and Xi Chen and Ting Chen},
  title        = {{Demystifying {DeFi} {MEV} Activities in {Flashbots} Bundle}},
  booktitle    = {Proceedings of the 2023 {ACM} {SIGSAC} Conference on Computer and
                  Communications Security, {CCS} 2023, Copenhagen, Denmark,
                  November 26-30, 2023},
  pages        = {165--179},
  publisher    = {{ACM}},
  year         = {2023},
  doi          = {10.1145/3576915.3616590}
}

@article{oz2025cross,
  title={{Cross-Chain Arbitrage: The Next Frontier of MEV in Decentralized Finance}},
  author={{\"O}z, Burak and Ferreira Torres, Christof and Schlegel, Christoph and Mazorra, Bruno and Gebele, Jonas and Rezabek, Filip and Matthes, Florian},
  journal={Proceedings of the ACM on Measurement and Analysis of Computing Systems},
  volume={9},
  number={3},
  pages={1--33},
  year={2025},
  publisher={ACM New York, NY, USA},
  doi = {10.1145/3771566}
}

@misc{dl_chains,
  title        = {{DefiLlama Chains}},
  author       = {{DefiLlama}},
  howpublished = {\url{https://defillama.com/chains}},
  note         = {Accessed: 2026-08-16},
  year         = {2026}
}

@misc{dl_dexs,
  title        = {{DefiLlama DEXs}},
  author       = {{DefiLlama}},
  howpublished = {\url{https://defillama.com/dexs}},
  note         = {Accessed: 2026-08-16},
  year         = {2026}
}

@inproceedings{SolmazOptimistic2025,
  author       = {Ozan Solmaz and
                  Lioba Heimbach and
                  Yann Vonlanthen and
                  Roger Wattenhofer},
  editor       = {Zeta Avarikioti and
                  Nicolas Christin},
  title        = {{Optimistic {MEV} in Ethereum Layer 2s: Why Blockspace Is Always in
                  Demand}},
  booktitle    = {7th Conference on Advances in Financial Technologies, {AFT} 2025,
                  Pittsburgh, PA, USA, October 8-10, 2025},
  series       = {LIPIcs},
  volume       = {354},
  pages        = {28:1--28:24},
  publisher    = {Schloss Dagstuhl - Leibniz-Zentrum f{\"{u}}r Informatik},
  year         = {2025},
  url          = {https://doi.org/10.4230/LIPIcs.AFT.2025.28},
  doi          = {10.4230/LIPICS.AFT.2025.28},
  bibsource    = {dblp computer science bibliography, https://dblp.org}
}

@article{gogol2025priority,
  title={{When Priority Fails: Revert-Based MEV on Fast-Finality Rollups}},
  author={Gogol, Krzysztof and Schneider, Manvir and Tessone, Claudio},
  journal={arXiv preprint arXiv:2506.01462},
  year={2025}
}

@article{sevim2026signals,
  title={{Signals and Spoils: Speculative Oracle Extractable Value in the Era of Cross-Chain Interoperability}},
  author={Sevim, Hasret Ozan and Ferreira Torres, Christof},
  journal={arXiv preprint arXiv:2606.03434},
  year={2026}
}

@article{WangBlockspace2026,
  author       = {Wenhao Wang and
                  Aditya Saraf and
                  Lioba Heimbach and
                  Kushal Babel and
                  Fan Zhang},
  title        = {{Blockspace Under Pressure: An Analysis of Spam {MEV} on High-Throughput
                  Blockchains}},
  journal      = {CoRR},
  volume       = {abs/2604.00234},
  year         = {2026},
  url          = {https://doi.org/10.48550/arXiv.2604.00234},
  doi          = {10.48550/ARXIV.2604.00234},
  eprinttype   = {arXiv},
  eprint       = {2604.00234},
  bibsource    = {dblp computer science bibliography, https://dblp.org}
}

@article{PahariSpeculative2026,
  author       = {Vabuk Pahari and
                  Johnnatan Messias and
                  Christof Ferreira Torres},
  title        = {{There Will Be Spam: Characterizing State-Invariant Transactions and
                  Speculative {MEV}}},
  journal      = {CoRR},
  volume       = {abs/2607.24172},
  year         = {2026},
  url          = {https://doi.org/10.48550/arXiv.2607.24172},
  doi          = {10.48550/ARXIV.2607.24172},
  eprinttype   = {arXiv},
  eprint       = {2607.24172},
  bibsource    = {dblp computer science bibliography, https://dblp.org}
}

@misc{dark_forest,
  title        = {{Ethereum is a Dark Forest}},
  author       = {{Dan Robinson, Georgios Konstantopoulos}},
  howpublished = {\url{https://www.paradigm.xyz/writing/ethereum-is-a-dark-forest}},
  note         = {Accessed: 2026-08-16},
  year         = {2026}
}

@inproceedings{daian2019flashboys20frontrunning,
  author    = {Philip Daian and Steven Goldfeder and Tyler Kell and Yunqi Li and
               Xueyuan Zhao and Iddo Bentov and Lorenz Breidenbach and Ari Juels},
  title     = {{{Flash Boys 2.0}: Frontrunning in Decentralized Exchanges, Miner
               Extractable Value, and Consensus Instability}},
  booktitle = {2020 {IEEE} Symposium on Security and Privacy, {SP} 2020,
               San Francisco, CA, USA, May 18-21, 2020},
  pages     = {910--927},
  publisher = {{IEEE}},
  year      = {2020},
  doi       = {10.1109/SP40000.2020.00040}
}

@misc{axiom,
  title        = {{Axiom}},
  author       = {{Axiom}},
  howpublished = {\url{https://axiom.trade}},
  note         = {Accessed: 2026-08-16},
  year         = {2026}
}

@misc{photon,
  title        = {{Photon}},
  author       = {{Photon}},
  howpublished = {\url{https://photon-sol.tinyastro.io}},
  note         = {Accessed: 2026-08-16},
  year         = {2026}
}

@misc{gmgn,
  title        = {{GMGN}},
  author       = {{GMGN}},
  howpublished = {\url{https://gmgn.ai}},
  note         = {Accessed: 2026-08-16},
  year         = {2026}
}

@misc{bullx,
  title        = {{BullX}},
  author       = {{BullX}},
  howpublished = {\url{https://bullx.io}},
  note         = {Accessed: 2026-08-16},
  year         = {2026}
}

@misc{trojan,
  title        = {{Trojan}},
  author       = {{Trojan}},
  howpublished = {\url{https://trojan.com}},
  note         = {Accessed: 2026-08-16},
  year         = {2026}
}

@misc{j_ultra,
  title        = {{Ultra Swap API Overview}},
  author       = {{Jupiter}},
  howpublished = {\url{https://developers.jup.ag/docs/ultra}},
  note         = {Accessed: 2026-08-16},
  year         = {2026}
}

@misc{sun2024tronmev,
  author       = {Justin Sun},
  title        = {{Statement on front-running transactions on the {TRON} network}},
  howpublished = {\url{https://x.com/justinsuntron/status/1830984536244789641}},
  note         = {Accessed: 2026-08-21},
  year         = {2024}
}

@misc{mempoolguru,
  author       = {{Mempool.guru}},
  title        = {{Mempool.guru}},
  howpublished = {\url{https://mempool.guru/}},
  note         = {Accessed: 2023-09-26},
  year         = {2023}
}

@misc{artifactrepo,
  author       = {Lioba Heimbach and Ozan Solmaz and Burak {\"O}z and Christof Ferreira Torres},
  title        = {{No Place to Hide: An Analysis on Protected Order Flow Sandwich Attacks}},
  howpublished = {\url{https://github.com/liobaheimbach/No-Place-to-Hide-An-Analysis-on-Protected-Order-Flow-Sandwich-Attacks}},
  note         = {Accessed: 2026-09-21},
  year         = {2026}
}

@misc{xatu-data,
  author       = {{ethPandaOps}},
  title        = {{Xatu Data}},
  year         = {2024},
  howpublished = {\url{https://ethpandaops.io/data/xatu/}},
  note         = {Accessed: 2026-08-21}
}

@misc{beaconchain-light,
  author       = {{beaconcha.in}},
  title        = {{beaconcha.in Light: Ethereum Mainnet Beacon Explorer}},
  year         = {2026},
  howpublished = {\url{https://light-mainnet.beaconcha.in/}},
  note         = {Powered by Dora; accessed: 2026-08-21}
}

@misc{jina-reader,
  author       = {{Jina AI}},
  title        = {{Reader API}},
  year         = {2024},
  howpublished = {\url{https://jina.ai/reader/}},
  note         = {Service endpoint: \url{https://r.jina.ai/};
                  accessed: 2026-08-21}
}

@misc{garimidi2025limits,
  author       = {Pranav Garimidi and Joseph Bonneau and Lioba Heimbach},
  title        = {{On the limits of encrypted mempools}},
  howpublished = {\url{https://a16zcrypto.com/posts/article/limits-encrypted-mempools/}},
  note         = {a16z crypto. Accessed: 2026-08-21},
  year         = {2025}
}

@misc{sfdp2024delist,
  author       = {Jordan Leech},
  title        = {{Solana Foundation removes certain operators from delegation program over malicious sandwich attacks}},
  howpublished = {\url{https://www.theblock.co/news/ecosystems/2024-06-10-solana-foundation-removes-certain-operators-from-delegation-program-over-malicious-sandwich-attacks-299244}},
  note         = {The Block. Accessed: 2026-08-21},
  year         = {2024}
}

@misc{marinade2025mip9,
  author       = {{Marinade}},
  title        = {{{MIP.9}: Blocklisting malicious validators from the Stake Auction Marketplace to combat sandwich attacks}},
  howpublished = {\url{https://forum.marinade.finance/t/mip-9-blocklisting-malicious-validators-from-stake-auction-marketplace-to-combat-sandwich-attacks/1716}},
  note         = {Accessed: 2026-08-21},
  year         = {2025}
}

@misc{jito2025blacklist,
  author       = {Hristina Vasileva},
  title        = {{Jito bans 15 additional validators after data emerges of widespread sandwich attacks}},
  howpublished = {\url{https://www.cryptopolitan.com/jito-bans-15-additional-validators-after-data-emerges-of-widespread-sandwich-attacks/}},
  note         = {Cryptopolitan. Accessed: 2026-08-21},
  year         = {2025}
}

@article{LiFairness2023,
  title={Transaction Fairness in Blockchains, Revisited},
  author={Li, Rujia and Hu, Xuanwei and Wang, Qin and Duan, Sisi and Wang, Qi},
  journal={IEEE Transactions on Dependable and Secure Computing},
  year={2025},
  publisher={IEEE}
}

@inproceedings{KelkarOrderFairness2020,
  title={Order-fairness for byzantine consensus},
  author={Kelkar, Mahimna and Zhang, Fan and Goldfeder, Steven and Juels, Ari},
  booktitle={Annual International Cryptology Conference},
  pages={451--480},
  year={2020},
  organization={Springer}
}

@inproceedings{KelkarThemis2023,
  author    = {Mahimna Kelkar and Soubhik Deb and Sishan Long and Ari Juels and Sreeram Kannan},
  title     = {Themis: Fast, Strong Order-Fairness in Byzantine Consensus},
  booktitle = {Proceedings of the 2023 ACM SIGSAC Conference on Computer and Communications Security},
  pages     = {475--489},
  year      = {2023},
  publisher = {ACM},
  doi       = {10.1145/3576915.3616658}
}

@inproceedings{KiayiasBounded2024,
  title={Ordering transactions with bounded unfairness: Definitions, complexity and constructions},
  author={Kiayias, Aggelos and Leonardos, Nikos and Shen, Yu},
  booktitle={Annual International Conference on the Theory and Applications of Cryptographic Techniques},
  pages={34--63},
  year={2024},
  organization={Springer}
}

@inproceedings{KavousiBlindPerm2023,
  author       = {Alireza Kavousi and
                  Duc Viet Le and
                  Philipp Jovanovic and
                  George Danezis},
  editor       = {Andrei Arusoaie and
                  Emanuel Onica and
                  Michael Spear and
                  Sara Tucci Piergiovanni},
  title        = {BlindPerm: Efficient {MEV} Mitigation with an Encrypted Mempool and
                  Permutation},
  booktitle    = {29th International Conference on Principles of Distributed Systems,
                  {OPODIS} 2025, Ia{\c{s}}i, Romania, December 3-5, 2025},
  series       = {LIPIcs},
  volume       = {361},
  pages        = {36:1--36:21},
  publisher    = {Schloss Dagstuhl - Leibniz-Zentrum f{\"{u}}r Informatik},
  year         = {2025},
  url          = {https://doi.org/10.4230/LIPIcs.OPODIS.2025.36},
  doi          = {10.4230/LIPICS.OPODIS.2025.36},
  bibsource    = {dblp computer science bibliography, https://dblp.org}
}

@inproceedings{BentovTesseract2019,
  author    = {Iddo Bentov and Yan Ji and Fan Zhang and Yunqi Li and Xueyuan Zhao and Lorenz Breidenbach and Philip Daian and Ari Juels},
  title     = {Tesseract: Real-Time Cryptocurrency Exchange Using Trusted Hardware},
  booktitle = {Proceedings of the 2019 ACM SIGSAC Conference on Computer and Communications Security},
  pages     = {1521--1538},
  year      = {2019},
  publisher = {ACM},
  doi       = {10.1145/3319535.3363221}
}

@misc{MonnotPEPC2022,
  author       = {Barnab{\'e} Monnot},
  title        = {Unbundling PBS: Towards Protocol-Enforced Proposer Commitments (PEPC)},
  year         = {2022},
  howpublished = {Ethereum Research},
  url          = {https://ethresear.ch/t/unbundling-pbs-towards-protocol-enforced-proposer-commitments-pepc/13879}
}

@misc{EthereumEIP7732,
  author       = {Ethereum Improvement Proposals},
  title        = {EIP-7732: Enshrined Proposer-Builder Separation},
  year         = {2024},
  howpublished = {Ethereum Improvement Proposals},
  url          = {https://eips.ethereum.org/EIPS/eip-7732}
}

@misc{ThieryFOCIL2024,
  author       = {Thomas Thiery and Francesco D'Amato and Julian Ma and Barnab{\'e} Monnot and Terence Tsao and Jacob Kaufmann and Jihoon Song},
  title        = {EIP-7805: Fork-Choice Enforced Inclusion Lists (FOCIL)},
  year         = {2024},
  howpublished = {Ethereum Improvement Proposals},
  url          = {https://eips.ethereum.org/EIPS/eip-7805}
}

@software{paradigm_cryo,
  author       = {{Paradigm}},
  title        = {cryo: Easy extraction of blockchain data to parquet, CSV, JSON, or Python dataframes},
  year         = {2023},
  url          = {https://github.com/paradigmxyz/cryo},
  publisher    = {GitHub},
  note         = {GitHub repository}
}

@software{evmole,
  author       = {{cdump}},
  title        = {EVMole: Extracts function selectors, arguments, state mutability and storage layout from EVM bytecode, even for unverified contracts},
  year         = {2026},
  url          = {https://github.com/cdump/evmole},
  publisher    = {GitHub},
  note         = {GitHub repository}
}

@misc{eip1967,
  author       = {Santiago Palladino and Francisco Giordano and Hadrien Croubois},
  title        = {EIP-1967: Proxy Storage Slots},
  year         = {2019},
  howpublished = {Ethereum Improvement Proposals},
  url          = {https://eips.ethereum.org/EIPS/eip-1967}
}

@misc{SUAVE2023,
title        = {SUAVE: A Single Unified Auction for Value Expression},
author       = {{Flashbots}},
year         = {2023},
howpublished = {\url{https://writings.flashbots.net/the-future-of-mev-is-suave}},
}

@inproceedings{BEATMEV2025,
  author    = {Jan Bormet and Sebastian Faust and Hussien Othman and Ziyan Qu},
  title     = {{BEAT-MEV}: Epochless Approach to Batched Threshold Encryption for {MEV} Prevention},
  booktitle = {34th USENIX Security Symposium (USENIX Security 25)},
  year      = {2025},
  pages     = {3457--3476},
  publisher = {USENIX Association},
  address   = {Seattle, WA},
  month     = aug
}

@misc{BEASTMEV2025,
  author       = {Jan Bormet and Arka Rai Choudhuri and Sebastian Faust and
                  Sanjam Garg and Hussien Othman and Guru-Vamsi Policharla and
                  Ziyan Qu and Mingyuan Wang},
  title        = {{BEAST-MEV}: Batched Threshold Encryption with Silent Setup for {MEV} Prevention},
  year         = {2025},
  howpublished = {Cryptology ePrint Archive, Paper 2025/1419},
  url          = {https://eprint.iacr.org/2025/1419}
}

@misc{EPRINT:DziFauLuh24,
      author = {Stefan Dziembowski and Sebastian Faust and Jannik Luhn},
      title = {Shutter Network: Private Transactions from Threshold Cryptography},
      howpublished = {Cryptology {ePrint} Archive, Paper 2024/1981},
      year = {2024},
      url = {https://eprint.iacr.org/2024/1981}
}

@misc{EPRINT:MomGorZha22,
      author = {Peyman Momeni and Sergey Gorbunov and Bohan Zhang},
      title = {{FairBlock}: Preventing Blockchain Front-running with Minimal Overheads},
      howpublished = {Cryptology {ePrint} Archive, Paper 2022/1066},
      year = {2022},
      url = {https://eprint.iacr.org/2022/1066}
}

@misc{FC:DHMW23,
      author = {Nico Döttling and Lucjan Hanzlik and Bernardo Magri and Stella Wohnig},
      title = {{McFly}: Verifiable Encryption to the Future Made Practical},
      howpublished = {Cryptology {ePrint} Archive, Paper 2022/433},
      year = {2022},
      url = {https://eprint.iacr.org/2022/433}
}

@misc{AC:CDKKN22,
      author = {Matteo Campanelli and Bernardo David and Hamidreza Khoshakhlagh and Anders Konring and Jesper Buus Nielsen},
      title = {Encryption to the Future: A Paradigm for Sending Secret Messages to Future (Anonymous) Committees},
      howpublished = {Cryptology {ePrint} Archive, Paper 2021/1423},
      year = {2021},
      url = {https://eprint.iacr.org/2021/1423}
}

@misc{EPRINT:GaiMelRom23,
      author = {Nicolas Gailly and Kelsey Melissaris and Yolan Romailler},
      title = {tlock: Practical Timelock Encryption from Threshold {BLS}},
      howpublished = {Cryptology {ePrint} Archive, Paper 2023/189},
      year = {2023},
      url = {https://eprint.iacr.org/2023/189}
}

@misc{EPRINT:CCNPS23,
      author = {Andrea Cerulli and Aisling Connolly and Gregory Neven and Franz-Stefan Preiss and Victor Shoup},
      title = {{vetKeys}: How a Blockchain Can Keep Many Secrets},
      howpublished = {Cryptology {ePrint} Archive, Paper 2023/616},
      year = {2023},
      url = {https://eprint.iacr.org/2023/616}
}

@misc{C:AgaFerPin25,
      author = {Amit Agarwal and Rex Fernando and Benny Pinkas},
      title = {Efficiently-Thresholdizable Batched Identity Based Encryption, with Applications},
      howpublished = {Cryptology {ePrint} Archive, Paper 2024/1575},
      year = {2024},
      url = {https://eprint.iacr.org/2024/1575}
}

@misc{USENIX:CGPW25,
      author = {Arka Rai Choudhuri and Sanjam Garg and Guru-Vamsi Policharla and Mingyuan Wang},
      title = {Practical Mempool Privacy via One-time Setup Batched Threshold Encryption},
      howpublished = {Cryptology {ePrint} Archive, Paper 2024/1516},
      year = {2024},
      url = {https://eprint.iacr.org/2024/1516}
}

@misc{USENIX:CGPP24,
      author = {Arka Rai Choudhuri and Sanjam Garg and Julien Piet and Guru-Vamsi Policharla},
      title = {Mempool Privacy via Batched Threshold Encryption: Attacks and Defenses},
      howpublished = {Cryptology {ePrint} Archive, Paper 2024/669},
      year = {2024},
      url = {https://eprint.iacr.org/2024/669}
}

@misc{EPRINT:SueAshShi24,
  author =       {Sora Suegami and Shinsaku Ashizawa and Kyohei Shibano},
  title =        {{Constant-Cost Batched Partial Decryption in Threshold Encryption}},
  howpublished = {Cryptology {ePrint} Archive, Paper 2024/762},
  url =          {https://eprint.iacr.org/2024/762},
  year =         {2024},
}

@misc{EPRINT:GWWW25,
      author = {Junqing Gong and Brent Waters and Hoeteck Wee and David J. Wu},
      title = {Threshold Batched Identity-Based Encryption from Pairings in the Plain Model},
      howpublished = {Cryptology {ePrint} Archive, Paper 2025/2103},
      year = {2025},
      url = {https://eprint.iacr.org/2025/2103}
}

@misc{EPRINT:ABDG+25,
      author = {Amit Agarwal and Kushal Babel and Sourav Das and Babak Poorebrahim Gilkalaye and Arup Mondal and Benny Pinkas and Peter Rindal and Aayush Yadav},
      title = {Weighted Batched Threshold Encryption with Applications to Mempool Privacy},
      howpublished = {Cryptology {ePrint} Archive, Paper 2025/2115},
      year = {2025},
      url = {https://eprint.iacr.org/2025/2115}
}

@misc{EPRINT:BonLauTas25,
      author = {Dan Boneh and Evan Laufer and Ertem Nusret Tas},
      title = {Threshold Batch Identity-based Encryption without Epochs},
      howpublished = {Cryptology {ePrint} Archive, Paper 2025/1254},
      year = {2025},
      url = {https://eprint.iacr.org/2025/1254}
}

@misc{EPRINT:FPTX25,
      author = {Rex Fernando and Guru-Vamsi Policharla and Andrei Tonkikh and Zhuolun Xiang},
      title = {{TrX}: Encrypted Mempools in High Performance {BFT} Protocols},
      howpublished = {Cryptology {ePrint} Archive, Paper 2025/2032},
      year = {2025},
      url = {https://eprint.iacr.org/2025/2032}
}

@misc{EPRINT:BNRT26,
      author = {Dan Boneh and Rohit Nema and Arnab Roy and Ertem Nusret Tas},
      title = {Efficient Batch Threshold Encryption Using Partial Fraction Techniques},
      howpublished = {Cryptology {ePrint} Archive, Paper 2026/674},
      year = {2026},
      url = {https://eprint.iacr.org/2026/674}
}

@Misc{EPRINT:XioZhaChe25,
  author =       "Saisi Xiong and
                  Jie Chen",
  title =        "{Efficient Batched {IBE} from Lattices in the Standard Model}",
  year =         2025,
  howpublished = "Cryptology ePrint Archive, Report 2025/2158",
  url =          "https://eprint.iacr.org/2025/2158",
}

@misc{passerat2026sok,
      author = {Jonathan Passerat-Palmbach},
      title = {{SoK}: Encrypted Mempools Through the {MEV} Lens},
      howpublished = {Cryptology {ePrint} Archive, Paper 2026/1643},
      year = {2026},
      doi = {10.4230/LIPIcs.AFT.2026.30},
      url = {https://eprint.iacr.org/2026/1643}
}

@misc{agarwal2026btx,
  author       = {Amit Agarwal and Sourav Das and Babak Poorebrahim Gilkalaye and
                  Peter Rindal and Victor Shoup},
  title        = {{{BTX}: Simple and Efficient Batch Threshold Encryption}},
  howpublished = {Cryptology ePrint Archive, Report 2026/754},
  url          = {https://eprint.iacr.org/2026/754},
  year         = {2026}
}

@inproceedings{ZhouHFT2021,
  author       = {Liyi Zhou and Kaihua Qin and Christof Ferreira Torres and Duc Viet Le and Arthur Gervais},
  title        = {{High-Frequency Trading on Decentralized On-Chain Exchanges}},
  booktitle    = {42nd {IEEE} Symposium on Security and Privacy, {SP} 2021},
  pages        = {428--445},
  publisher    = {{IEEE}},
  year         = {2021},
  doi          = {10.1109/SP40001.2021.00027}
}

@inproceedings{WangCyclic2022,
  author       = {Ye Wang and Yan Chen and Haotian Wu and Liyi Zhou and Shuiguang Deng and Roger Wattenhofer},
  title        = {{Cyclic Arbitrage in Decentralized Exchanges}},
  booktitle    = {Companion Proceedings of the Web Conference 2022, {WWW} 2022},
  pages        = {12--19},
  publisher    = {{ACM}},
  year         = {2022},
  doi          = {10.1145/3487553.3524201}
}

@inproceedings{HeimbachNonAtomic2024,
  author       = {Lioba Heimbach and Vabuk Pahari and Eric Schertenleib},
  title        = {{Non-Atomic Arbitrage in Decentralized Finance}},
  booktitle    = {45th {IEEE} Symposium on Security and Privacy, {SP} 2024},
  pages        = {3866--3884},
  publisher    = {{IEEE}},
  year         = {2024},
  doi          = {10.1109/SP54263.2024.00256}
}

@inproceedings{QinLiquidations2021,
  author       = {Kaihua Qin and Liyi Zhou and Pablo Gamito and Philipp Jovanovic and Arthur Gervais},
  title        = {{An Empirical Study of DeFi Liquidations: Incentives, Risks, and Instabilities}},
  booktitle    = {{ACM} Internet Measurement Conference, {IMC} 2021},
  pages        = {336--350},
  publisher    = {{ACM}},
  year         = {2021},
  doi          = {10.1145/3487552.3487811}
}

@inproceedings{GuptaPrivateOrderFlow2023,
  author       = {Tivas Gupta and Mallesh M. Pai and Max Resnick},
  title        = {{The Centralizing Effects of Private Order Flow on Proposer-Builder Separation}},
  booktitle    = {5th Conference on Advances in Financial Technologies, {AFT} 2023},
  series       = {LIPIcs},
  volume       = {282},
  pages        = {20:1--20:15},
  publisher    = {Schloss Dagstuhl - Leibniz-Zentrum f{\"{u}}r Informatik},
  year         = {2023},
  doi          = {10.4230/LIPIcs.AFT.2023.20}
}

@inproceedings{KursaweWendy2020,
  author       = {Klaus Kursawe},
  title        = {{Wendy, the Good Little Fairness Widget: Achieving Order Fairness for Blockchains}},
  booktitle    = {2nd {ACM} Conference on Advances in Financial Technologies, {AFT} 2020},
  pages        = {25--36},
  publisher    = {{ACM}},
  year         = {2020},
  doi          = {10.1145/3419614.3423263}
}

@inproceedings{ZhangByzantineOrdered2020,
  author       = {Yunhao Zhang and Srinath T. V. Setty and Qi Chen and Lidong Zhou and Lorenzo Alvisi},
  title        = {{Byzantine Ordered Consensus without Byzantine Oligarchy}},
  booktitle    = {14th {USENIX} Symposium on Operating Systems Design and Implementation, {OSDI} 2020},
  pages        = {633--649},
  publisher    = {{USENIX} Association},
  year         = {2020},
  url          = {https://www.usenix.org/conference/osdi20/presentation/zhang-yunhao}
}

@inproceedings{KelkarPermissionless2022,
  author       = {Mahimna Kelkar and Soubhik Deb and Sreeram Kannan},
  title        = {{Order-Fair Consensus in the Permissionless Setting}},
  booktitle    = {9th {ACM} {ASIA} Public-Key Cryptography Workshop, APKC@AsiaCCS 2022},
  pages        = {3--14},
  publisher    = {{ACM}},
  year         = {2022},
  doi          = {10.1145/3494105.3526239}
}

@inproceedings{CachinQuickOrderFairness2022,
  author       = {Christian Cachin and Jovana Micic and Nathalie Steinhauer and Luca Zanolini},
  title        = {{Quick Order Fairness}},
  booktitle    = {Financial Cryptography and Data Security, 26th International Conference, {FC} 2022},
  series       = {Lecture Notes in Computer Science},
  volume       = {13411},
  pages        = {316--333},
  publisher    = {Springer},
  year         = {2022},
  doi          = {10.1007/978-3-031-18283-9_15}
}

@inproceedings{ConstantinescuClock2023,
  author       = {Andrei Constantinescu and Diana Ghinea and Lioba Heimbach and Zilin Wang and Roger Wattenhofer},
  title        = {{A Fair and Resilient Decentralized Clock Network for Transaction Ordering}},
  booktitle    = {27th International Conference on Principles of Distributed Systems, {OPODIS} 2023},
  series       = {LIPIcs},
  volume       = {286},
  pages        = {8:1--8:20},
  publisher    = {Schloss Dagstuhl - Leibniz-Zentrum f{\"{u}}r Informatik},
  year         = {2023},
  doi          = {10.4230/LIPIcs.OPODIS.2023.8}
}

@inproceedings{StathakopoulouTEE2021,
  author       = {Chrysoula Stathakopoulou and Signe R{\"{u}}sch and Marcus Brandenburger and Marko Vukolic},
  title        = {{Adding Fairness to Order: Preventing Front-Running Attacks in {BFT} Protocols using TEEs}},
  booktitle    = {40th International Symposium on Reliable Distributed Systems, {SRDS} 2021},
  pages        = {34--45},
  publisher    = {{IEEE}},
  year         = {2021},
  doi          = {10.1109/SRDS53918.2021.00013}
}

@inproceedings{ZhangFlashFreezing2022,
  author       = {Haoqian Zhang and Louis{-}Henri Merino and Vero Estrada{-}Gali{\~{n}}anes and Bryan Ford},
  title        = {{Flash Freezing Flash Boys: Countering Blockchain Front-Running}},
  booktitle    = {42nd {IEEE} International Conference on Distributed Computing Systems Workshops, {ICDCS} Workshops 2022},
  pages        = {90--95},
  publisher    = {{IEEE}},
  year         = {2022},
  doi          = {10.1109/ICDCSW56584.2022.00026}
}

@inproceedings{EskandariSoK2019,
  author       = {Shayan Eskandari and Seyedehmahsa Moosavi and Jeremy Clark},
  title        = {{SoK: Transparent Dishonesty: Front-Running Attacks on Blockchain}},
  booktitle    = {Financial Cryptography and Data Security, {FC} 2019 International Workshops},
  series       = {Lecture Notes in Computer Science},
  volume       = {11599},
  pages        = {170--189},
  publisher    = {Springer},
  year         = {2019},
  doi          = {10.1007/978-3-030-43725-1_13}
}

@inproceedings{HeimbachKTW23,
  author       = {Lioba Heimbach and
                  Lucianna Kiffer and
                  Christof Ferreira Torres and
                  Roger Wattenhofer},
  editor       = {Marie{-}Jos{\'{e}} Montpetit and
                  Aris Leivadeas and
                  Steve Uhlig and
                  Mobin Javed},
  title        = {Ethereum's Proposer-Builder Separation: Promises and Realities},
  booktitle    = {Proceedings of the 2023 {ACM} on Internet Measurement Conference,
                  {IMC} 2023, Montreal, QC, Canada, October 24-26, 2023},
  pages        = {406--420},
  publisher    = {{ACM}},
  year         = {2023},
  url          = {https://doi.org/10.1145/3618257.3624824},
  doi          = {10.1145/3618257.3624824},
  bibsource    = {dblp computer science bibliography, https://dblp.org}
}

\clearpage
\appendix

\section{Data Collection}
\label{sec:appendix-data}
We provide a detailed description of our data collection pipeline in the following.
\subsection{EVM Chains}
\label{sec:data-evm}
We collect swap data from five EVM chains by decoding the respective log events, i.e., Ethereum, Base, Arbitrum, Monad, and Tron. Our coverage spans Uniswap~V2, V3 and V4 together with their identical forks, i.e., SushiSwap, PancakeSwap and SunSwap, as well as Aerodrome~V1 on Base, which gives us the largest automated market maker on each chain~\cite{dl_dexs}. We collect all this over each chain's RPC interface, along with the transaction metadata, fees, block builders, and per-sender transaction histories our analysis relies on.

For Ethereum pool availability and back-run analysis, we construct a catalog of pool creation and registration events. Using an Ethereum RPC and cryo~\cite{paradigm_cryo}, we scan 22 creation event topics covering Uniswap V1--V4-style deployments, Balancer V2 and V3, several Curve factory generations, DODO, Mavrick, Sushiswap Trident, Kyberswap, and Bancor V3. Uniswap V1 is scanned from its factory deployment at block 6,627,917. All others are scanned starting from 10,000,835, the canonical Uniswap V2 factory deployment block. The collection ends on June 30th, 2026. Of the 737,484 collected rows, 733,257 describe two-token pool creations, 581 describe multi-token pools, and 3,646 remain unresolved or excluded. Our pool-count and back-run analysis use only two token pools.

\subsubsection{Mempool Labeling on Ethereum}
For Ethereum, we use the Flashbots public mempool archive~\cite{flashbotsmempooldumpster} to assess whether a swap is public or private. We label a swap \textbf{public} when it was seen in the public mempool prior to inclusion, \textbf{private} when it was not and the archive has good data quality in the relevant hour, and \textbf{unknown} when it is absent during an hour with bad data quality. An hour has bad data quality when the archive holds fewer than half the transactions we expect for it, i.e., the median count for the same hour of day across the $\pm 10$-day window around it. To guard against false private classifications caused by gaps in the public archive, we additionally cross-check transactions labeled private
against an internal Flashbots transaction collection. Any transaction observed
there before inclusion is removed from the private set. Throughout, we treat only the swaps we label private as protected order flow. We note that this is the conservative choice, as it leaves every unknown swap outside the protected order flow and keeps us from labeling a public swap as private. The Flashbots archive begins in August 2023, so for the first weeks of our window we rely on mempool sightings obtained from the mempool.guru project instead~\cite{mempoolguru}. Appendix~\ref{sec:appendix-visibility} reports the resulting split.

\subsubsection{OFA Labeling on Ethereum}
For each identified victim transaction on Ethereum, we determine whether it used an OFA solution. OFA routing is an RPC-level decision made by the submitting user wallet or bot, so it cannot in general be recovered from blockchain data alone: a transaction sent to an OFA and one broadcast directly to the public mempool are, once included in a block, indistinguishable at the protocol level. Some OFA operators nonetheless publish auxiliary datasets---query interfaces, historical archives, or partial-disclosure hint streams---that let us reconstruct routing after the fact, though several of these channels are no longer actively maintained and their coverage windows do not always extend to the present. \Cref{tab:ofa-sources} summarizes, for each provider, the data source we rely on and the period it covers.

For MEV-Share specifically, three such sources are available: a Dune-indexed record submissions, Flashbots' public data browser S3 archive\footnote{Flashbots data browser: \url{https://flashbots-data.s3.us-east-2.amazonaws.com/index.html}}, and the event-stream API\footnote{MEV-Share event stream: \url{https://docs.flashbots.net/flashbots-mev-share/searchers/event-stream}}. We find that the three sources provide complementary coverage, with each capturing victims that are absent from the others. We therefore take the union of all three rather than relying on
any single source, which would otherwise undercount MEV-Share coverage. For MEVBlocker, we rely on the bundle data CoW Protocol uploads to Dune. For Blink, there are no public records of its order flow, so we instead use BuilderNet's internal bundle data, provided by Flashbots. For Merkle, we use bundle data on Dune uploaded by Flashbots, which covers a limited window.

\begin{table}[t]
\centering
\footnotesize
\resizebox{\columnwidth}{!}{%
\begin{tabular}{@{}l l@{}}
\toprule
\textbf{Data Source} & \textbf{Coverage Period} \\
\midrule
\textbf{MEV-Share} \\
\quad Dune (\texttt{dune.flashbots.dataset\_protect\_transactions})
& 2022-02 -- 2026-05 \\
\quad Flashbots Data Browser & 2022-10 -- 2026-03 \\
\quad Historical event-stream API & 2024-02 -- 2026-07 \\
\addlinespace
\textbf{MEVBlocker} \\
\quad Dune (\texttt{mevblocker.raw\_bundles}) & 2023-04 -- 2026-07 \\
\addlinespace
\textbf{Blink} \\
\quad Bundle data from BuilderNet & 2025-08 -- 2026-07 \\
\addlinespace
\textbf{Merkle} \\
\quad Dune (\texttt{dune.flashbots.merkle}) & 2025-01 -- 2026-03 \\
\bottomrule
\end{tabular}}
\caption{Data sources used to label Ethereum sandwich victims by OFA
provider, and the coverage period of each.}
\label{tab:ofa-sources}
\end{table}

\parhead{Coverage Limitations.}\label{par:ofa-coverage}
Our OFA attribution is incomplete. Coverage differs across providers and over time. In particular, our Blink data consists of bundles observed by BuilderNet and is available only from August 2025 onward, so we cannot identify earlier Blink submissions or bundles that were not received by BuilderNet. More generally, the external datasets used for all four OFAs may omit transactions even within their nominal coverage periods. We therefore treat OFA labels as positive evidence of exposure rather than exhaustive routing ground truth where the absence of a label does not imply that a transaction did not pass through an OFA.

\subsection{Reorged Blocks}
\label{sec:data-reorg}
We query Xatu~\cite{xatu-data} for Ethereum \texttt{chain\_reorg} events during our study period. These events are observer-local, so several nodes may report the same reorg. Grouping identical $(\mathit{slot},\mathit{oldHead},\mathit{newHead})$ tuples yields $63{,}842$ distinct transitions. These cover only what Xatu's nodes observed, so the network-wide count may be higher. From each old head, we follow parent roots to the eventually finalized canonical chain. We exclude $71$ roots that ultimately become canonical and compare each remaining execution-block hash with the canonical hash at the same height. This leaves $14{,}398$ unique, eventually non-canonical execution payloads, of which $14{,}397$ have complete ancestry. For all but one, the available Xatu metadata allow us to trace the corresponding beacon ancestry back to a finalized-canonical ancestor. For the remaining root, the parent-chain metadata stop before reaching such an ancestor.

We retrieve their transaction lists from Dora~\cite{beaconchain-light}, using Jina~\cite{jina-reader} as transport fallback. We validate each body's execution-block hash, height, parent hash and transaction count against Xatu. We recover \(9{,}493\) bodies (\(65.9\%\)), containing \(1{,}831{,}716\) transaction-inclusion occurrences. We were not able to obtain the remaining \(4{,}905\) bodies with publicly available data sources. %

\subsection{Solana}
\label{sec:data-solana}

We collect swap data from Solana by decoding each transaction's instructions and token-balance deltas, as Solana offers no event logs. Our coverage spans Raydium (V4, CLMM, CPMM), Orca Whirlpools, Meteora (DLMM, DAMM v2), Pump.fun (bonding curve, PumpSwap), SolFi v1, and Lifinity v2, which gives us the largest automated market makers on Solana~\cite{dl_dexs}. We collect all this through per-slot RPC calls, along with the slot leaders, the SPL token transfers, the transaction fees, the Jito tips, the application fees, and the routing programs.

\section{Per-Chain Detection Details}
\label{sec:detection-appendix}

The heuristics of Section~\ref{sec:detection-appendix-body} share one pipeline. This appendix records the parameters and adaptations that differ by chain.

\parhead{Amount Match (H1).}
On the EVM chains we set the tolerance to $\epsilon = 0.01$ on both sides, and a transaction that trades in both directions of the pool at once cannot serve as a leg. Solana runs detection with a relaxed lower bound, i.e., $A_{\mathrm{b,in}} \geq 0.5\,A_{\mathrm{f,out}}$, so that we do not miss a back leg cashed out over several transactions. We still require the exit to reach 99\% of the front output. It reaches it in one of two ways. Either the back legs inside the window sum to at least 99\%, or the wallet sells at least 99\% of the position later on, i.e., a staged exit.

\parhead{Victims (H1).}
A victim is any swap in the direction of the front leg by another wallet, and not itself a candidate leg, that sits strictly between the first front leg and the first back leg. A bot can wrap one victim in two nested round trips on the same pool. We then credit only the inner pair, so a bot never counts the same victim twice. Two different bots that each sandwich the same victim both count, since that is competition for one target. Solana is stricter. There a victim must trade at least 1\% of the front leg's size and execute at a worse price than the attacker's front leg, and we ignore SOL-paying fronts below 0.01~SOL as dust.

\parhead{Window and Types (H1).}
The EVM window spans five blocks between the front and the back leg. On Solana a leader produces four consecutive slots, so we bound the gap in \emph{leader rotations} instead, i.e., a back leg pairs with a front leg up to two leaders earlier, which lets an attack span three leaders in total. In both regimes a sandwich is \textbf{tight} when it consists of exactly three consecutive transactions in one block, or slot: the front leg, the victim, and the back leg. It is \textbf{within-block wide} when it stays inside one block with more transactions between the legs, and \textbf{cross-block wide} otherwise.

\parhead{Entity Resolution (H3).}
On the EVM chains we resolve every leg and victim to the account that signed it, and we retain bots of three kinds. A \emph{single sender} signs all of its legs itself. A \emph{recurring sender pair} splits them across two accounts, one for the front and one for the back transaction, that call the same address in every sandwich. A \emph{bot contract} is the receiver address of both transactions and the trader of the swap log, in every profitable sandwich we attribute to it.

We count only the sandwiches our detection links, and every candidate must clear the count and the rate floor. Note that on Ethereum we test the rate on protected flow alone. The density is the number of distinct attributed leg transactions divided by all swaps the accounts or the contract make in our study window. We attribute overlapping sandwiches to the candidate contract. A sender or a pair keeps its remaining sandwiches as a separate bot only when that remainder clears every floor. We label a pair by the bot address, while its membership and density stay pair-based.

On Solana the bot is the \texttt{trader} wallet for \textbf{vanilla} sandwiches. For \textbf{evasive} sandwiches the front wallet transfers the bought token to the back wallet inside the front transaction. We attribute the sandwich to the front wallet.

\parhead{Victim Visibility (Ethereum).}
On Ethereum we additionally label the visibility of a transaction in the mempool (cf.\ Appendix~\ref{sec:appendix-data}). A protected order flow sandwich then has at least one private victim transaction. We apply the full H3 filter to this subset, i.e., at least 100 protected sandwiches, at least 60\% of them profitable, and a density of at least 25\%. Density counts protected-sandwich legs over all study-window swap rows of the bot. We further require a private-victim share of 66\%, where the numerator counts only the sandwiches whose victims are all private and the denominator counts all of the bot's sandwiches.

\section{Threshold Sensitivity}
\label{sec:appendix-ablation}

The selection rule carries four floors, i.e., a total sandwich count of 100, a profit rate of 60\%, a density of 25\%, and on Ethereum a private-victim share of 66\%. In the following, we report how the retained population moves when each floor varies. Table~\ref{tab:ablation} sweeps density against the count floor for every chain, Table~\ref{tab:ablation-rate} sweeps the profit rate and the private-victim share, and Table~\ref{tab:ablation-epsilon} varies the amount-match tolerance $\epsilon$.

\parhead{Count.} The count floor gives the two ratio gates evidence to work with. An address with two profitable sandwiches out of two scores a perfect rate and a high density on what may well be coincidence. Lowering the floor to 50 leaves the Ethereum private scope and Tron unchanged, adds 37 addresses on public Ethereum, admits two 74-sandwich bots on Base, and adds 3{,}361 bots on Solana that together carry only 0.2 million further sandwiches. Raising it cuts real bots, e.g., at 500 Ethereum loses one of its seven bots and Tron four of its twelve. We set the floor at 100 because it sits exactly between these regimes, i.e., lowering it only adds marginal addresses and raising it starts to cost real bots. The two 74-sandwich Base bots are the price, i.e., they clear every other gate and plausibly attack, and we exclude them to guard against coincidence.

\parhead{Profit Rate.} The profit-rate floor asks whether a bot mostly succeeds. Ethereum retains eight entities at 0\% and 25\%, and seven at 50--75\%, wile Tron retains twelve throughout that range (Table~\ref{tab:ablation-rate}). On Base, 144 entities clear the count and density floors at 0\%, 14 at 25\%, 5 at 50\%, and the four reported entities at 60\%, whose rates run 93.9--100\%. At 75\%, a fifth sender qualifies because its overlapping contract no longer passes candidacy. On Solana bots spread continuously across the range, i.e., 10{,}248 pass at 0\% and 2{,}925 at 90\%, and no threshold separates two populations. We place the floor at 60\%, where the low-profit Base candidates are excluded while all reported EVM entities remain.

\parhead{Density.} The density floor separates dedicated sandwich bots from high-frequency traders whose two-way flow matches the sandwich structure incidentally. The Ethereum set is stable from 25\% to 60\% and admits one further entity at 10\%. Tron loses a 10{,}438-sandwich entity with a density of 31.7\% once the floor reaches 33\%, and Base loses the 239-sandwich campaign bot with a density of 51.5\% at 60\%. On Solana the floor does most of the selection and cuts 14{,}808 candidate bots to 8{,}631 at the operating point. We set the floor at 25\% because every entity we verified by hand clears it with a wide margin. On the EVM chains, the two lowest genuine densities in our study are the 31.7\% entity on Tron and the 51.5\% entity on Base named above.

\parhead{Private-Victim Share.} The private-victim share floor applies to Ethereum alone, where we observe victim visibility, and it separates the bots that specialize on protected order flow from public mempool bots with incidental private victims. This floor binds, i.e., 31 entities pass every other gate without a share requirement, 15 remain at 50\%, and 7 at 66\% (Table~\ref{tab:ablation-rate}). The floor further sits on a plateau. The set of seven is unchanged up to 80\% and the retained entities carry shares of 89.8--96.8\%, so two thirds selects the specialized population.

\parhead{Amount-Match Tolerance.} The tolerance $\epsilon = 0.01$ acts at detection time and defines which round trips count as a sandwich, whereas the four floors filter the detected entities afterwards. Table~\ref{tab:ablation-epsilon} re-filters the selected EVM sandwiches without rerunning detection or entity selection. Tightening the stored amount-match condition tenfold removes about 0.2\% of these attacks at most on the EVM chains. Solana runs with the relaxed bound of Appendix~\ref{sec:detection-appendix} and is not part of this sweep.

Taken together, every reported EVM bot clears the count and rate floors with a margin, and these floors exist to exclude degenerate candidates. The tolerance barely moves the counts. The two floors that select, i.e., the density on all chains and the private-victim share on Ethereum, both sit on plateaus of the sweep, below every verified bot and above the excluded population.

\begin{table}[t!]
  \centering
  \subcaptionbox{Ethereum (Private Scope)\label{tab:ablation-private}}{%
    {\scriptsize\setlength{\tabcolsep}{3pt}\begin{tabular}{@{}w{l}{30pt}*{4}{w{r}{44pt}}@{}}
\toprule
\textbf{Density $\geq$} & \textbf{SWs $\geq$50} & \textbf{SWs $\geq$100} & \textbf{SWs $\geq$500} & \textbf{SWs $\geq$1000} \\
\midrule
10\% & 9 / 31{,}095 & 8 / 31{,}018 & 6 / 30{,}248 & 4 / 28{,}457 \\
25\% & 7 / 30{,}607 & \textbf{7 / 30{,}607} & 6 / 30{,}248 & 4 / 28{,}457 \\
33\% & 7 / 30{,}607 & 7 / 30{,}607 & 6 / 30{,}248 & 4 / 28{,}457 \\
50\% & 7 / 30{,}607 & 7 / 30{,}607 & 6 / 30{,}248 & 4 / 28{,}457 \\
60\% & 7 / 30{,}607 & 7 / 30{,}607 & 6 / 30{,}248 & 4 / 28{,}457 \\
\bottomrule
\end{tabular}
}}

  \medskip
  \subcaptionbox{Ethereum (Public Scope)\label{tab:ablation-public}}{%
    {\scriptsize\setlength{\tabcolsep}{3pt}\begin{tabular}{@{}w{l}{30pt}*{4}{w{r}{44pt}}@{}}
\toprule
\textbf{Density $\geq$} & \textbf{SWs $\geq$50} & \textbf{SWs $\geq$100} & \textbf{SWs $\geq$500} & \textbf{SWs $\geq$1000} \\
\midrule
10\% & 278 / 2.74M & 237 / 2.74M & 125 / 2.71M & 96 / 2.69M \\
25\% & 232 / 1.79M & \textbf{195 / 1.79M} & 101 / 1.77M & 76 / 1.75M \\
33\% & 208 / 0.76M & 174 / 0.76M & 87 / 0.74M & 66 / 0.72M \\
50\% & 163 / 0.58M & 134 / 0.57M & 64 / 0.56M & 46 / 0.55M \\
60\% & 131 / 0.47M & 105 / 0.47M & 49 / 0.45M & 36 / 0.45M \\
\bottomrule
\end{tabular}
}}

  \medskip
  \subcaptionbox{Tron\label{tab:ablation-tron}}{%
    {\scriptsize\setlength{\tabcolsep}{3pt}\begin{tabular}{@{}w{l}{30pt}*{4}{w{r}{44pt}}@{}}
\toprule
\textbf{Density $\geq$} & \textbf{SWs $\geq$50} & \textbf{SWs $\geq$100} & \textbf{SWs $\geq$500} & \textbf{SWs $\geq$1000} \\
\midrule
10\% & 14 / 38{,}755 & 13 / 38{,}694 & 8 / 37{,}421 & 6 / 36{,}022 \\
25\% & 12 / 38{,}567 & \textbf{12 / 38{,}567} & 8 / 37{,}421 & 6 / 36{,}022 \\
33\% & 11 / 28{,}129 & 11 / 28{,}129 & 7 / 26{,}983 & 5 / 25{,}584 \\
50\% & 11 / 28{,}129 & 11 / 28{,}129 & 7 / 26{,}983 & 5 / 25{,}584 \\
60\% & 11 / 28{,}129 & 11 / 28{,}129 & 7 / 26{,}983 & 5 / 25{,}584 \\
\bottomrule
\end{tabular}
}}

  \medskip
  \subcaptionbox{Base\label{tab:ablation-base}}{%
    {\scriptsize\setlength{\tabcolsep}{3pt}\begin{tabular}{@{}w{l}{30pt}*{4}{w{r}{44pt}}@{}}
\toprule
\textbf{Density $\geq$} & \textbf{SWs $\geq$50} & \textbf{SWs $\geq$100} & \textbf{SWs $\geq$500} & \textbf{SWs $\geq$1000} \\
\midrule
10\% & 51 / 10{,}085 & 39 / 9{,}144 & 2 / 1{,}448 & 0 / 0 \\
25\% & 6 / 2{,}037 & \textbf{4 / 1{,}889} & 2 / 1{,}448 & 0 / 0 \\
33\% & 6 / 2{,}037 & 4 / 1{,}889 & 2 / 1{,}448 & 0 / 0 \\
50\% & 4 / 1{,}889 & 4 / 1{,}889 & 2 / 1{,}448 & 0 / 0 \\
60\% & 3 / 1{,}650 & 3 / 1{,}650 & 2 / 1{,}448 & 0 / 0 \\
\bottomrule
\end{tabular}
}}

  \medskip
  \subcaptionbox{Solana\label{tab:ablation-solana}}{%
    {\scriptsize\setlength{\tabcolsep}{3pt}\begin{tabular}{@{}w{l}{30pt}*{4}{w{r}{44pt}}@{}}
\toprule
\textbf{Density $\geq$} & \textbf{SWs $\geq$50} & \textbf{SWs $\geq$100} & \textbf{SWs $\geq$500} & \textbf{SWs $\geq$1000} \\
\midrule
10\% & 21{,}642 / 35.4M & 14{,}808 / 34.9M & 5{,}623 / 32.8M & 3{,}539 / 31.3M \\
25\% & 11{,}992 / 28.3M & \textbf{8{,}631 / 28.0M} & 3{,}648 / 26.9M & 2{,}345 / 26.0M \\
33\% & 9{,}178 / 24.5M & 6{,}449 / 24.3M & 2{,}843 / 23.5M & 1{,}822 / 22.7M \\
50\% & 5{,}016 / 18.1M & 3{,}582 / 18.0M & 1{,}508 / 17.6M & 970 / 17.2M \\
60\% & 3{,}086 / 15.7M & 2{,}297 / 15.6M & 883 / 15.3M & 558 / 15.0M \\
\bottomrule
\end{tabular}}}

  \caption{Retained population under variation of the density and sandwich-count floors, by chain. Note that public Ethereum only uses swap-log identities. Every cell is bots / sandwiches (M for millions), with the chosen operating point (density $\geq$ 25\%, at least 100 sandwiches) in bold. The profit-rate floor is held at 60\% throughout and Ethereum's private-victim share at its operating point of 66\%.
  Table~\ref{tab:ablation-rate} varies those two. Arbitrum and Monad are omitted, as neither retains a bot anywhere on the grid.}
  \label{tab:ablation}
\end{table}

\begin{table}[t!]
  \centering
  \resizebox{\columnwidth}{!}{% Generated by build_ablation_rate.py.
\begin{tabular}{@{}lrrrrrr@{}}
\toprule
\multicolumn{7}{l}{\textbf{\emph{Profit Rate} $\geq$}} \\
 & \textbf{0\%} & \textbf{25\%} & \textbf{50\%} & \textbf{60\%} & \textbf{75\%} & \textbf{90\%} \\
\midrule
Ethereum (Private) & 8 & 8 & 7 & \textbf{7} & 7 & 4 \\
Tron & 12 & 12 & 12 & \textbf{12} & 12 & 11 \\
Base & 144 & 14 & 5 & \textbf{4} & 5 & 4 \\
Solana & 10{,}248 & 10{,}153 & 9{,}162 & \textbf{8{,}631} & 6{,}545 & 2{,}925 \\
\midrule
\multicolumn{7}{l}{\textbf{\emph{Private-Victim Share} $\geq$ (Ethereum Only)}} \\
 & \textbf{0\%} & \textbf{50\%} & \textbf{66\%} & \textbf{80\%} & \textbf{90\%} & \textbf{95\%} \\
\midrule
Ethereum (Private) & 31 & 15 & \textbf{7} & 7 & 6 & 3 \\
\bottomrule
\end{tabular}
}
  \caption{Retained population under variation of the two floors Table~\ref{tab:ablation} holds fixed, with density and sandwich count at their operating point. Cells show the number of bots and the chosen floor is in bold.}
  \label{tab:ablation-rate}
\end{table}

\begin{table}[h]
  \centering
  \resizebox{\columnwidth}{!}{% Amount-match tolerance sensitivity. Generated by
% build_ablation_epsilon.py -- do not edit by hand.
\begin{tabular}{@{}lrrrr@{}}
\toprule
 & \textbf{$\epsilon$=0.001} & \textbf{$\epsilon$=0.0025} & \textbf{$\epsilon$=0.005} & \textbf{$\epsilon$=0.01} \\
\midrule
Ethereum (private) & 7 / 30{,}575 & 7 / 30{,}584 & 7 / 30{,}587 & \textbf{7 / 30{,}607} \\
Tron & 12 / 38{,}486 & 12 / 38{,}511 & 12 / 38{,}530 & \textbf{12 / 38{,}567} \\
Base & 4 / 1{,}889 & 4 / 1{,}889 & 4 / 1{,}889 & \textbf{4 / 1{,}889} \\
\bottomrule
\end{tabular}
}
  \caption{Post-filter sensitivity to the amount-match tolerance $\epsilon$ within the selected EVM population. Cells are represented bots / retained sandwiches and the chosen value is in bold.}
  \label{tab:ablation-epsilon}
\end{table}

\section{Ethereum Mempool Visibility}
\label{sec:appendix-visibility}

\begin{figure}[t]
  \centering
    \centering
    \includegraphics[width=\linewidth]{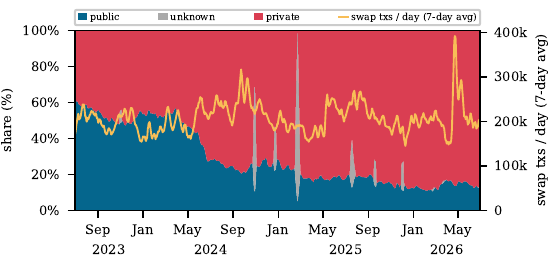}
  \caption{Ethereum mempool visibility per day. Shaded area, share by visibility (left axis, stacked to 100\%). Line, daily swap count (right axis) smoothed over seven days.}
  \label{fig:eth-visibility}
\end{figure}

For Ethereum, our analysis of protected order flow sandwiches covers only swaps that did not enter the public mempool prior to inclusion in a block. Figure~\ref{fig:eth-visibility} shows how the visibility of swaps evolves over our window. The public share starts at 60\% in mid 2023 and falls below 20\% by mid 2026. We note that our mempool data has periods of poor quality, i.e., the public sources do not reliably record what was gossiped in those hours. We mark the transactions of such hours as unknown, unless they were observed, and keep them outside the protected scope throughout the analysis, which is the conservative choice.

\section{Ethereum Public Mempool Sandwiches}
\label{sec:appendix-public}
\begin{figure}[ht]
  \centering
  \includegraphics[width=\linewidth]{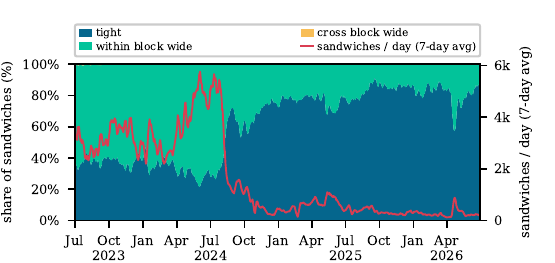}
  \caption{Public sandwich attacks per day on Ethereum. Shaded area, attack-type composition (left axis, stacked to 100\%). Line, daily count (right axis). Both smoothed over seven days.}
  \label{fig:eth-public-per-day}
\end{figure}

To illustrate the difference between protected order flow sandwiches and the public mempool sandwiches studied in prior work, we turn to the latter in the following. Figure~\ref{fig:eth-public-per-day} plots the daily number of \emph{public} Ethereum sandwiches and their split between tight and within-block wide. Only 0.2\% of public sandwiches span a block boundary. Within-block wide attacks are common at first, and from late 2024 onward the majority are tight. The daily count drops sharply over the same period.
Further, the wide attack classification is misleading as the associated sandwiches generally take the following structure:

\begin{center}\small
\begin{tabular}{@{}lll@{}}
\toprule
\textbf{Index} & \textbf{Role} & \textbf{Pool} \\
\midrule
170 & Front-Run & \emph{Both pools across one transaction} \\
171 & Victim & WETH/CTM \\
172 & Victim & USDT/H \\
173 & Back-Run & \emph{Both pools across one transaction} \\
\bottomrule
\end{tabular}
\end{center}

The attacker sends one front transaction and one back transaction that serve two pools at once, i.e., a bundle of two sandwiches, likely to save gas. Detection records one sandwich per pool, and each of them looks wide because the other pool's victim sits between its own legs. As the daily sandwich count falls, the opportunities to sandwich several pools in one block fall with it, which could explain why the wide pattern disappears.

Comparing this with Figure~\ref{fig:eth-per-day}, two things stand out. Public sandwiches are far more numerous than those on protected order flow, and a much larger share of them is tight. An attacker there sees the victim and bundles both legs around it and likely submits this bundle to an MEV auction, so the higher precision is what we would expect.

\begin{figure}[h]
  \centering
  \begin{subfigure}[t]{0.49\linewidth}
    \centering
    \includegraphics[width=\linewidth]{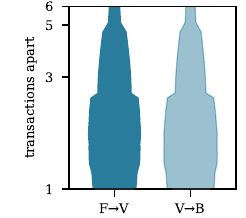}
    \caption{}
    \label{fig:eth-public-txs-apart}
  \end{subfigure}
  \hfill
  \begin{subfigure}[t]{0.49\linewidth}
    \centering
    \includegraphics[width=\linewidth]{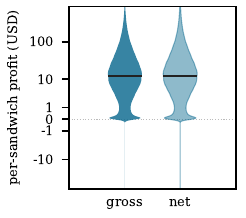}
    \caption{}
    \label{fig:eth-public-profit}
  \end{subfigure}
  \caption{Ethereum public-victim sandwiches (1{,}786{,}785 attacks, 195 bots). \subref{fig:eth-public-txs-apart}~Transaction-index distance from the front leg to the first victim (F$\rightarrow$V) and from that victim to the back leg (V$\rightarrow$B). \subref{fig:eth-public-profit}~Per-sandwich profit, gross and net of gas.}
  \label{fig:eth-public-appendix}
\end{figure}

Figure~\ref{fig:eth-public-txs-apart} shows the distance from the front transaction to the victim and from the victim to the back transaction. The two sides are symmetric, in line with the bundled example above, and the median is one transaction, i.e., the legs sit directly around the victim. Protected order flow sandwiches spread far wider (cf.\ Figure~\ref{fig:txs-apart}). Per-attack profit looks much the same across the two, at a public median of around \$10 (Figure~\ref{fig:eth-public-profit}) against a comparable median on protected flow (cf.\ Figure~\ref{fig:profit-violin}).

The difference lies in the share of attacks that turn a profit. Table~\ref{tab:public-profitability} reports the per-bot profitability ratio for the Ethereum public bots, at 98.9\% across all of them, against 91.9\% on protected flow. Public attacks act on full visibility of the victim and land exactly, whereas attacks on protected flow work from an incomplete picture (about the position of the transaction in the block) and carry the corresponding risk.

Finally, the number of bots differs dramatically, at 195 on the public mempool against seven bots run by a single entity on protected flow. The volume difference likely accounts for part of this, i.e., 1.79 million public attacks against 30,607 protected ones.

\begin{table}[h!]
  \centering

  \resizebox{0.85\linewidth}{!}{\begin{tabular}{@{}lrrr@{}}
\toprule
\textbf{Bot} & \textbf{Sandwiches} & \textbf{Profitable} & \textbf{Density} \\
\midrule
\href{https://etherscan.io/address/0x6b75d8af000000e20b7a7ddf000ba900b4009a80}{\texttt{0x6b7...a80}} & 985{,}291 & 98.7\% & 29.3\% \\
\href{https://etherscan.io/address/0x00000000a991c429ee2ec6df19d40fe0c80088b8}{\texttt{0x000...8b8}} & 218{,}842 & 100.0\% & 71.9\% \\
\href{https://etherscan.io/address/0x4736b02db015dcd1a57a69c889d073b100000000}{\texttt{0x473...000}} & 70{,}691 & 100.0\% & 50.3\% \\
\href{https://etherscan.io/address/0x429cf888dae41d589d57f6dc685707bec755fe63}{\texttt{0x429...e63}} & 64{,}332 & 91.2\% & 49.2\% \\
\href{https://etherscan.io/address/0x00000023c10000eecb940000b914cdfd76cc83d1}{\texttt{0x000...3d1}} & 52{,}632 & 100.0\% & 65.4\% \\
\href{https://etherscan.io/address/0x00fc00900000002c00be4ef8f49c000211000c43}{\texttt{0x00f...c43}} & 28{,}857 & 98.4\% & 34.7\% \\
\href{https://etherscan.io/address/0x8af9ca49688e52787f31742dc259002148efaa62}{\texttt{0x8af...a62}} & 24{,}819 & 100.0\% & 81.0\% \\
\href{https://etherscan.io/address/0x6e0064cb01008bcb00a91f00dc43e500a2ce00d6}{\texttt{0x6e0...0d6}} & 18{,}369 & 100.0\% & 86.4\% \\
\href{https://etherscan.io/address/0xe44a2179edab1e638eac86079fe4ed850ca77b6c}{\texttt{0xe44...b6c}} & 17{,}271 & 100.0\% & 28.8\% \\
\href{https://etherscan.io/address/0x00000000003b3cc22af3ae1eac0440bcee416b40}{\texttt{0x000...b40}} & 15{,}087 & 99.9\% & 36.4\% \\
\midrule
\textbf{All 195 Bots} & \textbf{1{,}786{,}785} & \textbf{98.9\%} & --- \\
\bottomrule
\end{tabular}
}
   \caption{Profitability ratio of Ethereum public-victim real bots (top ten by sandwich count). Profitable, share of the bot's sandwiches with a positive token-level round trip. Density, sandwich-leg share of the bot's swaps.}\label{tab:public-profitability}
\end{table}

\section{Sandwich Attack Structure on Solana}
\label{sec:appendix-sol-structure}

\newcommand{\soltx}[2]{\href{https://solscan.io/tx/#1}{\texttt{#2}}}
\newcommand{\solacct}[2]{\href{https://solscan.io/account/#1}{\texttt{#2}}}

\begin{table}[h]
  \centering
  \resizebox{\linewidth}{!}{%
  \begin{tabular}{@{}lllll@{}}
    \toprule
      \textbf{Slot} & \textbf{Role} & \textbf{Swapping Account} & \textbf{Amount} & \textbf{Transaction} \\
    \midrule
      329{,}378{,}888 & Front-Run & \solacct{FLg7qbsMkB1j5VvLNAqY3ZWb9RSVAX2EZhSai6d9w7kx}{FLg7qbsM} & 0.750000 SOL & \soltx{4Rdtny5TK6e8CSXbWyX5wX6Ydw59UgLmWFmzfg3FY9cqryaY4NTkP8okMa9uTX52U2y3JxYdkfRBSNbXVvVD7QFj}{4Rdtny5T} \\
      329{,}378{,}888 & \multicolumn{2}{l}{ Transfer \solacct{FLg7qbsMkB1j5VvLNAqY3ZWb9RSVAX2EZhSai6d9w7kx}{FLg7qbsM} $\to$ \solacct{7KGtS9vdJ3aMJGeG6sb2zfu9eS8KSVGCBaGodTkZgwbS}{7KGtS9vd}} & 130.15M tok & (same tx) \\
      329{,}378{,}889 & Victim & \solacct{F1M6SgP9Vt8uxjHDPWZntpgKwuSg6vh8ek98RpY9J9FP}{F1M6SgP9} & 0.010000 SOL & \soltx{2SnDsFPC8NrwBaCzi6WPvc8QkkRJqK5WsU5oFhFjxriAK3Nfoucc5LH5uS1gxnJeueHkkUzvXeKNaTJcJ5ZkJ8Af}{2SnDsFPC} \\
      329{,}378{,}890 & Back-Run & \solacct{7KGtS9vdJ3aMJGeG6sb2zfu9eS8KSVGCBaGodTkZgwbS}{7KGtS9vd} & 0.750830 SOL & \soltx{4QrrxiYy91n8LTRZUcFtMUtRu583RsLfzxTJS4WX37TPvRGFkfqcVsjhygrYkUQXzDH93TzQmcGqJCSFvNfaHtND}{4QrrxiYy} \\
    \bottomrule
  \end{tabular}}
  \caption{Cross-trader sandwich on Raydium CPMM pool \solacct{2AMKkPfxJXUFYeZmep18LHqxVvFcGWhHDhohse256b9P}{2AMKkPfx}. The front buy is executed by \solacct{FLg7qbsMkB1j5VvLNAqY3ZWb9RSVAX2EZhSai6d9w7kx}{FLg7qbsM} and the back sell by \solacct{7KGtS9vdJ3aMJGeG6sb2zfu9eS8KSVGCBaGodTkZgwbS}{7KGtS9vd}, which returns the full 130{,}151{,}136 tokens the front acquired. The front transaction itself hands those tokens to the back trader in a third instruction, which is the on-chain link between the two accounts. Profit 0.000830 SOL.}
  \label{tab:sol-ex-crosstrader}
\end{table}

\begin{table}[h]
  \centering
  \resizebox{\linewidth}{!}{%
  \begin{tabular}{@{}lllll@{}}
    \toprule
      \textbf{Slot} & \textbf{Role} & \textbf{Signer} & \textbf{Amount} & \textbf{Transaction} \\
    \midrule
      311{,}083{,}629 & Front-Run & \solacct{FnyFZ6v8RVuUgmvBetnh7ASdNcU5Dem3b8Vi6MDwk1hV}{FnyFZ6v8} & 37.13M tok & \soltx{44htaJt3YJkFESbVdg6WgekL72N2BtpkaHtxvhaxWaq2Lj2koQRpmER7HymXYmh7euocMfkivtipoDTUVV37yWW8}{44htaJt3} \\
      311{,}083{,}629 & Victim & \solacct{HNCpJeAqCXJ5WE3zKL6mCEbdTLwS1V8S4MRbdJHdzjLB}{HNCpJeAq} & 0.032 SOL & \soltx{5EzgxMZms42M6fkphKfqN4wdz2NhPYGtjZMmynwSAU72T42yzBWmUyVoFu5qz3MJ6NswvqPSQBS5D7T29iHPMoyc}{5EzgxMZm} \\
      311{,}083{,}630 & Back-Run 1 & \solacct{FnyFZ6v8RVuUgmvBetnh7ASdNcU5Dem3b8Vi6MDwk1hV}{FnyFZ6v8} & 18.57M tok & \soltx{5UH8CYfxTALPcRBGb5WukHk3waBF9gucnqQHnFZeD22RXWmeGDYAFqX9hWKU5JiMciKxcBiwp5W7PWxf835txXj5}{5UH8CYfx} \\
      311{,}083{,}632 & Back-Run 2 & \solacct{FnyFZ6v8RVuUgmvBetnh7ASdNcU5Dem3b8Vi6MDwk1hV}{FnyFZ6v8} & 18.57M tok & \soltx{64SBkK611Fo9x8YCBf2TVEGsGZws1shZpjMoizK7pBbNfQ6Ggrk71HCTqoRPD2mbsTLoTNwQDrSvcgi6q76UZicB}{64SBkK61} \\
    \bottomrule
  \end{tabular}}
  \caption{Split back leg on Pump AMM pool
  \solacct{6HKFa8i11WmX5DjKnq8FCgHFXEa6PiHqMQZtoNgfo56X}{6HKFa8i1}. The front buys 37{,}131{,}550 tokens for 2.180 SOL and each back leg sells exactly half of them, 18{,}565{,}775, so the two sells together return the full position. Profit 0.122 SOL.}
  \label{tab:sol-ex-splitleg}
\end{table}

In the following, we illustrate the shapes a sandwich attack takes on Solana with one example each, other than the vanilla sandwich attacks we also observe on the other chains.

\parhead{Separate Account.} Table~\ref{tab:sol-ex-crosstrader} shows an evasive attack, where the front and the back leg run on separate accounts and the front transaction hands the bought tokens to the back trader, which is the only on-chain link between the two. It nets 0.000830~SOL.

\parhead{Split Leg.} Table~\ref{tab:sol-ex-splitleg} shows a split back leg, where two sells unwind one buy and their summed input matches the front output, for 0.122~SOL.

\parhead{Staged Exit.} Table~\ref{tab:sol-ex-stagedexit} shows a staged exit, where the attacker returns 71.4\% of the position in the front's own slot and sells the remainder over the following 120 slots, closing 99.99\% of it for a net 20.767~SOL.

\begin{table}[h]
  \centering
  \resizebox{\linewidth}{!}{%
  \begin{tabular}{@{}lllll@{}}
    \toprule
      \textbf{Slot} & \textbf{Role} & \textbf{Signer} & \textbf{Amount} & \textbf{Transaction} \\
    \midrule
      426{,}772{,}197 & Front-Run & \solacct{2mqrindMAjJEQPLhroYWyiYPo5h9iAsahfdd4QtsjwdY}{2mqrindM} & 6.716 SOL & \soltx{HDwUyZyD9NEJvZTonVV3YZVTmft66CWoS5YQjcK51jwJdvZUYcyZDsMExdLEVZrRiTj5EwU7PdwqLhyrs7dmYu9}{HDwUyZyD} \\
      426{,}772{,}197 & Victim & \solacct{9mjtNH9vbXAHd3hJRcV5DHwiLkt7oHcf9wahUZZroT5N}{9mjtNH9v} & 64.0M tok & \soltx{47bpB79ZC9VFFPCfJ4v2ydLhNjLj87BdbcBbc4Qj5irEYpUQumSDyvY1fzMGzPAwFrEmv9dqgosKm22nEuuyxoxA}{47bpB79Z} \\
      426{,}772{,}197 & Victim & \solacct{5brv79eFZ2rGprXNvqgVJBkBptkkw8GJX1XydJyZLyAr}{5brv79eF} & 205.7M tok & \soltx{brrxoSyymbzeajD4kiUvrUN2Mr23iYpD4vZJA1UAj5skCqUB7dX6JsnnbKm5La9famB34y6Wsm6huT5RfSLqukb}{brrxoSyy} \\
      426{,}772{,}197 & Victim & \solacct{9UBhK7j17WTvUHer4JS56LtPtdfxJ9RStzH5dNk1Drn4}{9UBhK7j1} & 128.9M tok & \soltx{4yJL4pEbAHn5o7F2cRoTdUcHebK3gBhLcCtQfYNEjbHtz6XcMSczRzWebyT9r1jYFBTXoVkAHwA21ciuwnpj458r}{4yJL4pEb} \\
      426{,}772{,}197 & Back-Run 1 & \solacct{2mqrindMAjJEQPLhroYWyiYPo5h9iAsahfdd4QtsjwdY}{2mqrindM} & 9.595 SOL & \soltx{nUmuvpxzVfNxZAYTXWW1s4qWRhWLhPquwT7xJPiasSSkvz2s6zepBsMiFasFamkEv6ZVXfBTfDnphfR3qAQ7nti}{nUmuvpxz} \\
      426{,}772{,}197 & Back-Run 2 & \solacct{2mqrindMAjJEQPLhroYWyiYPo5h9iAsahfdd4QtsjwdY}{2mqrindM} & 9.218 SOL & \soltx{5ekUS4oCgiVjnHH1Kuog8d22TR7tHSTqyUj4zWxitowjPCVfV4idXnmjheHFyjbneyT1dXuVer5XS1ergoaCTcyK}{5ekUS4oC} \\
      426{,}772{,}206 & Back-Run 3 & \solacct{2mqrindMAjJEQPLhroYWyiYPo5h9iAsahfdd4QtsjwdY}{2mqrindM} & 1.269 SOL & \soltx{2ruuYaMK8rGi7JfPGECUQKuvwsWJ7XgDgBnFSHrLRf2DVp78kz7gR74mUXL6UPQnpe6URYZiyfWdaXHyCd1QpcbT}{2ruuYaMK} \\
      426{,}772{,}215 & Back-Run 4 & \solacct{2mqrindMAjJEQPLhroYWyiYPo5h9iAsahfdd4QtsjwdY}{2mqrindM} & 1.066 SOL & \soltx{59EBu3wdbnYjuxwfwvc7zMuSErWkFFVrpvZ5VBrZoX6tLERqNDjBnpsfFQtLF49HSt5PgcVffW4P9EcdrtMaLGrD}{59EBu3wd} \\
      426{,}772{,}226 & Back-Run 5 & \solacct{2mqrindMAjJEQPLhroYWyiYPo5h9iAsahfdd4QtsjwdY}{2mqrindM} & 1.055 SOL & \soltx{ssZ5cfZa93ee4dWu8RUGZJ3b5ays6UGDVhfStdxJJZyUc5TN8MT7Tq9pvk1Pjemy2fPYxQ7hYupYxTjC8K8ArbL}{ssZ5cfZa} \\
      426{,}772{,}237 & Back-Run 6 & \solacct{2mqrindMAjJEQPLhroYWyiYPo5h9iAsahfdd4QtsjwdY}{2mqrindM} & 0.826 SOL & \soltx{5v2ACmwpUVayUhw76Yjo5tyYV24N9ZahB27YLmJ5AzTSFrzkFE92rcGRtxRSoSqWkCTpWYbLR6A2CKkEhndTESU2}{5v2ACmwp} \\
      426{,}772{,}251 & Back-Run 7 & \solacct{2mqrindMAjJEQPLhroYWyiYPo5h9iAsahfdd4QtsjwdY}{2mqrindM} & 0.529 SOL & \soltx{2QTEp2DNupzJXvsFoK2UZW9GGTNQ6MerSyyXsgNmQUszNZHagqBvpnLGhSe6jEVHKR7UoLwEvKvRn4hAd786GeKu}{2QTEp2DN} \\
      426{,}772{,}266 & Back-Run 8 & \solacct{2mqrindMAjJEQPLhroYWyiYPo5h9iAsahfdd4QtsjwdY}{2mqrindM} & 0.490 SOL & \soltx{DjNA6uJDYLuy1PMsfddSjipF1BDNk1rMxtaWfxtUjjnCp9uyhPuttPY2rHHAPj7ijd67sRQj6RdzoWtt5zTjdWS}{DjNA6uJD} \\
      426{,}772{,}280 & Back-Run 9 & \solacct{2mqrindMAjJEQPLhroYWyiYPo5h9iAsahfdd4QtsjwdY}{2mqrindM} & 0.534 SOL & \soltx{5zDHGTEsDyVHMz6R8Jxognca1UgC7fzs5vTzf9aSLoeXDSqUoESr3DWdSjWCUcmCZoFwxU2m3JsZswjxcRKBjEhx}{5zDHGTEs} \\
      426{,}772{,}299 & Back-Run 10 & \solacct{2mqrindMAjJEQPLhroYWyiYPo5h9iAsahfdd4QtsjwdY}{2mqrindM} & 0.492 SOL & \soltx{4pbd4FbKXVy7QNLmh5SJW59Xp4mWQDPZPkk6Av4rRiEqBfDBCFoDacqsqYvh796GTiRqyG92G2oSJ5ytHSJfnGtS}{4pbd4FbK} \\
      426{,}772{,}317 & Back-Run 11 & \solacct{2mqrindMAjJEQPLhroYWyiYPo5h9iAsahfdd4QtsjwdY}{2mqrindM} & 2.409 SOL & \soltx{2Xk8o74J7RDShiYR6VqkGNoEbj4o9u8Nqwc4VGy8yCLFuouLVEvoc5UguB36DYXknoPvRCwBdgr4CGcvtRkPWtaQ}{2Xk8o74J} \\
    \bottomrule
  \end{tabular}}
  \caption{Staged exit by
  \solacct{2mqrindMAjJEQPLhroYWyiYPo5h9iAsahfdd4QtsjwdY}{2mqrindM}. The first two back legs land in the front's own slot and return 71.4\% of the position. The remaining nine decay over 120 slots until 99.99\% is sold, for 27.483 SOL out and 20.767 SOL net.}
  \label{tab:sol-ex-stagedexit}
\end{table}

We note that the first two examples most likely take this shape to escape detection, as each defeats a detector that pairs one buy with one sell on a single account within one slot or across neighboring slots~\cite{sandwichedme2025solana}. The staged exit is a split back leg that takes far longer to complete, and selling the position slowly may equally serve a better price.

\begin{figure}[h]
  \centering
  \includegraphics[width=\linewidth]{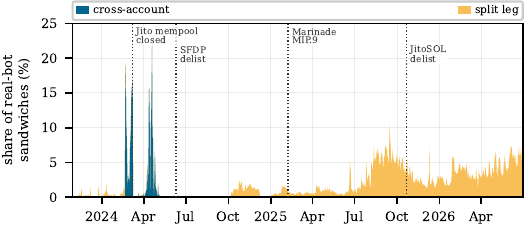}
  \caption{Daily share of sandwiches on Solana using an evasion mechanic: legs submitted by different accounts (cross-account) or a leg split over several transactions. Dotted rules mark interventions.}
  \label{fig:sol-evasive}
\end{figure}

Figure~\ref{fig:sol-evasive} shows the daily share of Solana sandwiches that rely on evasive mechanisms. The share rises sharply around the closure of the Jito mempool, most likely to avoid detection. From mid-2025, when the validators stop being the main source of exposure, split legs become significant and reach around 6\% of sandwiches by the end of our measurement period. Attackers likely use them to escape detection by the validators building the blocks. Note that we observe no such mechanisms at scale on the other chains.

\section{Attack Structure and Profit}
\label{sec:appendix-structure}

In the following, we take a closer look at attack structure and profit. Recall that protected order flow sandwich legs land less precisely than public mempool sandwiches on Ethereum. Figure~\ref{fig:txs-apart} shows that many transactions sit between the front-running leg, the victim, and the back-running leg. Only the swaps in the attacked pool matter, as everything else leaves the attack untouched. It is important to note that also swaps prior to the sandwich attack in the same pool matter, as the sandwich attack might encounter a different state than expected. As we cannot reliably determine what state the transaction expected, we focus on swaps in between the front and back leg in the same pool.

\begin{figure}[h]
  \centering
  \includegraphics[width=\linewidth]{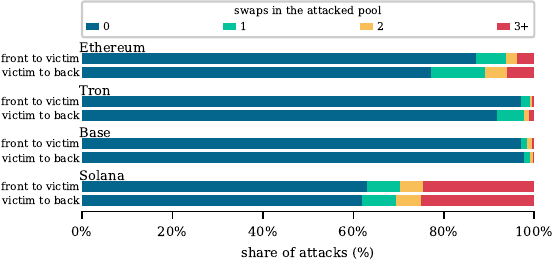}
  \caption{Swaps in the attacked pool between the front leg and the victim, and between the victim and the back leg, as a share of each chain's attacks. Counts of three or more are pooled into the last bucket.}
  \label{fig:pool-swaps-between}
\end{figure}

Figure~\ref{fig:pool-swaps-between} reports the number of such swaps between the legs, counted on each side of the first victim separately. We note that we take the first swap in the attacked direction as the victim, where in principle it could be the second or the third. Any swap in between means the attack does not execute as planned, and the interference can leave it unprofitable.

Ethereum, Tron and Base see no swap in between for more than 70\% of attacks, with Tron and Base above 85\%, and more than three is rare. Solana differs. Only 55.4\% of its attacks run free of interference, and 69.6\% carry at most three swaps on either side. Attacks on protected order flow are thus least exact on Solana.

Interference, i.e., any trade in the attacked pool other than the first victim between the legs, further separates the failed attacks from the successful ones. On Ethereum, 96.2\% of unprofitable attacks have a swap in the attacked pool between their legs, against 22.4\% of profitable ones. Tron shows the same gap at 91.0\% against 6.5\%, and Base at 88.7\% against 2.0\%.

\begin{figure}[h]
  \centering
  \includegraphics[width=\linewidth]{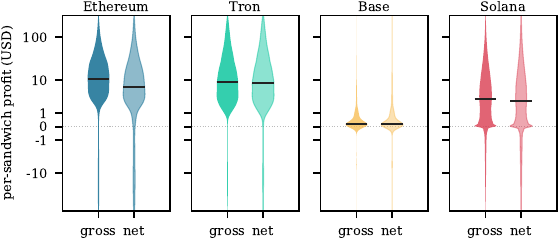}
  \caption{Per-sandwich profit distribution per chain, gross and net of gas.}
  \label{fig:profit-violin}
\end{figure}

Figure~\ref{fig:profit-violin} further shows the profit distribution per chain, gross and net of gas. Ethereum and Tron look alike, at a gross profit of around \$10 per attack. Base sits far below at \$0.16. Solana is smaller as well at \$3.22, and its distribution is considerably wider.

\section{Entity Clustering}
\label{sec:appendix-clustering}
We cluster bots performing sandwich attacks across the three EVM chains, Ethereum, Base, and Tron, where we observe sandwich attacks against protected order flow.

\subsection{Ethereum}
\label{sec:appendix-eth-clustering}

\begin{figure}[t]
    \centering
    \includegraphics[width=1.0\linewidth]{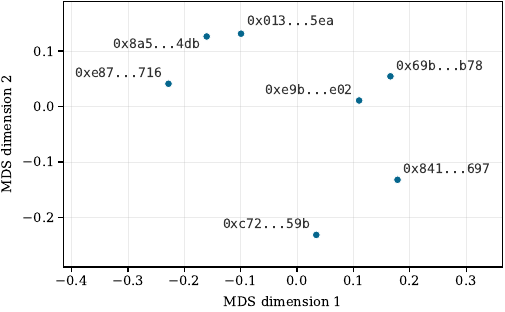}
    \caption{Identified protected order flow sandwich bot smart contracts on Ethereum, clustered based on bytecode and function selector similarity.}
    \label{fig:eth-bot-clustering}
\end{figure}

For Ethereum, we cluster the destination addresses of the identified sandwich transactions based on the bytecode associated with each address. We employ a two stage clustering approach. In the first stage, we cluster the bytecode of each sandwich contract. We first remove the Solidity CBOR metadata and disassemble the remaining bytecode into EVM opcodes. We then separate executable code from embedded data by removing all \texttt{PUSH} instructions and their associated data bytes. The resulting opcode sequences are subsequently partitioned into 3 grams, and DBSCAN is applied using Jaccard distance, with a similarity threshold of 75\%.
In the second stage, we leverage EVMole~\cite{evmole} to extract the function selectors embedded in each sandwich bot's bytecode. Within each bytecode based cluster, we then measure the similarity between contracts based on the overlap of their function selector sets. Specifically, we employ a 75\% set overlap similarity threshold and again apply DBSCAN to identify contracts implementing similar functionality.

The first, bytecode based stage yields two clusters: one containing three contracts and one containing four contracts. Applying the second stage merges the two clusters, resulting in a single final cluster, as shown in Figure~\ref{fig:eth-bot-clustering}. Based solely on the bytecode analysis, we can therefore infer that all seven identified bots likely belong to the same entity.
To validate our clustering results, we examine the bots' deployment history. The contract \href{https://etherscan.io/address/0xC72a44De529163B3EcD26E24F8EDA50d7943859b}{\texttt{0xc72...59b}} was deployed by \href{https://etherscan.io/address/0xA46cD8AD1Ce283ad5C9dc4AF79C9b51465cC004a}{\texttt{0xa46...04a}}, which was itself funded by \href{https://etherscan.io/address/0x928b59c7b3A313A522ca8e6538C95A7AF63012D9}{\texttt{0x928...2d9}}. Notably, \href{https://etherscan.io/address/0x928b59c7b3A313A522ca8e6538C95A7AF63012D9}{\texttt{0x928...2d9}} also deployed contracts belonging to the other six sandwich bots, including \href{https://etherscan.io/address/0x841b6fF0979F86bD76579A3cD03FA43a6A1bB697}{\texttt{0x841...697}}, for example. This funding and deployment relationship provides additional evidence that the seven bots are controlled by the same entity.

\subsection{Base}
\label{sec:appendix-base-clustering}

For Base, we apply the same methodology as for Ethereum. However, two of the four identified sandwich bot smart contracts are proxies implementing the EIP-1967 Transparent Proxy pattern~\cite{eip1967}. For \href{https://basescan.org/address/0xc9cC0E1ECf0AE9af9439933b0C38ee843977B41b}{\texttt{0xc9c...41b}}, we identify 16 distinct implementations, while \href{https://basescan.org/address/0xfcf103349cb306c0234c4171559150F385745A26}{\texttt{0xfcf...a26}} has 5 distinct implementations.
We retrieve the bytecode of all identified implementations and initially cluster the implementations associated with each proxy. We then incorporate these clusters into our aforementioned two-stage clustering procedure.

\begin{figure}[t]
    \centering
    \includegraphics[width=1.0\linewidth]{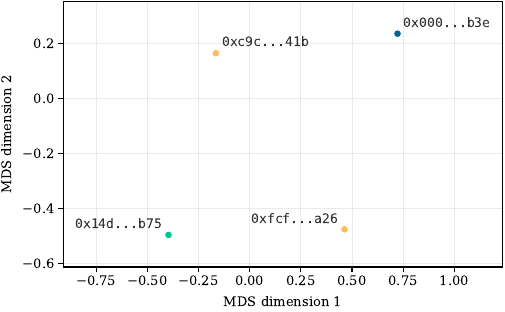}
    \caption{Identified protected order flow sandwich bot smart contracts on Base, clustered based on bytecode and function selector similarity.}
    \label{fig:base-bot-clustering}
\end{figure}

\figureautorefname{} \ref{fig:base-bot-clustering} shows that the two bots \href{https://basescan.org/address/0xc9cC0E1ECf0AE9af9439933b0C38ee843977B41b}{\texttt{0xc9c...41b}} and \href{https://basescan.org/address/0xfcf103349cb306c0234c4171559150F385745A26}{\texttt{0xfcf...a26}} have distinct bytecode but are are assigned to the same cluster via their function selectors, whereas the remaining two bots, \href{https://basescan.org/address/0x000000224f8364bB91259b24435881efc3752b3E}{\texttt{0x000...b3e}} and \href{https://basescan.org/address/0x14d676CBAaC800C5B0297cd1226770b5d2d2bb75}{\texttt{0x14d...b75}}, are assigned to distinct clusters. We manually validate these results by examining the deployment history of the four bots. Our analysis shows that \href{https://basescan.org/address/0x000000224f8364bB91259b24435881efc3752b3E}{\texttt{0x000...b3e}} and \href{https://basescan.org/address/0x14d676CBAaC800C5B0297cd1226770b5d2d2bb75}{\texttt{0x14d...b75}} were deployed by distinct accounts, while \href{https://basescan.org/address/0xc9cC0E1ECf0AE9af9439933b0C38ee843977B41b}{\texttt{0xc9c...41b}} and \href{https://basescan.org/address/0xfcf103349cb306c0234c4171559150F385745A26}{\texttt{0xfcf...a26}} were both deployed by \href{https://basescan.org/address/0xE59db59B17d59F0cA650DDA0eFc5f82683006300}{\texttt{0xe59...300}}. These findings confirm our clustering results and indicate that three distinct entities performed sandwich attacks against protected order flow on Base.

\subsection{Tron}
\label{sec:appendix-tron-funding}

For Tron, our clustering is different as all but one bot do not use a custom contract implementation but instead interact with pools directly from a EOA. Our clustering identifies nine of the twelve to be controlled by the same entity as indicated in Table~\ref{tab:tron-summary}. For this we have various forms of evidence that respectively link a subset of them: \emph{vanity addresses}, \emph{funding}, and \emph{pool partition}.

\parhead{Vanity Addresses.} The entity generated vanity addresses for four of the nine bots, all carrying the \texttt{0x1f0a} prefix. Three further wallets matching the prefix of the first by chance has probability $16^{-12}$, i.e., roughly 1 in $3\times10^{14}$. Note that MEV bots sometimes create recognizable addresses on purpose.

\begin{table}[h]
  \centering
  \resizebox{0.8\columnwidth}{!}{% First funding transfer per Tron persistent entity. Generated by
% build_tron_funding_table.py -- do not edit by hand.
\begin{tabular}{@{}llrl@{}}
\toprule
\textbf{Bot} & \textbf{First Funded} & \textbf{TRX} & \textbf{Funder} \\
\midrule
\href{https://tronscan.org/\#/address/TUE2Sq295N2jgRenGKC81AHE77jBNexTJc}{\texttt{0xc83...28b}} & --- & --- & (deployed contract) \\
\href{https://tronscan.org/\#/address/TEtPcNXwPj1PEdsDRCZfUvdFHASrJsFeW5}{\texttt{0x35e...9a1}} & 24-09-08 & 2,499 & \href{https://tronscan.org/\#/address/TAzsQ9Gx8eqFNFSKbeXrbi45CuVPHzA8wr}{\texttt{0x0b4...e24}} \\
\href{https://tronscan.org/\#/address/TLQgkULsFzihsvbCh5tLssYBRBhf87coLa}{\texttt{0x728...6e9}} & 25-02-21 & 3,000 & \href{https://tronscan.org/\#/address/TRXCenxBBhVdo2HzPKNtbQvBsRZz1e4Kbx}{\texttt{0xaa9...b1a}} \\
\href{https://tronscan.org/\#/address/TCoLAx59RB89dbmHrVhaLCJt9nACS2xd97}{\texttt{0x1f0...600}} & 25-03-22 & 600 & \href{https://tronscan.org/\#/address/TRXCenxBBhVdo2HzPKNtbQvBsRZz1e4Kbx}{\texttt{0xaa9...b1a}} \\
\href{https://tronscan.org/\#/address/TNdCoLAooNdYa27KGUnCqYtVsFXvJd3XFR}{\texttt{0x8ad...267}} & 25-03-18 & 1,500 & \href{https://tronscan.org/\#/address/TRXCenxBBhVdo2HzPKNtbQvBsRZz1e4Kbx}{\texttt{0xaa9...b1a}} \\
\href{https://tronscan.org/\#/address/TCoLazogrq7LH7dUHh8MMMpgJeUrPYGufM}{\texttt{0x1f0...f9e}} & 25-03-23 & 1,600 & \href{https://tronscan.org/\#/address/TRXCenxBBhVdo2HzPKNtbQvBsRZz1e4Kbx}{\texttt{0xaa9...b1a}} \\
\href{https://tronscan.org/\#/address/TCoLAwNYd4icfQCrk65evwt3cdJe5kKZqh}{\texttt{0x1f0...bd4}} & 25-03-19 & 1,300 & \href{https://tronscan.org/\#/address/TRXCenxBBhVdo2HzPKNtbQvBsRZz1e4Kbx}{\texttt{0xaa9...b1a}} \\
\href{https://tronscan.org/\#/address/TCoLa4ZPnWRQduaxYCG9iGVMX45KpFVird}{\texttt{0x1f0...e9f}} & 25-03-26 & 600 & \href{https://tronscan.org/\#/address/TRXCenxBBhVdo2HzPKNtbQvBsRZz1e4Kbx}{\texttt{0xaa9...b1a}} \\
\href{https://tronscan.org/\#/address/TEoG6ZYpWh7YSfGWx1rqcKu6TcZQMucoLA}{\texttt{0x34f...978}} & 25-03-20 & 1,000 & \href{https://tronscan.org/\#/address/TRXCenxBBhVdo2HzPKNtbQvBsRZz1e4Kbx}{\texttt{0xaa9...b1a}} \\
\href{https://tronscan.org/\#/address/TZBL7fQ3i4SbP3Usf5Fr3V5rWAstYfCoLA}{\texttt{0xfe9...ef4}} & 24-12-04 & 1,000 & \href{https://tronscan.org/\#/address/TXRRBBq6sCYPh6ZhtkPbBwFzHyYGCHMsKy}{\texttt{0xeb4...864}} \\
\href{https://tronscan.org/\#/address/TQU1hX6G71i9dzLGciU2xoryM9avE9coLA}{\texttt{0x9f0...70d}} & 25-03-23 & 1,000 & \href{https://tronscan.org/\#/address/TRXCenxBBhVdo2HzPKNtbQvBsRZz1e4Kbx}{\texttt{0xaa9...b1a}} \\
\href{https://tronscan.org/\#/address/TSGXM5pZKgfrddcoLAjMaYmDmz5D8Jyyy9}{\texttt{0xb2c...713}} & 25-02-16 & 800 & \href{https://tronscan.org/\#/address/TRXCenxBBhVdo2HzPKNtbQvBsRZz1e4Kbx}{\texttt{0xaa9...b1a}} \\
\bottomrule
\end{tabular}
}
  \caption{First inbound native TRX transfer to Tron persistent bots.
  \href{https://tronscan.org/\#/address/TRXCenxBBhVdo2HzPKNtbQvBsRZz1e4Kbx}{\texttt{0xaa9...b1a}} is the common funder of nine bots.}
  \label{tab:tron-funding}
\end{table}

\parhead{Funding.} The nine addresses we attribute to a single entity, shaded in Table~\ref{tab:tron-summary}, received their first inbound native transfer from \texttt{0xaa9...b1a} within 38 days of each other. These nine cover all four \texttt{0x1f0a} wallets and five wallets without that prefix. The four \texttt{0x1f0a} wallets were funded within eight days, but two wallets without the prefix were funded in between them. We note that the funding wallet paid out to 15 addresses in total, nine of them these entities. It therefore does not act as a general funding source such as an exchange, which makes the shared origin meaningful.

\begin{figure}[h]
  \centering
  \includegraphics[width=\linewidth]{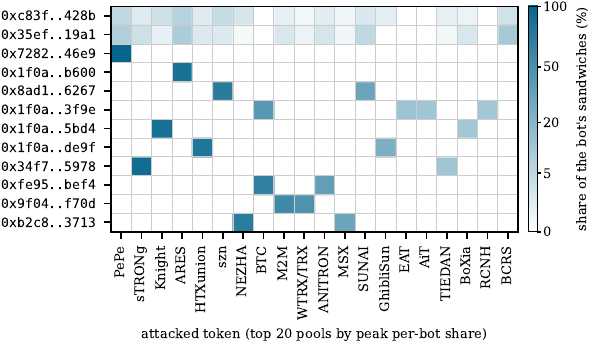}
  \caption{Tron sandwich attacks by bot and pool.}
  \label{fig:tron-bot-pool}
\end{figure}

\parhead{Pool Partition.}
The nine bots partition the pools they attack, sharing no pool across any of their 36 pairs (cf.\ Figure~\ref{fig:tron-bot-pool}). This partition suggests coordination between the entities through a joint operator.

\section{Origin Details}
\label{sec:appendix-origins}

This appendix contains the per-chain figures supporting Section~\ref{sec:origins}.

\subsection{Ethereum}
\label{sec:appendix-figures}

\subsubsection{Ruling Out Block Builders}
\label{sec:appendix-eth-builders}

Builders receive private transactions in plaintext, making the block-building stage a natural candidate for the exposure behind protected sandwiches on Ethereum. We test this by comparing each builder's sandwich share with its block share.

\begin{table}[h]
  \centering
  \resizebox{\columnwidth}{!}{\begin{tabular}{@{}lrrrrrrr@{}}
\toprule
\textbf{Builder} & \textbf{Blocks (\%)} & \textbf{25/11} & \textbf{25/12} & \textbf{26/01} & \textbf{26/02} & \textbf{26/03} & \textbf{26/04} \\
\midrule
Titan & 47.7 & 1.14 & 1.17 & 1.20 & 1.14 & 1.13 & 1.24 \\
BuilderNet (Flashbots) & 13.6 & 1.13 & 1.17 & 1.13 & 1.14 & 1.10 & 1.35 \\
Quasar & 15.0 & 1.05 & 0.96 & 0.87 & 0.88 & 0.99 & 0.67 \\
BuilderNet (Nethermind) & 5.7 & 1.04 & 1.16 & 1.23 & 1.26 & 1.28 & 1.11 \\
BuilderNet (Beaver) & 5.2 & 1.10 & 1.19 & 1.31 & 1.21 & 1.18 & 0.50 \\
beaverbuild & 1.4 & 1.29 & 0.67 & 1.29 & 1.35 & 1.37 & 0.42 \\
\midrule
\textbf{Sandwiches} & & \textbf{1,103} & \textbf{959} & \textbf{1,338} & \textbf{2,187} & \textbf{2,581} & \textbf{6,139} \\
\bottomrule
\end{tabular}
}\vspace{-4pt}
  \caption{Ethereum builder sandwich share divided by block share, per month. A ratio of one indicates no over- or under-representation.}
  \label{tab:eth-builder-excess}\vspace{-6pt}
\end{table}

Table~\ref{tab:eth-builder-excess} reports, for each builder, the ratio between its share of protected sandwiches and its share of blocks, restricted to the months with a substantial number of attacks. A ratio of one means the builder carries exactly as many sandwiches as its block share implies. No builder exhibits the clear over-representation that would be expected from a builder-specific exposure, in contrast to the substantial excess ratios observed for Solana leaders (cf.~Section~\ref{sec:solana-leaders}). Titan sits between 1.13 and 1.24 throughout while building 47.7\% of blocks, and the remaining builders move around one in both directions.

\subsubsection{OFA Attribution}
\label{sec:appendix-eth-ofa}
\Cref{tab:ofa-involvement-breakdown} reports the full distribution of observed
OFA label combinations among sandwich victims, together with their single-victim share. The largest individual category is Blink-only, accounting for 20.9\% of all victims, followed by the MEV-Share + MEVBlocker-only combination at 13.4\%. Overall, 33.2\% of all victims appear in multiple OFA datasets, corresponding to 53.0\% of attributed victims.

\begin{table}[htbp]
  \centering
  \begin{adjustbox}{max width=\linewidth}
  \begin{tabular}{@{}lrrrr@{}}
    \toprule
    \textbf{OFA Combination} & \textbf{Victims} & \textbf{\% of Total}
    & \textbf{\% of Attributed} & \textbf{Single-Victim} \\
    \midrule
    Unattributed & 14{,}702 & 37.26\% & -- & 43.53\% \\
    \midrule
    Blink & 8{,}259 & 20.93\% & 33.36\% & 89.85\% \\
    Share + Blocker & 5{,}280 & 13.38\% & 21.33\% & 88.58\% \\
    Share + Blink & 3{,}542 & 8.98\% & 14.31\% & 88.14\% \\
    Share & 1{,}727 & 4.38\% & 6.98\% & 29.53\% \\
    Blocker + Blink & 1{,}712 & 4.34\% & 6.91\% & 81.43\% \\
    Blocker & 1{,}553 & 3.94\% & 6.27\% & 67.42\% \\
    Share + Blocker + Blink & 1{,}430 & 3.62\% & 5.78\% & 92.87\% \\
    Share + Merkle & 611 & 1.55\% & 2.47\% & 90.83\% \\
    Share + Blocker + Blink + Merkle & 232 & 0.59\% & 0.94\% & 94.83\% \\
    Share + Blink + Merkle & 205 & 0.52\% & 0.83\% & 96.59\% \\
    Merkle & 104 & 0.26\% & 0.42\% & 56.73\% \\
    Share + Blocker + Merkle & 71 & 0.18\% & 0.29\% & 92.96\% \\
    Blink + Merkle & 21 & 0.05\% & 0.08\% & 90.48\% \\
    Blocker + Blink + Merkle & 11 & 0.03\% & 0.04\% & 54.55\% \\
    Blocker + Merkle & 1 & 0.00\% & 0.00\% & 100.00\% \\
    \midrule
    \textbf{Total} & \textbf{39{,}461} & \textbf{100.00\%} & -- & \\
    \bottomrule
  \end{tabular}
  \end{adjustbox}\vspace{-4pt}
  \caption{Observed OFA label combinations among victims of successful and
  unsuccessful private sandwich attempts. \emph{Share} and \emph{Blocker}
  denote MEV-Share and MEVBlocker, respectively. Each row reports the exact
  combination of OFA labels observed for a victim (e.g., \emph{Blink} denotes
  Blink only), together with the share for which it was the sole victim of the
  sandwich. \emph{Unattributed} denotes victims matched by none of the four datasets and is excluded from the \emph{\% of Attributed} column, which totals 24{,}759 victims.}\vspace{-6pt}
  \label{tab:ofa-involvement-breakdown}
\end{table}

\begin{figure}[htbp]
    \centering
    \includegraphics[width=1.0\linewidth]{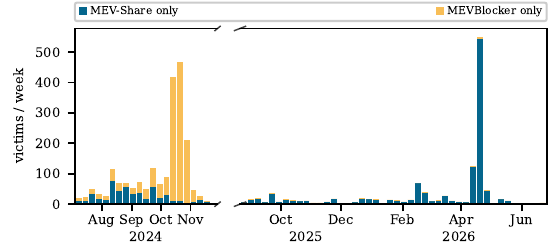}
    \caption{Weekly evolution of sandwich victims observed exclusively in MEV-Share
    or MEVBlocker.}
    \label{fig:mevshare_mevblocker}
\end{figure}

\Cref{fig:mevshare_mevblocker} shows the weekly evolution of victims observed only in MEV-Share or only in MEVBlocker. The temporal concentration of these single-OFA exposure categories is discussed in \Cref{sec:ofa-exposure-mechanisms}.

\subsubsection{Sandwiching Under the Risk of Back-Runs}
\label{sec:appendix-backrun}
A back-run and a sandwich feed on the same price displacement, so a back-run executed before the sandwicher's unwind takes away profit from that unwind. We quantify the tension from both sides which is relevant for the sandwiches we observe on OFA users (cf. \Cref{sec:ofa-attribution}). A two-pool benchmark derives when a sandwich survives such a back-run, and a counterfactual over the observed victims measures how often a profitable direct back-run exists at all.

\parhead{When Not to Sandwich?} In a within-block wide or a cross-block sandwich, an intervening back-runner may arbitrage away part of the victim and front-run-induced price distortion before the sandwicher unwinds, thereby reducing or even eliminating the sandwich's profit. Thus, a non-tight sandwich should decide under what conditions it's likely to succeed and, based on that, how to optimize it. We demonstrate this case in the simplest settings, where the sandwicher can decide based on the victim transaction, which can be back-run as well as sandwiched, the optimal front-run amount that would succeed under the optimal intervening back-runner. For simplicity, we assume that the victim, back-runner, and sandwicher are different identities that decide independently.

We consider a continuous, fee-only two-pool V2 benchmark. A victim swaps $X$ for $Y$ through pool~1, and the only alternative source-pool-disjoint route between the same tokens is pool~2. Let $(x_i,y_i)$ denote the $X$ and $Y$ reserves of pool $i$, let $f_i$ be its fee rate, and define $\gamma_i=1-f_i$. Before the front-run, we assume that the pools satisfy the gas-free marginal no-arbitrage condition
\[
\gamma_1\gamma_2
\leq
\frac{x_1y_2}{x_2y_1}
\leq
(\gamma_1\gamma_2)^{-1}.
\]
For an $X\!\rightarrow Y$ swap with input $\delta$, define
\[
\Phi_{\delta,\gamma}^{X\to Y}(x,y)
=
\left(
x+\delta,
\frac{xy}{x+\gamma\delta}
\right),
\]
with the reverse transition defined analogously.

We consider a sandwicher $\mathcal S$, a victim $\mathcal V$, and a back-runner $\mathcal B$, and condition on the realized same-block ordering
\[
\mathcal S_{\mathrm{front}}
\prec
\mathcal V
\prec
\mathcal B
\prec
\mathcal S_{\mathrm{unwind}}.
\]
The back-runner reads the live post-victim reserves at execution time and computes its trade on-chain~\cite{SolmazOptimistic2025,WangBlockspace2026, wu2026waitprobe}. For the theoretical optimum, we treat token amounts as continuous and ignore the back-runner's gas costs and capital constraints.%

Let $\delta_{\mathcal S}$ and $\delta_{\mathcal V}$ denote the $X$ inputs of the front-run and victim transactions through pool~1. After these transactions, its state is
\[
\Phi_{\delta_{\mathcal V},\gamma_1}^{X\to Y}
\left(
\Phi_{\delta_{\mathcal S},\gamma_1}^{X\to Y}(x_1,y_1)
\right)
=
(\widetilde x_1,\widetilde y_1).
\]
The back-runner cycles $X\!\rightarrow Y$ through pool~2 and then $Y\!\rightarrow X$ through pool~1. Let $\delta_{\mathcal B}$ denote the back-runner's input to pool~2. This trade produces
\[
q_{\mathcal B}(\delta_{\mathcal B})
=
\frac{\gamma_2y_2\delta_{\mathcal B}}
     {x_2+\gamma_2\delta_{\mathcal B}}
\]
units of $Y$. Define
\[
C=\gamma_1\gamma_2\widetilde x_1y_2,\qquad
D=x_2\widetilde y_1,\qquad
E=\gamma_2(\widetilde y_1+\gamma_1y_2).
\]
The back-runner's gross $X$-denominated profit is
\[
\Pi_{\mathcal B}(\delta_{\mathcal B})
=
\frac{\gamma_1\widetilde x_1q_{\mathcal B}(\delta_{\mathcal B})}
     {\widetilde y_1+\gamma_1q_{\mathcal B}(\delta_{\mathcal B})}
-\delta_{\mathcal B}
=
\frac{C\delta_{\mathcal B}}{D+E\delta_{\mathcal B}}
-\delta_{\mathcal B}.
\]
Writing $[z]_+=\max(z,0)$, its maximizing input is
\[
\delta_{\mathcal B}^*
=
\frac{[\sqrt{CD}-D]_+}{E},
\qquad
q_{\mathcal B}^*
=
\frac{\gamma_2y_2\delta_{\mathcal B}^*}
     {x_2+\gamma_2\delta_{\mathcal B}^*}.
\]
A positive-size gross-profitable back-run exists exactly when $C>D$. After this back-run, the pool states are
\[
(\overline x_1,\overline y_1)
=
\left(
\frac{\widetilde x_1\widetilde y_1}
     {\widetilde y_1+\gamma_1q_{\mathcal B}^*},
\widetilde y_1+q_{\mathcal B}^*
\right)
\]
and
\[
(\overline x_2,\overline y_2)
=
\left(
x_2+\delta_{\mathcal B}^*,
\frac{x_2y_2}
     {x_2+\gamma_2\delta_{\mathcal B}^*}
\right).
\]
The front-run leaves $\mathcal S$ holding $\delta'_{\mathcal S}=\gamma_1\delta_{\mathcal S}y_1/(x_1+\gamma_1\delta_{\mathcal S})$ units of $Y$. We consider a standard back-run-unaware sandwicher that unloads all of this inventory through pool~1 after $\mathcal B$. It receives $\delta''_{\mathcal S}=\gamma_1\delta'_{\mathcal S}\overline x_1/(\overline y_1+\gamma_1\delta'_{\mathcal S})$ units of $X$, giving gross profit
\[
P_{\mathcal S}
=
\delta''_{\mathcal S}-\delta_{\mathcal S}.
\]
For every fixed $\delta_{\mathcal S}>0$,
\[
P_{\mathcal S}>0
\quad\Longleftrightarrow\quad
\gamma_1^2y_1
\left(\overline x_1-\delta_{\mathcal S}\right)
>
\overline y_1
\left(x_1+\gamma_1\delta_{\mathcal S}\right).
\]
Here, $(\overline x_1,\overline y_1)$ depends on $\delta_{\mathcal S}$ through the victim state and $\mathcal B$'s best response. For a fixed front-run, any executed positive-size back-run reduces $\overline x_1$, increases $\overline y_1$, and therefore strictly lowers the proceeds of this fixed source-pool unwind. It may, but need not, eliminate the sandwich's gross profit.

Let $\mathcal D_{\mathcal S}$ denote the feasible front-run sizes imposed by the victim's slippage constraint and the sandwicher's available capital, and let $K_{\mathcal S}(\delta_{\mathcal S})$ denote the sandwicher's $X$-denominated execution and financing costs. Under this benchmark, the sandwicher should abstain when
\[
\sup_{\delta_{\mathcal S}\in\mathcal D_{\mathcal S}}
\left\{
P_{\mathcal S}(\delta_{\mathcal S})
-
K_{\mathcal S}(\delta_{\mathcal S})
\right\}
\leq 0.
\]
We leave this one-dimensional optimization and alternative back-run-aware unwind strategies outside our analysis.

\parhead{Back-Run Analysis.} The benchmark asks what a back-run costs the sandwicher. Our empirical counterfactual asks the prior question, whether the victim alone creates a profitable direct back-run at all, and, whenever the pre-victim state is conclusive, whether the victim created or enlarged the opportunity. It does not estimate the sandwicher's residual profit.

We restrict this analysis to the 31,259 victim transactions whose observed front-run occurs in the same block as the victim. For a victim in block $b_e$, the state at the end of block $b_e-1$ excludes both transactions and therefore provides a baseline uncontaminated by the observed front-run. We exclude the remaining 8,202 victims because their earlier-block front-run is already reflected in this state. We do not require the sandwicher's observed unwind to share the victim block because it is not executed in our victim-only counterfactual.

\parhead{Activity Convention.} We determine pool activity from on-chain state at the end of block $b_e-1$: positive reserves for V2, positive current-range liquidity for V3, positive \texttt{StateView} liquidity for V4, and corresponding protocol-specific checks for other venues. Unavailable state remains unknown rather than inactive. We include the source pool by construction. In four V3 cases, it is counted despite reporting zero current-range liquidity because the observed same-block front-run successfully accesses initialized liquidity. We exclude these four baseline states from numerical modeling.

\parhead{Counterfactual Execution.} For each event $e$, we identify the sandwiched source pool, the victim's swap direction $A_e\!\rightarrow B_e$, and every active alternative pool containing the same token pair. From the \texttt{Swap} log in the victim's receipt, we decode the exact amount of $A_e$ entering the source pool. Starting from the source pool's state at the end of block $b_e-1$, we apply this input only to the source pool and leave the alternatives unchanged. We recompute the victim's output because its observed output was produced after the front-run and therefore reflects a different state.\footnote{This is an observed-input source-leg replay rather than a full transaction replay: without the front-run, an exact-output or multihop transaction could produce a different source-leg input. Apart from the observed source-pool swap, we condition only on confirmed chain state through block $b_e-1$, thereby omitting transactions earlier in block $b_e$ and information that a searcher might obtain from OFA order flow or other pending transactions. These restrictions isolate the victim's marginal effect under a reproducible counterfactual.}

For every active alternative pool $p$, we attempt to simulate a direct cycle that trades $A_e$ for $B_e$ through $p$ and then trades the resulting $B_e$ back to $A_e$ through the source pool. For V2/V2 routes, we compute the exact integer profit-maximizing input separately in the pre- and post-victim states. For routes containing V3 or V4, we search candidate inputs using exact integer swaps with initialized-tick traversal and retain the most profitable input found. A route is \emph{victim-created} when the pre-victim state is certified to admit no gross-profitable exact-input trade and the post-victim search finds a profitable input. A route that was already profitable before the victim is \emph{victim-expanded} when the victim strictly increases its profit. For V2/V2 routes we compare the exact pre- and post-victim optima, and for concentrated-liquidity routes we replay the profitable pre-victim input after the victim. When the pre-victim search is inconclusive, we retain any post-victim opportunity we find but leave its victim-effect classification open. When the post-victim search is inconclusive, the route stays inconclusive rather than being recorded as unprofitable.

We evaluate only this victim-induced direction. An $A_e\!\rightarrow B_e$ victim makes $B_e$ relatively scarcer in the source, potentially making it profitable to buy $B_e$ from an alternative and sell it into the source. We do not analyze arbitrage in the reverse direction because its execution before the source-pool unwind would move the source further in the victim's direction and improve, rather than erode, the proceeds of that unwind.

\parhead{Execution and Validation.} V2 routes use exact integer arithmetic with historically verified fees. We exclude 1,138 candidate routes across 804 victims whose fee semantics cannot be verified. We model pools created by the official Uniswap V3 factory and official hookless, static-fee Uniswap V4 pools using their historical price, liquidity, and tick state. We exclude other V3-style pools and hooked or dynamic-fee V4 pools because their exact execution semantics require additional implementation-specific modeling. Native ETH is treated as WETH for token matching and valuation.

The victim receipt must contain exactly one source-pool swap in the recorded direction. We reject a route if source decoding is ambiguous or its historical state or fee cannot be established. We also reject it if the victim changes the candidate alternative, since applying the victim only to the source would then combine incompatible pool states. One exact positive post-victim execution is sufficient to establish an opportunity. We conclude that no profitable direct route exists only when every active direct alternative has a conclusive post-victim result and none is profitable. Unsupported or post-victim-inconclusive routes prevent a negative conclusion. Pre-victim inconclusiveness suppresses only the victim-created or victim-expanded label. It does not invalidate an exact positive post-victim result. This conclusion concerns only the modeled direct $A_e$--$B_e$ alternatives and does not imply that no profitable multihop route exists.

\parhead{Gas and Valuation.} Gross profit is initially denominated in the victim's input token. When that token is WETH, conversion to ETH is exact. When the victim's output token is WETH, we approximate the input token's ETH value using the source pool's pre-victim reserve ratio. We do not assign a gas-adjusted conclusion to non-WETH pairs. We assume 150,000 gas for V2/V2 routes, 200,000 for V2/V3 routes, and 250,000 for V3/V3 routes or any route containing V4. Let $G_{e,p}$ denote this route-specific gas usage, $g_e$ the victim transaction's effective gas price in wei per gas, and $\widehat{\Pi}^{\,\mathrm{WETH}}_{e,p}$ the route's gross profit expressed in wei. We price the hypothetical back-run at one wei per gas above the victim's realized effective gas price:
\[
\widehat{\Pi}^{\mathrm{net}}_{e,p}
=
\widehat{\Pi}^{\,\mathrm{WETH}}_{e,p}
-
G_{e,p}(g_e+1).
\]
This is a profitability screen rather than an ordering model or an exact net-profit estimate: actual gas usage may differ, and the calculation omits builder payments, financing costs (e.g., flashloans), and failed attempts.

\begin{table}[t]
  \centering
  \footnotesize

  \label{tab:appendix-backrun-results}
  \begin{tabular}{@{}lr@{}}
    \toprule
    \textbf{Measure} & \textbf{Victim Transactions} \\
    \midrule
    Same-block cohort & 31,259 \\
    At least one active direct alternative & 9,886 \\
    At least one conclusive post-victim route & 7,977 \\
    All active alternatives conclusive post-victim & 6,374 \\
    Post-victim gross-positive route & 3,905 \\
    At least one victim-created route & 2,272 \\
    At least one victim-expanded route & 1,859 \\
    Gross-positive and ETH-evaluable & 3,863 \\
    Profitable after the gas screen & 484 \\
    \bottomrule
  \end{tabular}\vspace{-4pt}
  \caption{Same-block victim-only direct-back-run results.}\vspace{-10pt}
\end{table}

\parhead{Results.} Among the 9,886 victims with an active direct alternative, at least one post-victim route has a conclusive result in 7,977 cases (80.7\%), while every active direct alternative has a conclusive post-victim result in 6,374. The event-level result remains inconclusive for 2,504 victims, primarily because the victim changes an alternative, the V2 fee cannot be verified, execution requires an unsupported protocol, noncanonical V3 implementation, or hooked or dynamic-fee V4 pool, or the post-victim search is inconclusive. The latter includes 56 candidate routes for which the post-victim marginal return is positive but no profitable exact integer input is found.

At zero gas cost, we find a gross-profitable route for at least 3,905 victims (12.5\% of the cohort). Of these, 2,272 have at least one victim-created route and 1,859 have at least one victim-expanded route. Overall, 3,895 victims have at least one victim-created or victim-expanded route. 236 exhibit both classifications on different alternatives. Of the remaining ten gross-positive victims, four have an inconclusive pre-victim search and therefore receive no victim-effect classification, while six have only pre-existing routes for which we find no verified expansion.

Profit is ETH-valued for 3,863 positive cases and is strongly right-skewed: the median, 90th percentile, and 99th percentile of the best detected gross profit are $2.63{\times}10^{-5}$, $1.37{\times}10^{-3}$, and $1.52{\times}10^{-2}$ ETH. The gas screen removes 3,379 of these 3,863 opportunities (87.5\%), leaving 484 profitable cases, or 1.5\% of the cohort. Thus, based on our methodology, detected direct opportunities occur for 12.5\% of the cohort before gas and 1.5\% under our gas screen. For routes containing V3 or V4, every reported positive result is supported by exact integer swap execution, but the search may miss another profitable input or understate its maximum profit.

Future work should derive back-run-aware sandwich strategies and extend the empirical analysis to multihop routes.

\subsection{Tron}
\label{sec:appendix-tron-producers}

Tron implements first-come-first-served transaction ordering in the client, so a block producer running a modified ordering policy is a natural candidate for the observed sandwiches on protected order flow. We test this by comparing each producer's share of sandwiches with its share of blocks.

\begin{figure}[h]
  \centering
  \includegraphics[width=\columnwidth]{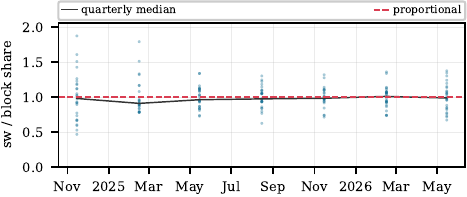}
  \caption{Sandwich production per Tron block producer, relative to the blocks it produces. Each UTC calendar quarter, a producer's share of that quarter's sandwiches divided by its share of that quarter's blocks. A ratio of 1.0 indicates sandwiching in proportion to block production. Quarters with fewer than 300 sandwiches are omitted, as are producers making up less than 2\% of a quarter's blocks.}
  \label{fig:tron-leader-excess}
\end{figure}

Figure~\ref{fig:tron-leader-excess} shows each producer's share of sandwiches relative to its share of blocks. The ratio ranges from 0.47 to 1.88, with 90\% of leader-quarters between 0.72 and 1.34, and no producer sustains an elevated share across quarters. Attacks thus land almost uniformly across producers, making producer-specific misordering an unlikely explanation for the observed sandwiches.

\subsection{Base}
\label{sec:appendix-base-details}
In the following, we detail the behavior of the four sandwich attack bots on Base, which form three entities.

\parhead{\href{https://basescan.org/address/0x000000224f8364bb91259b24435881efc3752b3e}{\texttt{0x000...b3e}}.} The evidence for this bot is most consistent with pre-inclusion visibility, either through provider-local transaction-pool exposure or an equivalent private pending-order-flow source. Its activity is confined to 13 to 28 August 2023, which overlaps Base's disclosures that some shared RPC offerings exposed transactions from their local pending pools, along with the corresponding Base and op-geth mitigations~\cite{base_incident_report_1,base_tx_pool_leak,base_incident_report_2,op_tx_pool_leak} (cf.\ Figure~\ref{fig:base-000-incident}). Our detector output holds 202 candidates, 200 of which have a positive gross return. Among the 199 same-block candidates, the back-run reproduces the victim transaction's complete fee values in every case, i.e., the transaction type, the effective gas price, \texttt{maxFeePerGas}, and \texttt{maxPriorityFeePerGas}. These copies span 128 distinct fee tuples, which rules out a single fixed fee configuration.

\parhead{\href{https://basescan.org/address/0x14d676cbaac800c5b0297cd1226770b5d2d2bb75}{\texttt{0x14d...b75}}.} Our detector flags 239 sandwich sequences for this bot, 234 of which follow one pattern. The victims and the tokens they trade share their funding. In 232 of the 234 cases the target token's deployer was funded directly by \href{https://basescan.org/address/0xc43f317ed4d81cbbfe2c9c98b4cc6f303519f078}{\texttt{0xc43...078}}, and 233 of the 243 victim transactions originate from three EOAs funded by \href{https://basescan.org/address/0xb01dbb5310c0ccd0ff594bc222a89ad7734d242b}{\texttt{0xb01...42b}}. Both hubs trace upstream to one funding root, \href{https://basescan.org/address/0xea7117b187ff0b33fc4d14a6c1f148407b585259}{\texttt{0xea7...259}}. Every occurrence follows the same loop, i.e., \href{https://basescan.org/address/0xc43f317ed4d81cbbfe2c9c98b4cc6f303519f078}{\texttt{0xc43...078}} funds a token deployer, the deployer creates the token and seeds the pool, \href{https://basescan.org/address/0x14d676cbaac800c5b0297cd1226770b5d2d2bb75}{\texttt{0x14d...b75}} front-runs, the buyer EOAs execute their buys, and \href{https://basescan.org/address/0x14d676cbaac800c5b0297cd1226770b5d2d2bb75}{\texttt{0x14d...b75}} follows with the back-run. Table~\ref{tab:base-14d-cycle} shows one cycle of the loop. The token creators and the victims therefore operate as one infrastructure. We observe no on-chain funding link between \href{https://basescan.org/address/0x14d676cbaac800c5b0297cd1226770b5d2d2bb75}{\texttt{0x14d...b75}}, or its deployer and beneficiary \href{https://basescan.org/address/0xaea4dbb2191594bdc77fc587b1289039662508fd}{\texttt{0xaea...8fd}}, and the victim-side funding lineage. We therefore do not attribute it to the same operator and classify it as an opportunistic actor exploiting an automated campaign across multiple blocks rather than a conventional intra-block sandwich bot. %

\parhead{\href{https://basescan.org/address/0xc9cc0e1ecf0ae9af9439933b0c38ee843977b41b}{\texttt{0xc9c...41b}} and \href{https://basescan.org/address/0xfcf103349cb306c0234c4171559150f385745a26}{\texttt{0xfcf...a26}}.}
These two contracts are successive deployments of one managed execution system. Both are ERC-1967 proxies controlled by \href{https://basescan.org/address/0xe59db59b17d59f0ca650dda0efc5f82683006300}{\texttt{0xe59...300}}, they share upgrade infrastructure, and they reuse execution EOAs. All 21 EOAs sending \href{https://basescan.org/address/0xfcf103349cb306c0234c4171559150f385745a26}{\texttt{0xfcf...a26}} front-run legs had previously executed through \href{https://basescan.org/address/0xc9cc0e1ecf0ae9af9439933b0c38ee843977b41b}{\texttt{0xc9c...41b}}. Their operating periods do not overlap, which points to an operational migration rather than two independent bots (cf.\ \Cref{sec:appendix-base-clustering}).

The detector identifies 1{,}448 candidates, 750 for \href{https://basescan.org/address/0xc9cc0e1ecf0ae9af9439933b0c38ee843977b41b}{\texttt{0xc9c...41b}} and 698 for \href{https://basescan.org/address/0xfcf103349cb306c0234c4171559150f385745a26}{\texttt{0xfcf...a26}}, of which 1{,}402 have a positive gross profit before gas and 1{,}445 sit within one block. 18 \href{https://basescan.org/address/0xc9cc0e1ecf0ae9af9439933b0c38ee843977b41b}{\texttt{0xc9c...41b}} candidates contain two victim transactions
, which yields 1{,}463 transaction-level fee comparisons. In 1{,}274 of them (87.1\%) the effective-gas-price deltas are $(g_F-g_V,g_B-g_V)=(+1,0)$~wei. The relation holds for all 698 \href{https://basescan.org/address/0xfcf103349cb306c0234c4171559150f385745a26}{\texttt{0xfcf...a26}} comparisons and for 576 of the 765 \href{https://basescan.org/address/0xc9cc0e1ecf0ae9af9439933b0c38ee843977b41b}{\texttt{0xc9c...41b}} comparisons.

The victim flow is highly structured. We tie 1{,}109 of the 1{,}466 distinct victim transactions (75.6\%) to seven funding-linked cohorts spanning 64 wallets, linked through direct funding, a common dispersal transaction, or multi-hop ancestry. The four largest cohorts contribute 481 transactions from ten wallets tracing to \href{https://basescan.org/address/0xea7117b187ff0b33fc4d14a6c1f148407b585259}{\texttt{0xea7...259}}, 284 from eight DUAL wallets funded together with two dispersal transactions (one funds them with ETH for gas \href{https://basescan.org/tx/0x76a9c861a5d15ed6b9bc3a4a5c37c142af3ec6230ad4ed92792ec7b7aabc1929}{\texttt{0x76a...929}}, the other with DUAL token \href{https://basescan.org/tx/0xf73eb901c2aa0a3dbb946cc0ccc212f5067028d149db62d1e95ed4b65ea54ac7}{\texttt{0xf73...ac7}}), 129 from four wallets funded directly by \href{https://basescan.org/address/0xe6e7d3c6379ad80de02f26ccc72605d0f70d5201}{\texttt{0xe6e...201}}, and 121 from fifteen wallets tracing to \href{https://basescan.org/address/0x24f1ce82cfd505a88fba5a23111abde99a799999}{\texttt{0x24f...999}}. Three smaller cohorts add 61, 20, and 13 transactions, the last being 13 AMETA buyers whose five immediate funding branches converge at \href{https://basescan.org/address/0xfbf566c3598b637d22b0a1dcf9423d0ca049ffec}{\texttt{0xfbf...fec}}. The two largest cohorts carry 69.0\% of the funding-linked flow and the four largest 91.5\%. Funding-linked flow is also more common under \href{https://basescan.org/address/0xfcf103349cb306c0234c4171559150f385745a26}{\texttt{0xfcf...a26}} (632/698, 90.5\%) than under \href{https://basescan.org/address/0xc9cc0e1ecf0ae9af9439933b0c38ee843977b41b}{\texttt{0xc9c...41b}} (477/768, 62.1\%).

Funding ancestry alone does not demonstrate common control. Its significance comes from the execution regularities that accompany it. All 284 sandwiched DUAL transactions trade the same WETH/DUAL pool, arrive on a five-to-six-block cadence, set zero minimum output, and use the same effective gas price in 280/284 cases. The four wallets funded by \href{https://basescan.org/address/0xe6e7d3c6379ad80de02f26ccc72605d0f70d5201}{\texttt{0xe6e...201}} execute their 129 sandwiched swaps through the Aerodrome Router using the same function, one-hop USDC/LIQ route, pool, self-recipient form, zero minimum output, and fee tier, on a 92--95-block cadence. The \href{https://basescan.org/address/0x24f1ce82cfd505a88fba5a23111abde99a799999}{\texttt{0x24f...999}}, \href{https://basescan.org/address/0xedacb9f7c684706848e51eeca8bd885d2f9304ba}{\texttt{0xeda...4ba}}, and \href{https://basescan.org/address/0xb73f09ef07bae0056f5771cac317493a39a8fb96}{\texttt{0xb73...b96}} cohorts use the public Uniswap V2 Router, so router reuse alone tells us little there. What matters is the conjunction of cohort-level funding links with regular execution, and every transaction within each cohort uses the same gas price. These regularities are consistent with coordinated or automated systems.

The \href{https://basescan.org/address/0xea7117b187ff0b33fc4d14a6c1f148407b585259}{\texttt{0xea7...259}} cohort also connects this activity to the launch factory previously targeted by \href{https://basescan.org/address/0x14d676cbaac800c5b0297cd1226770b5d2d2bb75}{\texttt{0x14d...b75}}. Its ten buyer wallets make 481 purchases across 71 newly created tokens. On a separate branch, \href{https://basescan.org/address/0xea7117b187ff0b33fc4d14a6c1f148407b585259}{\texttt{0xea7...259}} funds \href{https://basescan.org/address/0xc43f317ed4d81cbbfe2c9c98b4cc6f303519f078}{\texttt{0xc43...078}}, which supplies 13.02~ETH to each token creator. Each creator deploys a token at nonce zero, provides liquidity, later removes it, and returns the ETH to \href{https://basescan.org/address/0xc43f317ed4d81cbbfe2c9c98b4cc6f303519f078}{\texttt{0xc43...078}}. The median intervals from funding to deployment, deployment to liquidity, and liquidity to first purchase are 8, 7, and 61 blocks. Roughly fourteen months earlier, \href{https://basescan.org/address/0x14d676cbaac800c5b0297cd1226770b5d2d2bb75}{\texttt{0x14d...b75}} targeted the same funding and launch lifecycle in 232 of its 234 core cases. The two campaigns share no token addresses, and \href{https://basescan.org/address/0x14d676cbaac800c5b0297cd1226770b5d2d2bb75}{\texttt{0x14d...b75}} has a separate observed funding lineage, so the connection is the reoccurrence of a predictable token launch cycle rather than a shared operator.

\begin{figure}[hbtp]
  \centering
  \includegraphics[width=\linewidth]{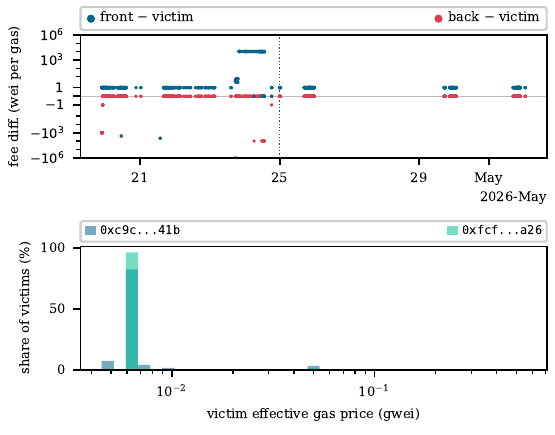}
  \caption{Distribution of differences of fees set by attacker and victim as well as full fee distribution of the victim transactions for bots \href{https://basescan.org/address/0xc9cc0e1ecf0ae9af9439933b0c38ee843977b41b}{\texttt{0xc9c...41b}} and \href{https://basescan.org/address/0xfcf103349cb306c0234c4171559150f385745a26}{\texttt{0xfcf...a26}}.
  }
  \label{fig:c9-fcf-fee}\vspace{-6pt}
\end{figure}

These findings suggest prediction and pre-positioning are plausible for structured cohorts, but not always. Of \href{https://basescan.org/address/0xfcf103349cb306c0234c4171559150f385745a26}{\texttt{0xfcf...a26}}'s 698 victim transactions, 667 use one of three dominant effective-gas-price settings: 6.0, 6.3, or 51 million wei. The remaining 31 transactions come from 22 senders and use 24 distinct effective gas prices, nine call targets and selectors, and 14 pools. Thirty lie outside the seven funding-linked cohorts, while one belongs to the \href{https://basescan.org/address/0xea7117b187ff0b33fc4d14a6c1f148407b585259}{\texttt{0xea7...259}} cohort. Nevertheless, everyone is bracketed with the exact $(+1,0)$ relation (cf.\ Figure~\ref{fig:c9-fcf-fee}). Alignment, this specific to the transaction, is difficult to derive from a static schedule and remains consistent with either pre-sequencing transaction visibility or a shared off-chain backend. The on-chain evidence neither distinguishes between these mechanisms nor establishes common control between the funding cohorts and the executor.

\parhead{Multi-victim cases.}
Multi-victim cases are rare in our Base dataset, with most sandwiches involving a single candidate victim. In the 23 multi-victim cases, both ordinary purchases and atomic arbitrage transactions can be labeled as victims. These cases require a more nuanced definition of a \textit{victim}: an atomic arbitrage can remain profitable and still be harmed if the sandwich front-run reduces its profit. Using the arbitrage detection methodology of Solmaz et al.~\cite{SolmazOptimistic2025}, we identify ten atomic arbitrage transactions among the victims in these cases. Such instances may be more common on high-throughput, low-cost L2s, such as Base, where inexpensive execution supports optimistic on-chain searching~\cite{SolmazOptimistic2025,wu2026waitprobe}.

\begin{figure}[b]
  \centering
  \includegraphics[width=\linewidth]{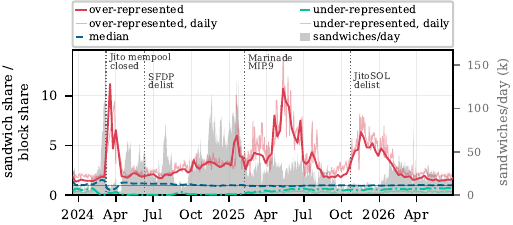}
  \caption{Sandwich production per Solana block producer, relative to the blocks it produces. Each week, a leader's share of that week's sandwiches divided by its share of that week's blocks, with the ten most over- and under-represented leaders also shown daily. Dotted rules mark interventions against sandwiching, shading daily attack volume.}
  \label{fig:sol-leader-conc}\vspace{-6pt}
\end{figure}

\subsection{Solana}
\label{sec:appendix-solana-details}
\subsubsection{Leader Concentration}
In the following, we detail how sandwich attacks distribute across Solana's leaders (cf.\ Section~\ref{sec:solana-leaders}).

Figure~\ref{fig:sol-leader-conc} tracks each leader's share of sandwiches relative to its share of blocks on a weekly basis, with the ten most over- and the ten most under-represented leaders shown daily. The two groups start far apart, and both move towards one over time. The suppressing leaders do so from the first quarter of 2025 and the over-represented ones follow from the first quarter of 2026. We note that these curves cover the extremes of the distribution alone.

\begin{figure}[h]
  \centering
  \includegraphics[width=\linewidth]{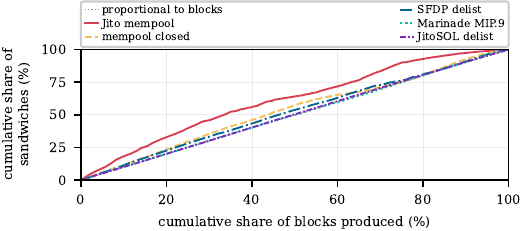}
  \caption{Cumulative share of Solana sandwiches against cumulative share of blocks produced, per intervention era. The diagonal is production in proportion to block share.}
  \label{fig:sol-leader-conc-lorenz}\vspace{-6pt}
\end{figure}

Figure~\ref{fig:sol-leader-conc-lorenz} plots the Lorenz curves per era, i.e., the cumulative share of sandwiches against the cumulative share of blocks produced, with leaders ordered from most to least attacked. Distance from the diagonal measures how much of the attack volume sits with few leaders. Under the Jito mempool the curve rises steeply, i.e., the top ten leaders alone hold 26.4\% of the sandwiches on 15.7\% of the blocks, and each era's curve sits closer to the diagonal than the last. By the Marinade MIP.9 era, the curve lies close to the diagonal, indicating that sandwiches are distributed across leaders approximately in proportion to the blocks they produce.

\begin{figure}[h]
  \centering
  \includegraphics[width=\linewidth]{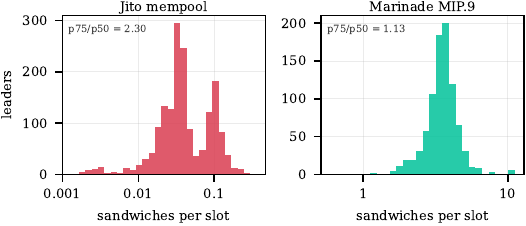}
  \caption{Per-leader sandwich rate, leaders with at least 5{,}000 blocks
  in the era. Jito's mempool was opt-in per validator, and the rate is
  correspondingly bimodal during the Jito mempool era. In later eras, the distribution becomes unimodal as the distinction between leader groups disappears.}
  \label{fig:sol-leader-bimodal}
\end{figure}

Figure~\ref{fig:sol-leader-bimodal} shows the distribution of the per-leader sandwich rate in each era, restricted to leaders with at least 5{,}000 blocks. Under the Jito mempool the distribution carries two peaks, i.e., leaders fall into two behaviors, which follows from the mempool being opt-in per validator. The ratio of the 75th to the 50th percentile lies between 2 and 3 during the Jito mempool era and falls to 1.13 afterwards, indicating that per-leader sandwich rates become substantially more homogeneous.

\subsubsection{Application Concentration}
We further investigate the concentration of sandwich attacks across applications on Solana (cf.\ Section~\ref{sec:solana-apps}).

\begin{figure}[hbtp]
  \centering
  \includegraphics[width=\linewidth]{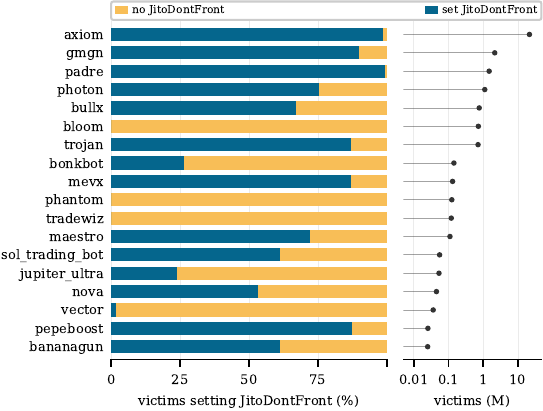}
  \caption{Application-level concentration of protected order flow sandwich victims on Solana,
  since \texttt{jitodontfront} launched (2025--04--17). For each application, we report the share of sandwiched victims carrying the
flag (left) and the total number of sandwiched transactions (right). Three of the four applications with the largest victim volumes set the flag on at least 90\% of their victims.}
  \label{fig:sol-apps}
\end{figure}

Figure~\ref{fig:sol-apps} pairs each application's adoption of \texttt{jitodontfront} with the volume of its sandwiched transactions since the flag launched on 17~April 2025. Adoption is close to binary per application, since the application sets the flag rather than the user. Axiom sets it on 98.5\% of its sandwiched victims and Padre on 99.4\%, whereas Bloom, TradeWiz and Phantom never set it. Axiom dominates by volume, and three of the four most sandwiched applications, Axiom, GMGN, and Padre, set the flag on at least 90\% of their victims. We recall that the flag only forces a flagged transaction to sit first in any Jito bundle, i.e., it rules out front-running inside a bundle and leaves every other path to the leader untouched. Setting the flag therefore does not by itself guarantee protection outside the Jito bundle path.

Table~\ref{tab:sol-app-excess-sole} repeats the excess-ratio analysis of Figure~\ref{fig:sol-app-share-excess}, restricting the numerator to single-victim sandwiches, for which there is no ambiguity over which enclosed transaction was targeted. The excess ratio decreases for every over-represented application, by roughly half for Axiom and Padre, but remains well above one. Jupiter Ultra rises to 2.0 in this subset, while unattributed flow remains below one in both analyses.

\begin{table}[hbtp]
  \centering
  \resizebox{\columnwidth}{!}{\begin{tabular}{@{}lrrrr@{}}
\toprule
\textbf{App} & \textbf{Quarters} & \textbf{Sole SWs} & \textbf{Excess} & \textbf{Excess (sole)} \\
\midrule
Axiom & 6 & 619,987 & 18.9 & 8.1 \\
Photon & 10 & 1,375,685 & 11.1 & 7.2 \\
BullX & 9 & 472,404 & 10.7 & 5.4 \\
GMGN & 9 & 395,763 & 10.6 & 8.5 \\
Padre & 5 & 40,149 & 8.8 & 3.3 \\
Trojan & 10 & 728,267 & 5.6 & 5.0 \\
BonkBot & 11 & 827,141 & 5.4 & 3.9 \\
Jupiter Ultra & 12 & 1,628,708 & 0.7 & 2.0 \\
\midrule
Unattributed & 12 & 10,498,623 & 0.5 & 0.7 \\
\bottomrule
\end{tabular}
}
  \caption{Excess ratio (victim share over swap share) per application on Solana, as the median over supported quarters, over all victims and over sole-victim sandwiches only (Sole SWs: number of such victims).}
  \label{tab:sol-app-excess-sole}
\end{table}

\end{document}